\documentclass[aps,pra,twocolumn, notitlepage,superscriptaddress,10pt]{revtex4-2}
\usepackage{stylesetting}
\usepackage{booktabs}
\usepackage{graphicx}
\usepackage{multirow}
\usepackage{enumitem}
\usepackage{adjustbox}

\usepackage[utf8]{inputenc}
\usepackage{amsmath,amsfonts,amssymb}
\usepackage{hyperref,cleveref}
\usepackage{listings,xcolor,booktabs,array}
\usepackage{pgfplots}
\usepackage{qcircuit}
\pgfplotsset{compat=1.18}
\usepackage{array}

\lstdefinelanguage{QASM3}{
  keywords={OPENQASM, qubit, bit, include, gate, measure, reset, barrier, if, else, creg, qreg, let, const, angle, int, float, duration, stretch, delay, box},
  sensitive=true,
  morecomment=[l]{//},
  morestring=[b]",
}
\begin{document}

\preprint{APS/123-QED}

\title{A Three-Layer Architecture for Fault-Tolerant Quantum Computing}

\author{Zhirao Wang}
\affiliation{Center on Frontiers of Computing Studies, Peking University, Beijing 100871, China}
\affiliation{School of Computer Science, Peking University, Beijing 100871, China}

\author{Zhou You}
\affiliation{Center on Frontiers of Computing Studies, Peking University, Beijing 100871, China}
\affiliation{School of Computer Science, Peking University, Beijing 100871, China}

\author{Yiming Huang}
\affiliation{Center on Frontiers of Computing Studies, Peking University, Beijing 100871, China}
\affiliation{School of Computer Science, Peking University, Beijing 100871, China}

\author{Tianyi Li}
\affiliation{Center on Frontiers of Computing Studies, Peking University, Beijing 100871, China}
\affiliation{School of Computer Science, Peking University, Beijing 100871, China}

\author{Ying Li}%
\email{yli@gscaep.ac.cn}
\affiliation{Graduate School of China Academy of Engineering Physics, Beijing 100193, China}

\author{Xiao Yuan}
\email[]{xiaoyuan@pku.edu.cn}
\affiliation{Center on Frontiers of Computing Studies, Peking University, Beijing 100871, China}
\affiliation{School of Computer Science, Peking University, Beijing 100871, China}

\author{Yuan Yao}
\email[]{yuan.yao@pku.edu.cn}
\affiliation{Center on Frontiers of Computing Studies, Peking University, Beijing 100871, China}
\affiliation{School of Computer Science, Peking University, Beijing 100871, China}

\date{\today}

\begin{abstract}
Fault tolerance is an indispensable prerequisite for constructing large-scale universal quantum computers. Drawing philosophies from classical computer architecture, this paper presents a hardware-agnostic three-layer high-level architectural framework for generic fault-tolerant quantum computation. Guided by the real execution workflows of fault-tolerant quantum algorithms, the proposed model is decoupled from specific physical qubit hardware platforms and quantum error correction codes, serving as a universal abstract standard rather than a platform-specific implementation scheme. Special attention is devoted to the intermediate Fault-Tolerance Layer, which serves as the architectural bridge between application-level logical programs and hardware-level execution. We systematically characterize its five internal components, the interfaces and data exchanged among them, and the execution, correction, and adaptation paths that together enable logical synthesis, fault-tolerant resources management, decoding, and runtime fault-tolerant control. An end-to-end example is further provided to illustrate the full-stack operating pipeline of fault-tolerant quantum algorithms under this framework. Given the increasing emphasis on modular, heterogeneous, and cross-layer fault-tolerant quantum systems, our architecture provides a unified foundational
model for organizing such designs.
\end{abstract}

\maketitle

\section{Introduction}
Fault tolerance, which describes the ability to execute reliable logical computations even when individual physical qubits suffer random noise errors, is a fundamental requirement for practical large-scale quantum computing~\cite{shorFaulttolerantQuantum1996, preskill1998reliable,campbellRoadsFaulttolerant2017}. Substantial relevant progress has been made in quantum algorithms,  quantum error correction(QEC) codes, and hardware platforms. Yet practical large-scale quantum computing requires not advances in these components in isolation, but coordination amongs them since different QEC codes and platforms exhibit distinct advantages and trade-offs~\cite{terhal2015quantum,mohseni2024build}.

Classical computing provides a vital point of reference, which has achieved remarkable hierarchical through decades of evolution, ranging from software, operating systems and circuit technologies~\cite{david2017computer}. However quantum computing has not yet reached an equally mature architectural stage. As noted by Ding and Chong, current quantum computing more resembles the early era of classical computing when hardware strongly constrained algorithmic realization and tight full-stack co-design is required~\cite{dingQuantumComputer2022,zhangClassicalArchitecture2023, koboriLSQCAResourceEfficient2025}. Hence hierarchical paradigms from classical computing cannot be adopted outright.

This does not imply quantum computing must be monolithic. The crux lies in adpoting the methodological discipline of classical architecture but tracing the actual fault tolerant execution trajectory of quantum algorithms. Unlike in classical architecture where error handling merely serves as a localized component, quantum architecture must account for how high-level programs are synthesized into logical operations~\cite{fowler2012surface}, how resources are allocated among magic-state factories, ancilla pools and system qubits~\cite{gidney2021factor,gidneyMagicState2024a}, how logical instructions are translated into hardware-executable operations~\cite{litinski2019game}, and how measurement outcomes and device drift are incorporated into runtime decision-making~\cite{ryan2021realization, acharyaQuantumError2025}. To rephrase, fault tolerance is an underlying subsystem permeating the whole architecture and needs to co-operate with all full-stack modules covering top-level software to bottom-level hardware~\cite{jones2012layered}.

In this work, we propose a three-layer architecture for fault-tolerant quantum computing, with particular emphasis on a meticulously designed fault-tolerance layer. Following the fault-tolerant execution trajectory of quantum algorithms, we characterize this layer as the coordinator between applications and hardware, identify its functions, define interfaces upwards and downwards, and illustrate its role through an end-to-end example. Notably, our architecture is agnostic to underlying physical hardware and specific QEC codes. Instead of delving into specific implementations, we draw a universal framework for fault-tolerant quantum computing inspired by paradigms from classical computer systems.

The rest of the paper is organized as follows. \Cref{sec:related} reviews the fundamental background of fault tolerance and summarizes key prior research in this field. \Cref{sec:overall} presents the overall architecture of our proposal, alongside its core design motivations and principles. \Cref{sec:internal} elaborates on the detailed implementation of the central fault-tolerance layer. \Cref{sec:hamiltonian_example} demonstrates the end-to-end workflow of the proposed architecture via a complete illustrative example of fault-tolerant algorithm execution. \Cref{sec:discussion} discusses open challenges and promising directions for future extension of our framework. Finally, \Cref{sec:conclusion} concludes this work.

\section{Background \& Related Work}\label{sec:related}
\subsection{QEC fundamentals on leading physical platforms}

Practical quantum computing requires fault tolerance, and this fact has direct architectural consequences~\cite{shor1995scheme}. Thus for architects intuitive comprehension of fault-tolerance’s architectural role and current development landscape outweighs exhaustive mathematical derivations.
Most fault-tolerant schemes are built on the stabilizer formalism~\cite{gottesmanClassQuantum1996}, where a logical code space is defined as the joint $+1$ eigenspace of a commuting set of Pauli operators. In practice, this means that a logical qubit is encoded into many physical qubits and must be maintained through repeated syndrome extraction and decoding. CSS codes~\cite{calderbankGoodQuantum1996, steaneSimpleQuantum1996} are especially important because they allow $X$- and $Z$-type errors to be handled separately, which simplifies both code construction and decoding.

A second architectural fact is that universal fault-tolerant computation is not obtained for free. By the Eastin-Knill theorem~\cite{eastinRestrictionsTransversal2009}, no nontrivial quantum code can support a universal set of transversal gates. As a result, while some logical gates may be relatively easy to realize fault-tolerantly, non-Clifford gates typically require additional resources, most commonly magic-state distillation and injection~\cite{bravyiUniversalQuantum2005,bravyi2012magic}. For architecture design, this means that fault tolerance is not only a matter of protecting logical memory, but also of coordinating the ancillary resources and control flow required for non-Clifford execution.

Finally, the realization of fault-tolerant quantum computing is strongly shaped
by the underlying hardware platform. Although the fundamental principles of
quantum error correction are shared across implementations, practical
fault-tolerant architectures depend critically on hardware characteristics such
as connectivity, coherence time, gate fidelity, measurement model, reset
mechanisms, and qubit mobility. Consequently, superconducting circuits~\cite{koch2007charge,ivezicIBMUnveils2023,ivezicIBMUnveils2024,acharyaQuantumError2025,gaoEstablishingNew2025,somoroffMillisecondCoherence2023,nesterovCnotGates2022}, neutral-atom arrays~\cite{bluvsteinQuantumProcessor2022a,bluvsteinLogicalQuantum2024a}, and trapped-ion systems~\cite{ciracQuantumComputations1995,monroeDemonstrationFundamental1995,winelandQuantumInformation2003,pinoDemonstrationTrappedion2021,wangSingleIon2021,mosesRaceTrackTrappedIon2023} should not be viewed merely as different physical realizations of a
common fault-tolerant stack. Their distinct hardware capabilities lead to
different architectural trade-offs and resource-management strategies. Quantum
error correction is therefore not only a coding problem but also a central
architectural consideration that links hardware capabilities to the design of
fault-tolerant execution. In the following, we examine these hardware platforms
individually and discuss the architectural implications of their fault-tolerant
operation.

\subsubsection{Superconducting circuits.}

Superconducting circuits currently provide the most mature route toward large-scale on-chip integration among leading quantum hardware platforms. Transmon-based systems, in particular, have demonstrated rapid progress in fabrication scalability, control integration, and system-level engineering, with IBM's Condor processor reaching more than one thousand monolithically integrated qubits on a single chip~\cite{ivezicIBMUnveils2023}. More recent production-scale processors, including IBM's Heron~r3 (156 qubits), Google's Willow (105 qubits), and Zuchongzhi~3.0 (105 qubits), report single-qubit gate fidelities above $99.9\%$ and two-qubit gate fidelities above $99.5\%$~\cite{ivezicIBMUnveils2024,acharyaQuantumError2025,gaoEstablishingNew2025}. At the device level, simultaneous optimization of single-qubit control, two-qubit entangling gates, and readout has also been demonstrated in transmon systems, with representative fidelities reaching $99.98\%$, $99.93\%$, and $>99.94\%$, respectively~\cite{marxer999Fidelity2026}. Meanwhile, fluxonium-based devices have emerged as a promising alternative within the superconducting platform. Their long coherence times and strong level anharmonicity support high-fidelity control, and recent experiments have reported both single-qubit and two-qubit fidelities above $99.9\%$. This makes fluxonium an increasingly relevant candidate for future fault-tolerant superconducting architectures~\cite{somoroffMillisecondCoherence2023,nesterovCnotGates2022}.

From an architectural perspective, however, the most important feature of superconducting platforms is not only their fidelity trajectory, but their geometry. Their two-dimensional planar layout with predominantly nearest-neighbor coupling strongly favors QEC schemes based on local stabilizer checks. This makes topological codes—most prominently the surface code—the natural architectural baseline for fault-tolerant superconducting systems. In the rotated planar surface code, physical qubits are arranged on a 2D lattice and alternating stabilizer checks detect $X$- and $Z$-type errors separately; a distance-$d$ logical patch requires on the order of $d^2$ physical qubits and has a circuit-level threshold around $1\%$ under depolarizing noise assumptions~\cite{kitaevFaulttolerantQuantum2003,raussendorfFaultTolerantQuantum2007,fowlerSurfaceCodes2012}. For architects, the significance of the surface code is that it couples naturally to planar layouts, repeated local measurements, and measurement-driven logical operations such as lattice surgery.

Recent experiments have strengthened this surface-code-centered picture. Krinner \emph{et al.} demonstrated repeated QEC on a distance-three surface code using 17 physical qubits, achieving a logical error probability of approximately $3\%$ per cycle~\cite{krinnerRealizingRepeated2022}. Google Quantum AI later demonstrated below-threshold scaling on distance-5 and distance-7 surface codes using the Willow processor, showing that increasing code distance suppressed logical error rates and that the distance-7 logical qubit outlived all constituent physical qubits~\cite{acharyaQuantumError2025}. IBM, in parallel, has pursued the heavy-hexagon code as a
surface-code-related subsystem-code variant adapted to heavy-hex
connectivity~\cite{chamberlandTopologicalSubsystem2020}, and has experimentally
demonstrated multi-round error correction for a distance-three logical qubit
on a heavy-hexagon superconducting lattice~\cite{sundaresanDemonstratingMultiround2023}. More recently, USTC demonstrated below-threshold surface-code error correction
on a superconducting processor using all-microwave leakage suppression, achieving
logical-error suppression with a distance-seven surface-code memory~\cite{heExperimentalQuantum2025}. These results reinforce the architectural view that superconducting fault tolerance is currently driven by planar locality, repeated syndrome extraction, and measurement-based logical operations.

At the same time, superconducting platforms also expose one of the clearest architectural tensions in fault-tolerant design. Surface-code-like schemes offer favorable locality, mature decoding methods, and a natural match to fixed-layout hardware, but they carry large physical-qubit overhead. High-rate quantum low-density parity-check (qLDPC) codes are therefore being actively explored as a potential route to reducing this overhead, in some cases by an order of magnitude relative to surface-code baselines~\cite{bravyiHighthresholdLowoverhead2024a,yoderTourGross2025}. However, many qLDPC constructions require nonlocal or long-range stabilizer checks that are difficult to realize efficiently on fixed planar superconducting hardware. The resulting tension is architectural rather than merely code-theoretic: locality favors surface-code-style organizations, whereas qubit efficiency motivates qLDPC-style organizations. Resolving this tension—whether through modular layouts, additional communication structures, code switching, or new hardware-aware code constructions—remains one of the central architectural challenges for superconducting fault-tolerant quantum computing.

\subsubsection{Neutral atom arrays.}

Neutral-atom quantum computing uses optical tweezers to trap individual atoms-typically rubidium, strontium, or ytterbium-with qubit states encoded in electronic or nuclear spin levels and entangling gates mediated by the Rydberg blockade effect~\cite{everedHighfidelityParallel2023b}. Architecturally, the defining feature of this platform is reconfigurability: tweezer arrays can be rearranged dynamically during computation, allowing qubits to be moved, reordered, and brought into interaction on demand. This lifts the fixed nearest-neighbor constraint of planar superconducting layouts and enables more flexible interaction graphs, resource movement, and zone-based execution models~\cite{bluvsteinQuantumProcessor2022a}.

These properties have direct implications for fault tolerance. First, atom loss can often be detected directly on the device and converted into located erasure errors, which substantially improves decoding performance and can raise effective fault-tolerance thresholds relative to purely unlocated Pauli noise~\cite{wuErasureConversion2022}. Second, long-range and dynamically reconfigurable connectivity makes it easier to contemplate code families and execution strategies that are difficult to realize on fixed-lattice hardware, including high-rate qLDPC codes and constant-overhead fault-tolerant architectures~\cite{pecorariHighrateQuantum2025,xuConstantoverheadFaulttolerant2024}. For architects, this means that neutral-atom systems naturally support a style of fault-tolerant organization in which encoded data, interaction regions, readout zones, and replenishment resources can be orchestrated more dynamically than in static planar devices.

The experimental progress of the platform has been correspondingly rapid. In 2024, Bluvstein \emph{et al.} demonstrated a programmable logical quantum processor with surface codes scaled to distance $d=7$, color-code operation at break-even fidelity, and 48 logical qubits encoded in $[[8,3,2]]$ blocks that outperformed physical qubits on algorithmic benchmarks~\cite{bluvsteinLogicalQuantum2024a}. Reichardt \emph{et al.} demonstrated logical computation with 24-qubit GHZ states and 28-qubit Bernstein–Vazirani circuits~\cite{reichardtLogicalComputation2024}, while Chow \emph{et al.} implemented circuit-based leakage-to-erasure conversion with detection accuracy above $93\%$~\cite{chowCircuitBasedLeakagetoErasure2024}. In 2025, Rodriguez \emph{et al.} reported the first experimental logical magic-state distillation on a neutral-atom platform, thereby completing the basic ingredients of a universal fault-tolerant gate set~\cite{salesrodriguezExperimentalDemonstration2025}. Taken together, these results suggest that neutral-atom systems are not only a promising physical platform, but also a particularly rich architectural regime for fault-tolerant computation, where movement, erasure handling, and flexible connectivity can all be incorporated into the execution model.

\subsubsection{Trapped-ion systems.}

Trapped-ion quantum computing naturally spans two closely related architectural regimes: \emph{single-chain} systems and \emph{QCCD}-style multi-zone systems. In the single-chain regime, quantum information is encoded in internal electronic or hyperfine states of ions confined by electromagnetic potentials in ultra-high-vacuum environments, with multi-qubit interactions mediated by shared motional modes of the ion chain~\cite{ciracQuantumComputations1995,monroeDemonstrationFundamental1995,winelandQuantumInformation2003}. This regime is distinguished by exceptionally high-fidelity control, long coherence times, and effectively all-to-all connectivity within a chain. State-of-the-art trapped-ion systems have demonstrated single-qubit gate fidelities above $99.999\%$, two-qubit gate fidelities of $99.97(1)\%$, and state-preparation-and-measurement errors as low as $10^{-6}$~\cite{löschnauerScalableHighFidelity2025,anHighFidelity2022}. Hyperfine qubits can remain coherent from seconds to hours, with coherence exceeding one hour in specialized settings~\cite{wangSingleIon2021}. These properties make trapped ions particularly attractive for compact fault-tolerant demonstrations and for code families that benefit from nonlocal stabilizer measurements.

For scalable fault-tolerant architectures, trapped ions are increasingly organized in the quantum charge-coupled device (QCCD) model, where ions are shuttled among multiple trapping zones for storage, interaction, and readout~\cite{pinoDemonstrationTrappedion2021,mosesRaceTrackTrappedIon2023}. This introduces a second architectural regime: rather than relying on a single shared chain, the system uses movement and zone specialization to preserve high-fidelity interactions while scaling the number of fault-tolerant resourcess. From an architectural perspective, QCCD adds a new degree of freedom absent in fixed-lattice platforms: movement can be traded against locality constraints, enabling more flexible implementations of logical operations, syndrome extraction, and protected communication.

These platform characteristics have direct implications for fault-tolerant design. Because trapped-ion systems offer effectively nonlocal connectivity together with high-fidelity control, they are especially well suited to compact fault-tolerant schemes such as Steane and color-code constructions, as well as to architectures that rely on nonlocal stabilizer measurements, logical teleportation, and lattice-surgery-like logical interactions. More broadly, the platform favors designs in which logical overhead can be reduced by exploiting connectivity and movement, rather than by forcing all fault-tolerant computation into a fixed local geometry.

Experimentally, trapped-ion platforms---and Quantinuum systems in particular---have produced a sequence of important fault-tolerant milestones. Ryan-Anderson \emph{et al.} demonstrated real-time fault-tolerant quantum error correction in the Steane code in 2021~\cite{ryan-andersonRealizationRealTime2021}. Egan \emph{et al.} subsequently demonstrated fault-tolerant control of an error-corrected qubit~\cite{eganFaulttolerantControl2021}. In 2022, Postler \emph{et al.} demonstrated a universal set of fault-tolerant logical operations, including a logical $T$ gate via magic-state injection, on logical qubits encoded in color codes~\cite{postlerDemonstrationFaulttolerant2022}. More recently, Ryan-Anderson \emph{et al.} demonstrated high-fidelity logical teleportation using transversal protocols and lattice surgery on a trapped-ion quantum computer~\cite{ryan-andersonHighfidelityFaulttolerant2024}, while Daguerre \emph{et al.} experimentally demonstrated high-fidelity logical magic states prepared through code switching~\cite{daguerreExperimentalDemonstration2025}. Together, these results make trapped ions one of the clearest current examples of a platform where fault-tolerant architecture, logical operations, and fault-tolerant algorithmic execution are beginning to connect end-to-end.

\subsubsection{Overall}
Taken together, these platforms show that fault-tolerant quantum computer architecture cannot be derived from a single preferred code or a single preferred hardware model. Superconducting systems currently favor locally checkable codes such as the surface code and heavy-hexagon variants, but face a persistent architectural trade-off between locality and qubit efficiency. Neutral-atom platforms naturally support more dynamic execution models, where qubit movement, reconfigurable geometry, and erasure-aware operation enable different resource organizations and higher-rate codes, at the cost of new system-level challenges in transport scheduling, zone coordination, and replenishment. Trapped-ion platforms, with their exceptional fidelities and effectively all-to-all connectivity, relax many of the locality constraints that dominate planar devices, but replace them with a different set of architectural questions involving ion movement, zone scheduling, readout orchestration, and multi-region fault-tolerant execution.

For architecture design, the key conclusion is therefore not that one platform is universally superior, but that each platform pushes fault-tolerant execution toward a different organization of resources, logical operations, and runtime feedback. A useful systems architecture must be able to accommodate this diversity without collapsing into platform-specific ad hoc design. This is precisely why an explicit fault-tolerance layer is valuable: it provides a stable architectural abstraction in which platform-specific QEC strategies can be expressed, coordinated, and exposed through standardized interfaces, while preserving a common structure across the overall quantum computing stack.


\subsection{Existing system stack proposals}
\label{subsec:existing-stacks}

Existing proposals for the quantum computing system can be broadly divided into five groups. At the top of the stack are programming languages and software development kits which focus on expressing quantum algorithms independently of specific hardware backends. Below them are compiler stacks and toolchains, including intermediate representations synthesized from high-level languages and resource estimator. A third line of work targets fault-tolerant compilation more directly, translating logical circuits into code-specific fault-tolerant execution patterns and allocating the fault-tolerant resources. A fourth layer focuses on real-time decoding and relevent microarchitecture, emphasizing low-latency control and feedback in the QEC loop. Besides, there are proposals aiming to address cross-platform abstraction interfaces and these counts as the fifth class. Table \ref{tab:stack-comparison} contrasts the architectural coverage of these existing quantum stack solutions.Collectively, these efforts either focus on the co-design of limited closely coupled layers, or are highly bound to particular fault-tolerance schemes and physical platforms. We will further elaborate on each candidate work below.

\begin{table*}[t]
\centering
\caption{Comparison of coverage of existing quantum computing system stack proposals}
\label{tab:stack-comparison}
\renewcommand{\arraystretch}{1.15}
\small
\begin{adjustbox}{max width=\linewidth}
\begin{tabular}{@{} c c c c c c c @{}}
\toprule
\parbox[c]{4.0cm}{\textbf{System}} &
\parbox[c]{3.0cm}{\textbf{Programming \& \\ Optimization}} &
\parbox[c]{3.0cm}{\textbf{Resource \\ Estimation}} &
\parbox[c]{3.0cm}{\textbf{FT Compilation \& Allocation}} &
\parbox[c]{3.0cm}{\textbf{Cross-Platform}} &
\parbox[c]{3.0cm}{\textbf{Real-Time Decoding}} &
\parbox[c]{3.0cm}{\textbf{Runtime Feedback}} \\
\midrule
\parbox[c]{4.0cm}{Q\#~\cite{svoreEnablingScalable2018}}
& \Large $\checkmark$ & \Large $\checkmark$ & \Large $\bigcirc$ & \Large $\checkmark$ & \Large $\times$ & \Large $\times$ \\
\parbox[c]{4.0cm}{qiskit~\cite{javadi2024quantum}}
& \Large $\checkmark$ & \Large $\bigcirc$ & \Large $\bigcirc$ & \Large $\checkmark$ & \Large $\times$ & \Large $\checkmark$ \\
\parbox[c]{4.0cm}{PennyLane~\cite{bergholmPennyLaneAutomatic2022}}
& \Large $\checkmark$ & \Large $\checkmark$ & \Large $\bigcirc$ & \Large $\checkmark$ & \Large $\times$ & \Large $\times$ \\
\parbox[c]{4.0cm}{Cirq~\cite{ho2018announcing}}
& \Large $\checkmark$ & \Large $\bigcirc$ & \Large $\bigcirc$ & \Large $\times$ & \Large $\times$ & \Large $\times$ \\
\parbox[c]{4.0cm}{Quipper~\cite{green2013quipper}}
& \Large $\checkmark$ & \Large $\checkmark$ & \Large $\times$ & \Large $\times$ & \Large $\times$ & \Large $\times$ \\
\parbox[c]{4.0cm}{ProjectQ~\cite{steiger2018projectq}}
& \Large $\checkmark$ & \Large $\checkmark$ & \Large $\times$ & \Large $\times$ & \Large $\times$ & \Large $\times$ \\
\parbox[c]{4.0cm}{BenchQ~\cite{zapatabenchqfull}}
& \Large $\times$ & \Large $\checkmark$ & \Large $\bigcirc$ & \Large $\checkmark$ & \Large $\times$ & \Large $\times$ \\
\parbox[c]{4.0cm}{Quilc~\cite{smith2020open}}
& \Large $\checkmark$ & \Large $\times$ & \Large $\bigcirc$ & \Large $\times$ & \Large $\times$ & \Large $\times$ \\
\parbox[c]{4.0cm}{Flexion~\cite{yinFlexionAdaptive2025}}
& \Large $\times$ & \Large $\checkmark$ & \Large $\bigcirc$ & \Large $\times$ & \Large $\times$ & \Large $\times$ \\
\parbox[c]{4.0cm}{TKET~\cite{sivarajah2021tuket}}
& \Large $\times$ & \Large $\bigcirc$ & \Large $\checkmark$ & \Large $\checkmark$ & \Large $\times$ & \Large $\times$ \\
\parbox[c]{4.0cm}{Lattice Surgery Compile~\cite{watkins2024high}}
& \Large $\times$ & \Large $\checkmark$ & \Large $\checkmark$ & \Large $\times$ & \Large $\times$ & \Large $\times$ \\
\parbox[c]{4.0cm}{Chipmunq~\cite{wegmannChipmunqFaultTolerant2026}}
& \Large $\times$ & \Large $\bigcirc$ & \Large $\checkmark$ & \Large $\times$ & \Large $\times$ & \Large $\times$ \\
\parbox[c]{4.0cm}{TopQAD~\cite{schererAutomatedDesign2026}}
& \Large $\times$ & \Large $\checkmark$ & \Large $\checkmark$ & \Large $\times$  & \Large $\times$ & \Large $\times$ \\
\parbox[c]{4.0cm}{RISC-Q~\cite{liu2026scalableopensourceqecsubmicrosecond}}
& \Large $\times$ & \Large $\times$ & \Large $\bigcirc$ & \Large $\checkmark$ & \Large $\bigcirc$ & \Large $\bigcirc$ \\
\parbox[c]{4.0cm}{QECool, QuLATIS~\cite{uenoQECOOLOnLine2021,uenoQULATISQuantum2022a}}
& \Large $\times$  & \Large $\times$  & \Large $\times$  & \Large $\times$  & \Large $\checkmark$ & \Large $\checkmark$ \\
\parbox[c]{4.0cm}{MQSS~\cite{burgholzer2026munich}}
& \Large $\checkmark$ & \Large $\bigcirc$ & \Large $\bigcirc$ & \Large $\checkmark$ & \Large $\times$ & \Large $\bigcirc$ \\
\parbox[c]{4.0cm}{IBM Quantum Starling~\cite{yoderTourGross2025}}
& \Large $\checkmark$ & \Large $\checkmark$ & \Large $\bigcirc$ & \Large $\times$ & \Large $\checkmark$ & \Large $\checkmark$ \\
\parbox[c]{4.0cm}{QIR-EE~\cite{wong2025cross}}
& \Large $\times$ & \Large $\bigcirc$ & \Large $\bigcirc$ & \Large $\checkmark$ & \Large $\times$ & \Large $\times$ \\
\parbox[c]{4.0cm}{CUDA-Q~\cite{caldwell2025platform}}
& \Large $\checkmark$ & \Large $\bigcirc$ & \Large $\bigcirc$ & \Large $\checkmark$ & \Large $\times$ & \Large $\times$ \\

\bottomrule
\multicolumn{7}{@{}c}{ {\Large $\checkmark$}: Fully covered \quad {\Large $\bigcirc$}: Partially covered or platform-locked \quad {\Large $\times$}: Not covered}\\
\end{tabular}
\end{adjustbox}
\end{table*}

\subsubsection{Programming languages, software development kits}
The first class of proposals includes works that provide logical representations of algorithms, together with associated toolkits for development. These frameworks typically expose high-level programming interfaces and translate high-level quantum programs into standardized intermediate representations such as OpenQASM\cite{cross2022openqasm} or QIR~\cite{qirbook-full} to coordinate with lower layer. Microsoft's Q\# is a standalone domain-specific language (DSL) distinguished by its native integration with the Azure Quantum Resource Estimator for automatic physical resource derivation~\cite{svoreEnablingScalable2018}. IBM's Qiskit centers on a transpiler for automatic qubit mapping and routing across diverse backends, complemented by Qiskit Runtime for session management and machine-status feedback~\cite{javadi2024quantum}. Xanadu's PennyLane originated as a quantum machine learning framework and has since expanded into general-purpose quantum programming. It provides an \texttt{estimator} module for industrial-scale resource estimation and the Catalyst JIT compiler, with broad hardware coverage through its plugin architecture~\cite{bergholmPennyLaneAutomatic2022}. Google's Cirq offers fine-grained NISQ circuit control and automatic ancilla management via \texttt{QubitManager}~\cite{ho2018announcing}. Quipper is a Haskell-embedded DSL with a built-in resource estimator for scalable circuit description and complexity analysis~\cite{green2013quipper}. ProjectQ is an extensible compiler framework with a \texttt{ResourceCounter} backend and automatic qubit-lifecycle management across multiple hardware backends~\cite{steiger2018projectq}.

Combined with the overview in Table~\ref{tab:stack-comparison}, we observe that while these frameworks are primarily designed to express algorithmic logic and most support native resource estimation at the algorithm layer, they universally lack automated fault-tolerant compilation pipelines, and few integrate native fault-tolerant resource allocation or end-to-end real-time error correction capabilities.

\subsubsection{Intermediate representations}

The second class goes deeper away user, including intermediate representations synthesized from high-level languages and resource estimator. These tools sit below the programming frameworks, translating logical circuits into lower-level representations or estimating the physical resources required for fault-tolerant execution. BenchQ leverages the Azure Quantum Resource Estimator to derive physical requirements including qubit count, T-factory overhead and runtime across superconducting and trapped-ion modalities, and incorporates a graph-state compiler for estimation preprocessing~\cite{zapatabenchqfull}. Quilc is Rigetti's quantum compiler, which automatically maps, routes, and schedules gates from Quil to native processor instructions through a multi-pass optimization pipeline~\cite{smith2020open}. Flexion is a prototype toolchain cited in surface-code resource-comparison studies, providing partial compilation and logical-resource analysis~\cite{yinFlexionAdaptive2025}.

These toolchains exhibit a narrower scope than the programming frameworks: BenchQ decouples estimation from compilation, Quilc decouples compilation from fault tolerance and resource awareness, and Flexion remains insufficiently documented to assess architectural completeness. None closes the gap between logical algorithm specification and fault-tolerant compilation.

\subsubsection{Fault-tolerant compilation and execution}

The third class focuses on fault-tolerant compilation and design automation, translating logical algorithms into error-corrected physical implementations. Quantinuum's TKET features a retargetable optimization pipeline and support for diverse backends including IBM, Rigetti, IonQ, and Quantinuum. Its hardware-agnostic intermediate representation enables aggressive circuit optimization, though fault-tolerant compilation and real-time decoding remain outside its scope~\cite{sivarajah2021tuket}. The Lattice Surgery Compiler translates arbitrary quantum circuits into surface code lattice surgery operations, automatically managing patch allocation, ancilla routing, and magic state factory placement, and outputs a resource estimate of space-time volume and cycle counts~\cite{watkins2024high}. Chipmunq is a hardware-aware compiler for modular chiplet superconducting architectures, mapping surface-code circuits onto chiplet networks through partitioning, sequencing, and fidelity-aware routing~\cite{wegmannChipmunqFaultTolerant2026}. TopQAD is a system-level design suite comprising a compiler, an assembler for physical layout refinement, and a resource estimator that optimizes space-time trade-offs and configures magic state distillation factories~\cite{schererAutomatedDesign2026}.

These tools remain offline frameworks: the Lattice Surgery Compiler requires a hardware-specific backend for physical execution, Chipmunq is bound to chiplet architectures without runtime integration, and TopQAD operates at the design exploration stage. None incorporates real-time decoding or runtime machine-state feedback.

\subsubsection{Hardware microarchitecture and real-time decoding}
The fourth class comprises control microarchitectures and real-time decoding systems that bridge compiled instructions with physical hardware execution. RISC-Q is a generator for quantum control system-on-chips based on the open RISC-V instruction set, producing classical processors that orchestrate qubit control pulses through custom instructions and memory-mapped I/O. It enables rapid prototyping of control electronics for superconducting, trapped-ion, and neutral-atom platforms by parameterizing peripheral modules such as radio-frequency signal generators and analog-to-digital converters~\cite{liu2026scalableopensourceqecsubmicrosecond}. QECool is an online decoding hardware to decode surface code syndromes within each QEC cycle~\cite{uenoQECOOLOnLine2021}. Its distributed architecture assigns one decoding unit per ancilla qubit, using spike-signal propagation to perform minimum-weight perfect matching in parallel across the lattice. QuLATIS extends this approach to lattice surgery scenarios, optimizing the decoding pipeline for patch operations, multi-qubit measurements, and the higher-dimensional syndrome spaces that arise during logical gate execution~\cite{uenoQULATISQuantum2022a}.


These systems are essential because they make repeated syndrome extraction and feedback physically realizable. However, they approach the problem from the low levels of the stack upward. Their focus is on low-latency control, decoder implementation, and microarchitectural support for QEC cycles, rather than on the broader coordination of application structure, logical synthesis, non-Clifford resource provisioning, and runtime resource adaptation across the full system stack. As a result, they solve a crucial component of fault-tolerant execution, but do not by themselves define the higher-level architectural layer that integrates fault-tolerant services with the rest of the system.

\subsubsection{Full-stack architecture}

The fifth class represents the most comprehensive proposals, aiming to unify programming, compilation, resource management, and hardware execution across multiple platforms. These architectures come closest to the integrated design target of this work. MQSS is an open-source full-stack platform comprising a front-end language interface, a multi-level compiler, an HPC-integrated scheduler, and the Quantum Device Management Interface (QDMI) for vendor-agnostic hardware access~\cite{burgholzer2026munich}. Its modular design anticipates fault-tolerant computing through support for mid-circuit measurements and varied qubit encodings, yet automatic QEC compilation and real-time decoding are still under development. IBM Quantum Starling is a planned modular fault-tolerant quantum computer system targeting 2029, featuring a full-stack architecture from algorithm compilation through logical qubit execution to real-time syndrome decoding~\cite{yoderTourGross2025}. Its heavy-hexagon qLDPC code design reduces physical qubit overhead by approximately ninety percent compared to surface codes, and the system integrates magic state distillation, module-to-module classical communication, and hardware-aware compilation. QIR-EE is a cross-platform runtime developed by Oak Ridge National Laboratory, executing QIR programs through LLVM and the XACC framework across IBM, IonQ, and Quantinuum backends~\cite{wong2025cross}. While it provides a unified execution layer, it offers no fault-tolerant compilation, no resource estimation, and no runtime hardware-state feedback. NVIDIA's CUDA-Q is unique in targeting \emph{quantum supercomputing}, bridging quantum processing units (QPUs) with classical GPU acceleration~\cite{caldwell2025platform}. It support microsecond-latency syndrome decoding via the NVQLink interconnect and offers the most extensive hardware-backend coverage, including IonQ, Quantinuum, IQM, OQC, Pasqal, and QuEra.

These architectures advance toward integration but remain fragmented: MQSS spans the full stack but leaves fault tolerance incomplete; Starling promises the most complete integration yet remains a future system; QIR-EE provides cross-platform execution without fault-tolerant or resource-aware capabilities. CUDA-Q relies on external tools for resource analysis. None of them spans the entire pipeline from high-level algorithm, fault-tolerant compilation to real-time decoding.

\subsection{The gap}
\label{subsec:the-gap}

Existing system proposals have already established many important building blocks for fault-tolerant quantum computing, but they typically emphasize either \emph{platform specialization} or \emph{function specialization}. 

On the one hand, several systems are tightly co-designed around a particular hardware platform and code family. IBM's stack targets superconducting qubits with qLDPC-based architectures; FT-QuMA focus on superconducting platforms with surface-code-style execution; TISCC and Flexion are developed around trapped-ion QCCD assumptions; neutral-atom ones, including QuEra and Harvard's processor, remain experimental demonstrations without unified compiler or framework. These efforts demonstrate that deep co-design can be highly effective, but they also make software abstractions, decoding strategies, and hardware interfaces difficult to transfer across platforms.

On the other hand, many existing proposals remain focused on a specific portion of the stack. Programming frameworks such as Q\#, Qiskit, and PennyLane provide strong abstractions for algorithm expression and logical circuit construction, but leave fault-tolerant execution largely to lower layers. Compiler stacks and FT compilers such as MQSS, TKET, and Jabalizer improve logical lowering and fault-tolerant circuit generation, but usually produce static schedules with limited runtime coordination. Microarchitectures such as FT-QuMA and RISC-Q address low-latency control and decoding, but assume that logical schedules and resource assignments have already been fixed. Hardware roadmaps and modular architectures such as IBM Starling and QuIRC, meanwhile, address physical organization and connectivity, but do not define a general abstraction for logical fault-tolerant execution. As a result, the interfaces between adjacent layers often remain implicit or narrowly tailored to a particular design point.

The gap, therefore, is not the absence of useful frameworks, but the absence of a sufficiently comprehensive architectural framework that connects these pieces into a coherent end-to-end stack. What remains under-specified is a common architectural description of how algorithmic intent is progressively refined into fault-tolerant execution, how resources are allocated and re-allocated across encoded memory, ancilla pools, and non-Clifford factories, how runtime feedback propagates across layer boundaries, and how different modules can evolve without forcing redesign of the entire system.

\section{Overall Three-Layer Architecture}\label{sec:overall}
Motivated by the gap discussed in the preceding section, this work proposes a comprehensive layered architecture centered on a dedicated fault‑tolerance layer between the application stack and the hardware backend. By defining standardized interfaces between adjacent layers and among internal modules, it provides a unified framework in which software toolchains, fault-tolerance strategies, and hardware platforms can be coordinated more systematically than in existing layer-specific or platform-specific designs.

\begin{figure*}[t]
    \centering
    \includegraphics[width=0.75\textwidth]{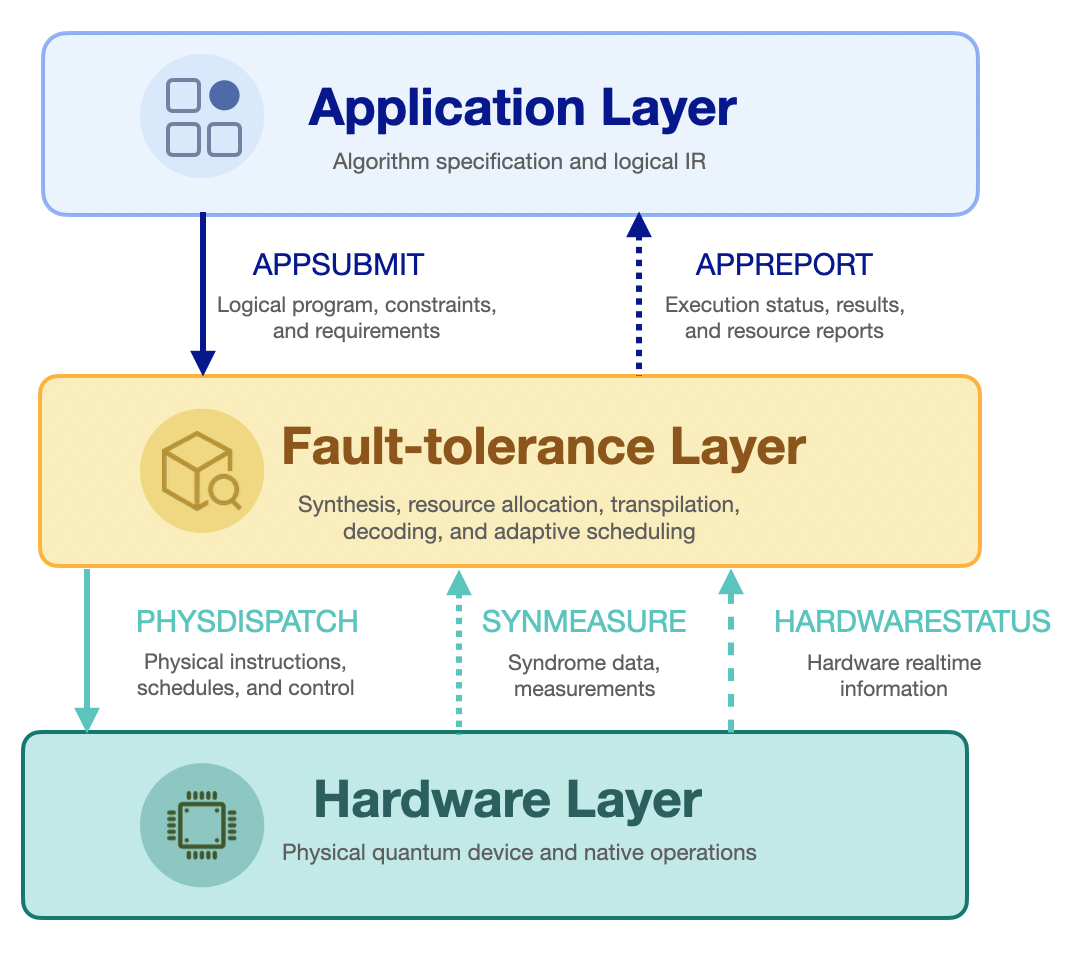}
    \caption{
    Overall three-layer architecture. The Application Layer is responsible for algorithm specification and logical intermediate representation (IR) generation. The Fault-Tolerance Layer is the architectural center of the stack and coordinates synthesis, fault-tolerant resources allocation, transpilation, decoding, and adaptive scheduling. The Hardware Layer abstracts the physical quantum device and native operations. The downward interfaces are \textsc{AppSubmit}, which carries logical program descriptions and constraints from the Application Layer to the Fault-Tolerance Layer, and \textsc{PhysDispatch}, which carries backend-executable physical instructions from the Fault-Tolerance Layer to the Hardware Layer. The upward interfaces are \textsc{AppReport}, which returns execution status, results, and resource reports to the application side, \textsc{SynTelemetry}, which returns syndrome and measurement data for decoding, and \textsc{HardwareStatus}, which returns device-level runtime information for higher-level adaptation.}
    \label{fig:qecframe}
\end{figure*}

\subsection{Three-Layer Stack Overview}
\label{subsec:three-layer}
The proposed architecture is organized as a three-layer stack, shown in Fig.~\ref{fig:qecframe}. This division reflects the natural separation among three distinct responsibilities in practical quantum computing: expressing the user problem at the algorithmic level, coordinating its fault-tolerant realization, and executing the resulting fault-tolerant operations on a concrete physical device. Rather than treating fault tolerance as a backend implementation detail, our architecture elevates it to an explicit middle layer that mediates between application intent and hardware execution.

The architecture comprises three principal layers:
\paragraph{Application Layer.}
The Application Layer is the uppermost tier and concerns primarily with user interaction. It accepts application-level inputs like algorithms in Q\# or circuit descriptions in openQASM~3 and translates them into a hardware-agnostic logical circuit description $\mathcal{C}_{\mathrm{log}}$. This layer reasons in terms of algorithm structure, logical gates, and problem semantics, but makes no commitment to how fault tolerance is realized: it does not choose QEC codes, allocate fault-tolerant resources, manage ancillas or non-Clifford factories, or reason about calibration drift and hardware diagnostics. It remains application-facing and problem-facing throughout.
\paragraph{Fault-Tolerance Layer.}
The Fault-Tolerance Layer is the architectural center of the stack and the main systems contribution of this work. It is responsible for coordinating all of the issues that arise when a logical quantum program is to be executed fault-tolerantly, including logical synthesis, fault-tolerant resources allocation, logical-to-physical lowering, decoding, correction handling, and runtime adaptation. These functions are implemented by modular subsystems that communicate through explicit data objects.

\paragraph{Hardware Layer.}
The Hardware Layer is the lowest tier and provides an abstracted view of the underlying physical quantum device. Its purpose is to expose execution and feedback capabilities in a standardized form, while hiding backend-specific details of how those capabilities are physically realized. The role of the Hardware Layer is therefore not to erase hardware diversity, but to abstract it: superconducting, neutral-atom, trapped-ion, and other backends may implement the same architectural contracts in different ways while preserving a common interface to the upper layers.

\subsection{Cross-Layer Communication}
\label{subsec:interfaces}

The architecture uses five cross-layer communication channels. These channels are protocol specifications rather than implementation mandates. Concrete realizations may use shared memory, remote procedure calls (RPC), message queues, or other communication mechanisms, depending on deployment topology. Their purpose is to make the layer boundaries explicit by specifying what information crosses them, in what direction, and with what semantics.

\paragraph{\textsc{AppSubmit} (Application $\rightarrow$ Fault-Tolerance).}
This interface carries the logical description of the computation from the Application Layer into the Fault-Tolerance Layer. Its primary payload is a hardware-agnostic logical circuit or program representation, together with optional execution constraints including target error rate, latency budget, qubit budget, or application-level optimization goals. Conceptually, \textsc{AppSubmit} defines the contract by which the application requests fault-tolerant execution without prescribing how that execution is realized.

\paragraph{\textsc{AppReport} (Fault-Tolerance $\rightarrow$ Application).}
This interface returns application-visible execution information from the Fault-Tolerance Layer to the Application Layer. Its payload may include progress notifications, logical completion signals, final measurement results, resource-utilization reports, and execution summaries. It may also carry escalation alerts when the requested execution target cannot be met under current runtime and resource conditions, for example due to insufficient available resources, persistent hardware degradation, or failure to satisfy a target reliability threshold. In this sense, \textsc{AppReport} allows the Application Layer to observe outcomes and revise computation goals without direct exposure to low-level hardware detail.

\paragraph{\textsc{PhysDispatch} (Fault-Tolerance $\rightarrow$ Hardware).}
This interface carries backend-executable execution commands from the Fault-Tolerance Layer into the Hardware Layer. Its payload typically consists of batches of physical instructions or backend-level control objects, together with timing constraints, measurement schedules, synchronization requirements, and other execution metadata required by the device. The interface does not prescribe how these commands are generated internally within the Fault-Tolerance Layer; it only specifies the downward contract by which fault-tolerant execution is dispatched to the hardware backend.

\paragraph{\textsc{SynMeasure} (Hardware $\rightarrow$ Fault-Tolerance).}
This interface returns syndrome and measurement data from the Hardware Layer into the Fault-Tolerance Layer. Its payload includes syndrome measurement streams, readout data, and other measurement-side information needed for interpreting the progress of error correction during execution. The role of \textsc{SynMeasure} is focused on the fast reaction path: it supplies the measurement information from which correction or frame-update information can be inferred.

\paragraph{\textsc{HardwareStatus} (Hardware $\rightarrow$ Fault-Tolerance).}
This interface returns execution and device level status information from the Hardware Layer into the Fault-Tolerance Layer. Its payload may include calibration drift indicators, gate-fidelity degradation, coherence-time variation, transport failures, execution stalls, or other backend-level runtime diagnostics that do not directly enter the syndrome-measurement stream but may require higher-level adaptation. In this sense, \textsc{HardwareStatus} is distinct from \textsc{SynMeasure}: the latter carries measurement data for correction, whereas \textsc{HardwareStatus} carries device-state information for runtime adaptation.

These cross-layer channels define the external architectural contract of the three-layer stack. \textsc{AppSubmit} carries application intent into the fault-tolerant stack; \textsc{PhysDispatch} carries executable commands into the hardware backend; \textsc{SynMeasure} and \textsc{HardwareStatus} return two distinct classes of hardware-derived information for fast correction and slower adaptation, respectively; and \textsc{AppReport} closes the loop back to the Application Layer by exposing execution outcomes, resource conditions, and escalation events. This explicit interface structure allows the architecture to remain modular, extensible, and platform-agnostic at the system level.

\section{Five Internal Components of the Fault-Tolerance Layer}\label{sec:internal}

Following the three-layer stack overview and the cross-layer communication model, we now turn to the internal organization of the Fault-Tolerance Layer. This layer is decomposed into five components: \textit{Synthesizer}, \textit{Resource Allocator}, \textit{Transpiler}, \textit{Decoder}, and \textit{Controller}. These modules operate around a dual-ISA abstraction: a logical instruction stream $\mathrm{ISA}_L$ that captures fault-tolerance-ready logical operations, and a physical instruction stream $\mathrm{ISA}_P$ that drives the hardware layer. Their interaction follows carefully designed dataflows that separate forward refinement, fast correction, and slow adaptation. Table~\ref{tab:ft-layer-modules} concisely summarizes the role of each module.

\subsection{Internal Module Data Flow Graph}
\label{subsec:ft-layer-internal}

\begin{figure*}[t]
    \centering
    \includegraphics[width=0.7\textwidth]{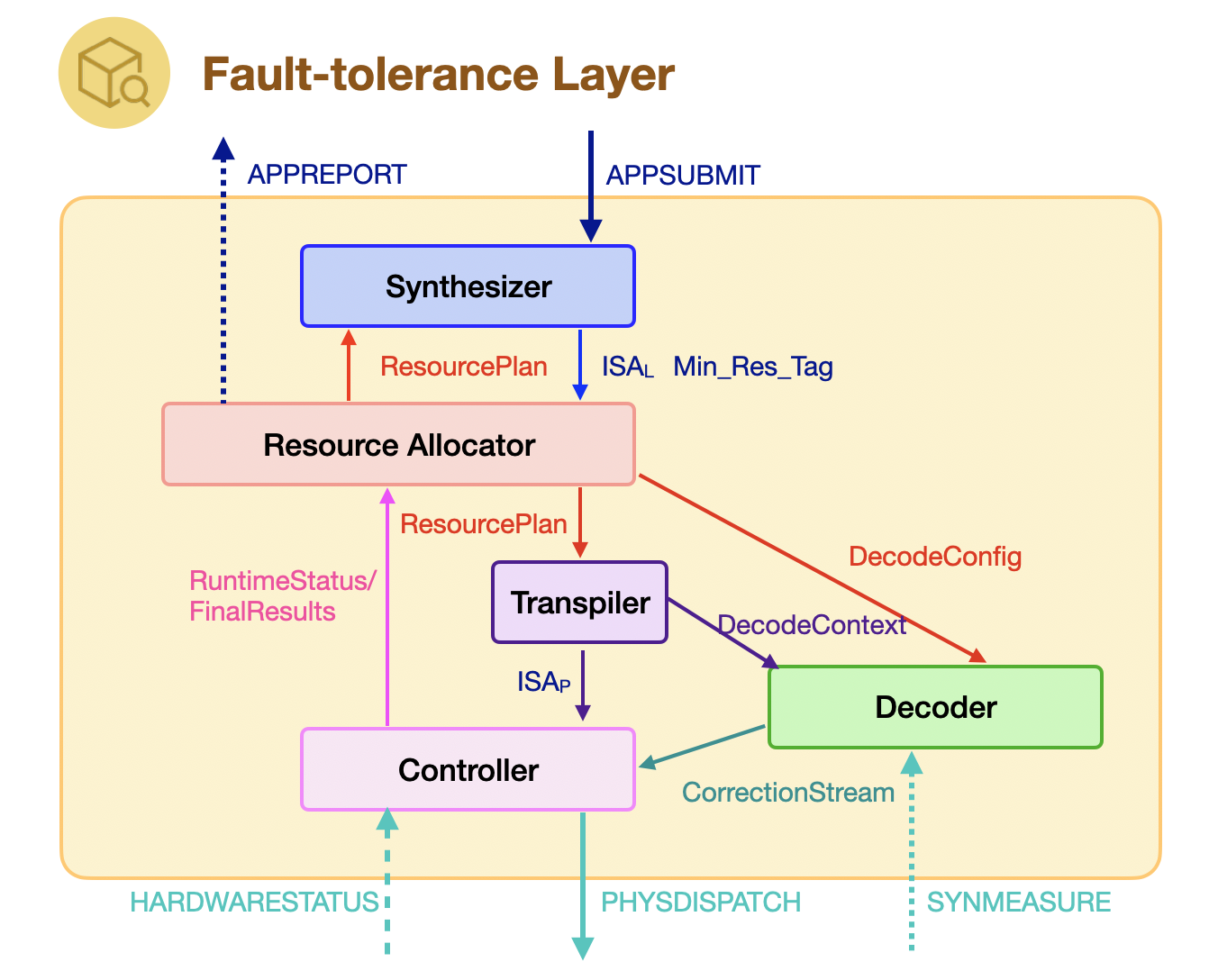}
    \caption{
    Internal module of the Fault-Tolerance Layer. The layer is decomposed into five functional modules: Synthesizer, Resource Allocator, Transpiler, Decoder, and Controller. The arrows indicate the principal internal data flows among these modules. In particular, the Synthesizer emits the logical instruction stream $\mathrm{ISA}_L$ to the Resource Allocator and receives \textit{Budget} hints in return; the Resource Allocator emits a fault-tolerant \textit{ResourcePlan} to the Transpiler and a \textit{DecodeConfig} object to the Decoder; the Transpiler emits backend-executable instructions $\mathrm{ISA}_P$ to the Controller and a \textit{DecodeContext} object to the Decoder; the Decoder emits a \textit{CorrectionStream} to the Controller; and the Controller returns \textit{RuntimeStatus} to the Resource Allocator. Cross-layer interfaces are also shown: \textsc{AppSubmit} enters from the Application Layer, while \textsc{PhysDispatch}, \textsc{SynMeasure}, and \textsc{HardwareStatus} connect the layer to the Hardware Layer.
    }
    \label{fig:ftlayer-internal}
\end{figure*}

\begin{table*}[t]
\renewcommand{\arraystretch}{1.5}
\centering
\caption{Summary of the five modules in the Fault-Tolerance Layer.}
\label{tab:ft-layer-modules}
\small
\begin{tabular}{cccc}
\toprule
\textbf{Module} & \textbf{Input} & \textbf{Output} & \textbf{Function} \\
\midrule
Synthesizer
&
\parbox[c]{3.0cm}{$\mathcal{C}_{\mathrm{log}}$, program constraints}
&
\parbox[c]{3.0cm}{$\mathrm{ISA}_L$ stream, Min\_Res\_Tag}
&
\parbox[c]{8.0cm}{Produces a fault-tolerance-ready logical instruction stream under budget-aware synthesis.}
\\[4pt] \hline

Resource Allocator
&
\parbox[c]{3.0cm}{\vspace{2pt} $\mathrm{ISA}_L$, Min\_Res\_Tag, \textit{RuntimeStatus}}
&
\parbox[c]{3.0cm}{\vspace{2pt} \textit{ResourcePlan} $\mathcal{R}$, \textit{DecodeConfig}}
&
\parbox[c]{8.0cm}{\vspace{2pt} Selects a fault-tolerant resources plan that satisfies the logical error budget and minimizes resource usage.}
\\[4pt] \hline

Transpiler
&
\parbox[c]{3.0cm}{\textit{ResourcePlan} $\mathcal{R}$}
&
\parbox[c]{3.0cm}{$\mathrm{ISA}_P$, \textit{DecodeContext}}
&
\parbox[c]{8.0cm}{\vspace{2pt} Lowers the fault-tolerant plan into backend-executable physical instructions and decoding context.}
\\[4pt] \hline

Decoder
&
\parbox[c]{3.0cm}{\vspace{2pt} \textsc{SynMeasure}, \textit{DecodeConfig}, \textit{DecodeContext}\vspace{2pt} }
&
\parbox[c]{3.0cm}{\textit{CorrectionStream}}
&
\parbox[c]{8.0cm}{Converts syndrome data into a fast correction stream for runtime execution.}
\\[4pt] \hline 

Controller
&
\parbox[c]{3.0cm}{\vspace{2pt} $\mathrm{ISA}_P$, \textit{CorrectionStream}, \textsc{HardwareStatus}}
&
\parbox[c]{3.0cm}{\textsc{PhysDispatch}, \textit{RuntimeStatus}}
&
\parbox[c]{8.0cm}{Merges physical instructions with correction feedback and coordinates runtime dispatch.}
\\
\bottomrule
\end{tabular}
\end{table*}

The Fault-Tolerance Layer is organized around explicit module boundaries rather than implicit compiler-style call chains. Figure~\ref{fig:ftlayer-internal} should therefore be read first as a \emph{structural} diagram: it identifies the five internal components and the principal data objects they exchange. At this stage, the emphasis is not yet on the overall runtime semantics, but on how responsibility is partitioned inside the layer. The Synthesizer and Resource Allocator jointly define the fault-tolerant execution plan, the Transpiler lowers that plan into backend-executable instructions while also producing decoding context, the Decoder translates syndrome information into correction information, and the Controller coordinates runtime execution while also providing slower-timescale status updates back into the planning path.

\subsection{Dual-ISA and Internal Data Objects}
\label{subsec:dual-isa-ftlayer}
The internal organization of the layer is built around a small number of explicit data objects. Their purpose is to decouple the five components while still preserving a coherent end-to-end execution path.

The most important abstraction is the dual-ISA structure. The first level, $\mathrm{ISA}_L$, is the fault-tolerance-ready logical instruction stream produced by the Synthesizer. It captures the logical computation in a form suitable for fault-tolerant execution, but remains independent of any specific backend realization. The second level, $\mathrm{ISA}_P$, is the backend-executable physical instruction stream produced by the Transpiler. It represents the point at which fault-tolerant logical execution is translated into concrete hardware actions. This dual-ISA separation creates a stable abstraction boundary: the Synthesizer and Resource Allocator operate above it, while the Hardware Layer consumes only the lower-level physical form.

Beyond the two instruction abstractions, the layer uses several auxiliary objects:
\begin{itemize}
    \item Interfaces
        \begin{itemize}[leftmargin=0pt]
             \item \textsc{AppSubmit}: The interface takes as input the hardware-agnostic logical circuit from the Application Layer, optionally supplemented by additional execution information;
            \item \textsc{PhysDispatch}: The interface forwards batched physical instructions and execution commands from the Controller to the Hardware Layer;
            \item \textsc{SynMeasure}: The interface recieves syndrome and measurement data returned by the Hardware Layer;
            \item \textsc{HardwareStatus}: The interface receives device-level runtime status, which is sent by the Hardware Layer to the Controller.
        \end{itemize}
    \item Internal objects
        \begin{itemize}[leftmargin=0pt]
            \item $\mathcal{C}_{\mathrm{log}}$: the hardware-agnostic logical circuit received from the Application Layer;
            \item \textit{Budget}: Resource and feasibility hints returned by the Resource Allocator to the Synthesizer;
            \item $\mathrm{ISA}_L$ and $\mathrm{ISA}_P$: The logical instruction stream and physical instruction stream;
            \item $\mathcal{R}$ (\textit{ResourcePlan}): A fault-tolerant resources and scheduling plan emitted by the Resource Allocator;
            \item \textit{DecodeConfig}: Decoder-side configuration information emitted by the Resource Allocator;
            \item \textit{DecodeContext}: Execution- and layout-specific decoding context emitted by the Transpiler;
            \item \textit{CorrectionStream}: Fast-path correction or Pauli-frame information emitted by the Decoder;
            \item \textit{RuntimeStatus}: Slower-timescale runtime summaries emitted by the Controller to the Resource Allocator.
        \end{itemize}

\end{itemize}

Architecturally, these objects fall into three categories: planning objects ($\mathcal{C}_{\mathrm{log}}$, \textit{Budget}, $\mathrm{ISA}_L$, and $\mathcal{R}$), runtime execution objects ($\mathrm{ISA}_P$, \textsc{PhysDispatch}, and \textit{CorrectionStream}), and runtime feedback objects (\textsc{SynMeasure}, \textsc{HardwareStatus}, \textit{DecodeConfig}, \textit{DecodeContext}, and \textit{RuntimeStatus}). This categorization is useful because it anticipates the three higher-level paths introduced later in this section.

\subsection{Synthesizer}
\label{subsec:circuit-synthesizer}

\begin{figure*}[t]
    \centering
    \includegraphics[width=\linewidth]{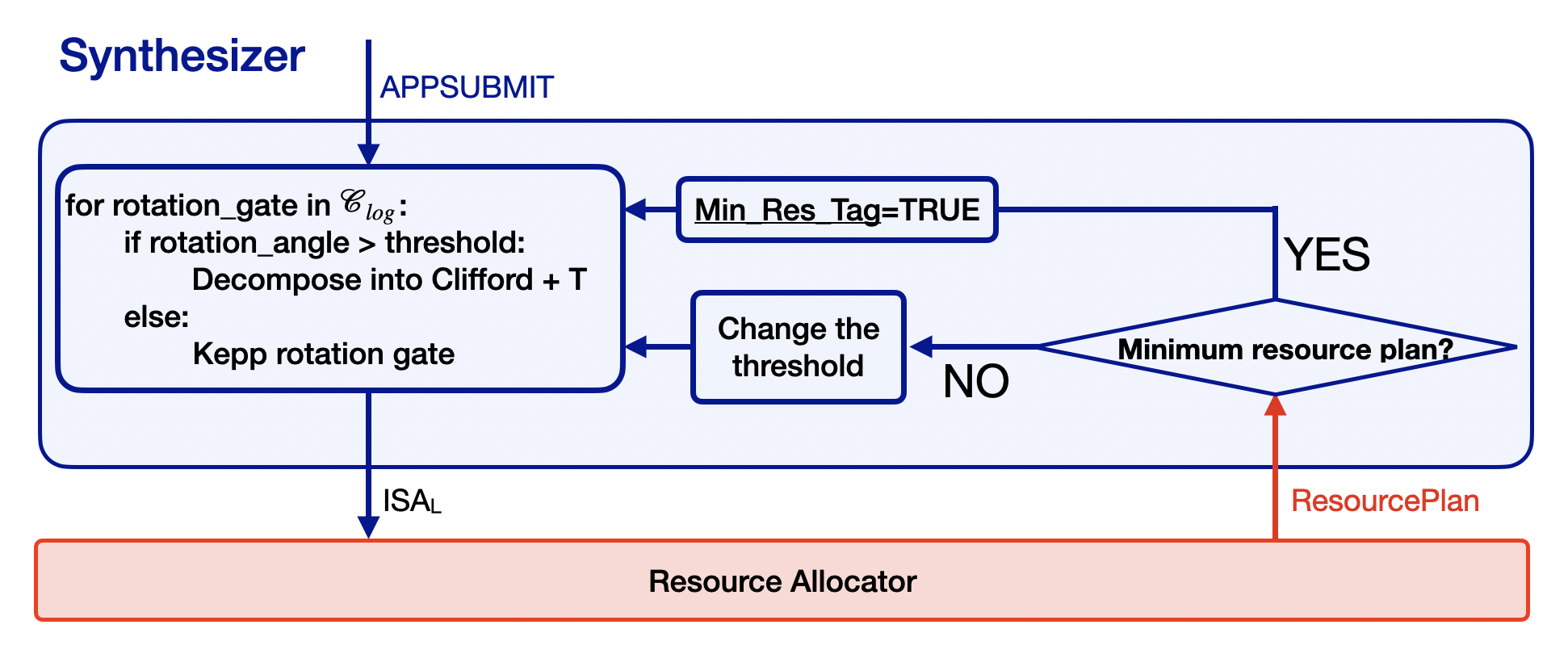}
    \caption{
    Internal logic of the Synthesizer. The module receives the logical circuit through \textsc{AppSubmit} and applies a threshold-based synthesis rule to each rotation gate. For a given threshold $\theta_{\mathrm{th}}$, a rotation is either kept as a direct logical rotation primitive or decomposed into a Clifford+$T$ representation. The resulting candidate logical instruction stream $\mathrm{ISA}_L$ is then checked against the current resource budget by the Resource Allocator. If the candidate is feasible, the current threshold is accepted and the resulting $\mathrm{ISA}_L$ is finalized; otherwise, the threshold is adjusted and synthesis is repeated.
    }
    \label{fig:synthesizer}
\end{figure*}

The Synthesizer is the entry point of the Fault-Tolerance Layer. It receives the
hardware-agnostic logical circuit $\mathcal{C}_{\mathrm{log}}$ and associated
application-level constraints from the Application Layer through \textsc{AppSubmit}.
Its role is to transform the application-level logical program into a
fault-tolerance-ready logical instruction stream $\mathrm{ISA}_L$, together with synthesis
metadata such as non-Clifford demand summaries, approximation annotations, and
strategy choices.

Architecturally, the Synthesizer determines \emph{how} the logical computation should
be expressed over the available fault-tolerant logical primitives, without yet
deciding \emph{where} or \emph{with what fault-tolerant resources} the program will be executed. In the present design, this decision is centered on non-Clifford rotation
gates. For each rotation gate with angle $\theta$ appearing in
$\mathcal{C}_{\mathrm{log}}$, the Synthesizer applies a threshold-based synthesis rule:
\begin{align*}
\label{synthesis-rule}
    &|\theta| > \theta_{\mathrm{th}}
\;\Rightarrow\;
\text{decompose into Clifford+}T, \\
&|\theta| \le \theta_{\mathrm{th}}
\;\Rightarrow\;
\text{keep as a logical rotation primitive}.
\end{align*}
Here, $\theta_{\mathrm{th}}$ is a continuously adjustable synthesis parameter that controls the trade-off between direct logical rotations and magic-state-intensive Clifford+$T$ decompositions.

As illustrated in Fig.~\ref{fig:synthesizer}, the Synthesizer operates in a optimization loop with the Resource Allocator. For a current threshold value $\theta_{\mathrm{th}}$, the Synthesizer first generates a candidate decomposition of the logical circuit and emits the corresponding candidate instruction stream $\mathrm{ISA}_L$ to the Resource Allocator. The Allocator then act as a cost-function oracle to evaluates the resource consumption of the candidate. If the consumption has reached a minimum, the current synthesized instruction stream becomes the finalized $\mathrm{ISA}_L$ used by downstream modules. Otherwise the Allocator returns the resource cost to the Synthesizer which adjusts $\theta_{\mathrm{th}}$ via an optimization procedure (e.g., gradient-based updates) until the minimum is reached. In this sense, the threshold is not chosen once and for all, but iteratively optimized through continuous interaction with the resource layer.

Finally the Synthesizer outputs a resource-aware logical program representation. It preserves application-level intent while selecting a realizable non-Clifford strategy regime, and it provides the downstream layers with an explicit logical instruction stream together with the metadata needed for fault-tolerant resources allocation.

\subsection{Resource Allocator}
\label{subsec:resource-allocation}

\begin{figure*}[t]
    \centering
    \includegraphics[width=\linewidth]{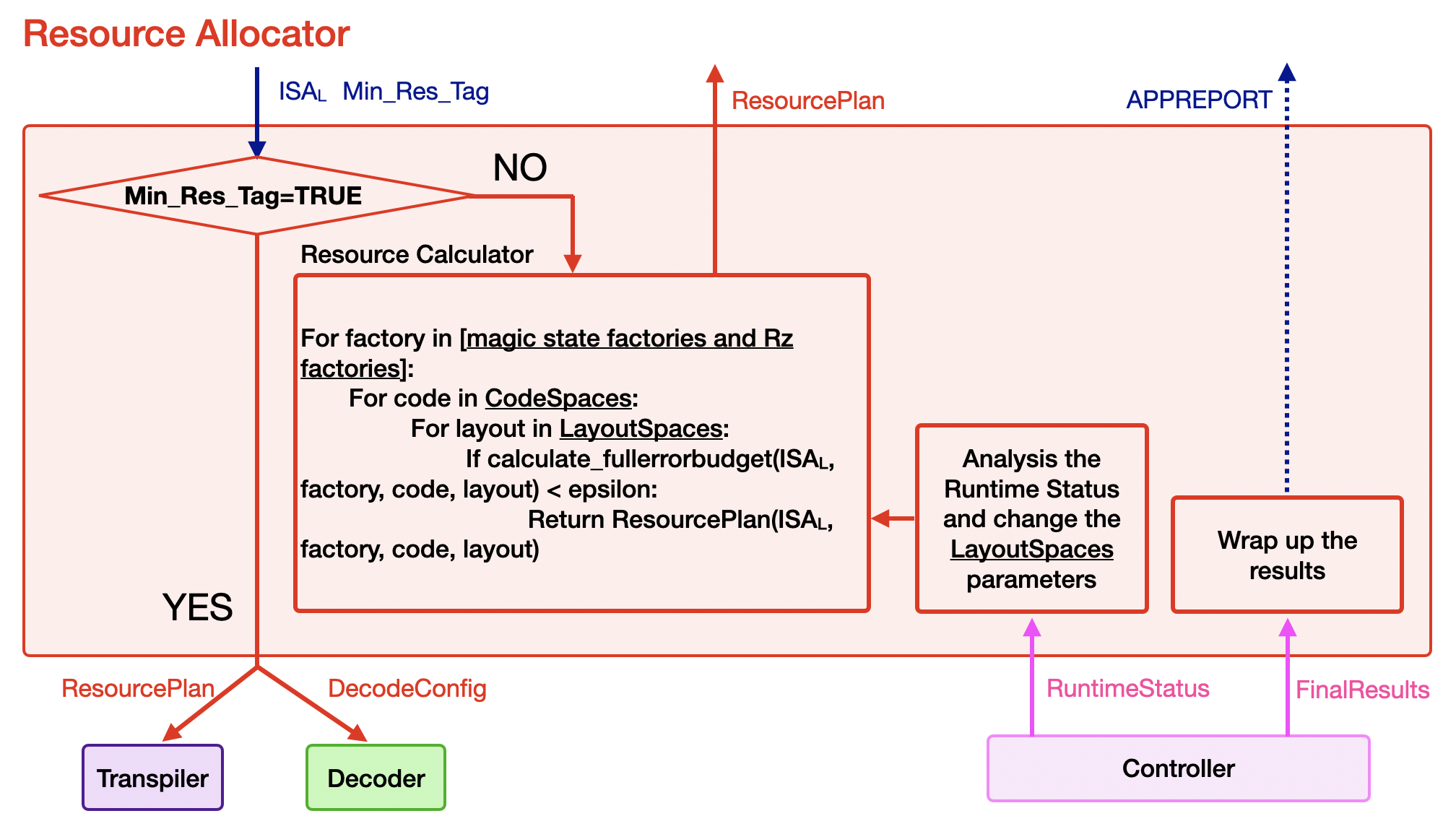}
    \caption{
    Internal logic of the Resource Allocator. The allocator receives a candidate
    logical instruction stream $\mathrm{ISA}_L$ from the Synthesizer and explores a joint
    design space over factory configurations, code realizations, and layout
    candidates. For each candidate, it selects code parameters from the target
    reliability budget, evaluates factory throughput and layout constraints, and
    estimates the resulting spacetime cost. Feasible realizations are retained as
    valid \textit{ResourcePlan} candidates, and the minimum-cost plan is selected.
    The resulting \textit{ResourcePlan} is sent to the Transpiler, while the
    corresponding \textit{DecodeConfig} is sent to the Decoder.
    }
    \label{fig:allocator}
\end{figure*}

The Resource Allocator determines how the candidate logical program produced by the Synthesizer can be realized in fault-tolerant form, with associated resource consumption and an error estimate for that realization, as summarized in Fig.~\ref{fig:allocator}. In the present architecture, the Synthesizer emits a
candidate logical instruction stream $\mathrm{ISA}_L$ under a current synthesis
threshold, and the Allocator evaluates whether that candidate is feasible under
the available resource and reliability constraints. 

This interaction preserves a clear separation of concerns. The Synthesizer is
responsible for selecting logical realizations, while the Resource Allocator is
responsible for evaluating their resource feasibility. The feedback channel
allows these decisions to be coordinated without merging the two functions into
a single monolithic component.

At the resource-planning level, the Allocator searches jointly over three design
dimensions: factory configuration, code realization, and layout organization. A
candidate realization is represented by
\begin{equation}
(\mu,c,\ell)
\in
\mathcal{F}\times\mathcal{K}\times\mathcal{L}(c,\mu),
\label{eq:alloc-search-space}
\end{equation}
where $\mu\in\mathcal{F}$ denotes a factory configuration, $c\in\mathcal{K}$
denotes a code realization, and $\ell\in\mathcal{L}(c,\mu)$ denotes a layout
candidate compatible with the selected code, factory footprint, routing
geometry, and backend connectivity. The dependence of $\mathcal{L}$ on $c$ and
$\mu$ is important: the available layouts are constrained by patch size,
routing geometry, factory placement, and hardware topology.

The Allocator begins from a run-level algorithmic failure budget
$\epsilon_{\mathrm{alg}}$. This budget is divided into a storage component and
an operation component,
\begin{equation}
\epsilon_{\mathrm{alg}}
=
\epsilon_{\mathrm{store}}
+
\epsilon_{\mathrm{op}},
\qquad
\epsilon_{\mathrm{store}}
=
\alpha\,\epsilon_{\mathrm{alg}},
\qquad
\epsilon_{\mathrm{op}}
=
(1-\alpha)\,\epsilon_{\mathrm{alg}},
\label{eq:error-budget-split}
\end{equation}
where $\alpha\in[0,1]$ is a configurable allocation parameter. The storage
budget determines the required protection strength of the code, while the
operation budget constrains logical operations, resource-state preparation,
injection protocols, and approximation error.

For a concrete example, consider the rotated planar surface code. A commonly
used scaling model for the logical memory error per cycle is
\begin{equation}
p_{\mathrm{L}}^{\mathrm{sc}}(d,p)
\approx
A
\left(
\frac{p}{p_{\mathrm{th}}^{\mathrm{sc}}}
\right)^{\lfloor(d+1)/2\rfloor},
\label{eq:surface-logical-error}
\end{equation}
where $p$ is the physical error rate,
$p_{\mathrm{th}}^{\mathrm{sc}}$ is the surface-code threshold,
and $A$ is a code- and decoder-dependent prefactor.

If the execution contains $N_{\mathrm{mem}}$ simultaneously protected logical
blocks stored for $D_{\mathrm{mem}}$ effective fault-tolerant cycles, then the
storage contribution is estimated as
\begin{equation}
\epsilon_{\mathrm{store}}^{\mathrm{est}}
=
N_{\mathrm{mem}}
D_{\mathrm{mem}}
p_{\mathrm{L}}^{\mathrm{sc}}(d,p).
\label{eq:storage-error}
\end{equation}

The code distance is chosen such that
\begin{equation}
\epsilon_{\mathrm{store}}^{\mathrm{est}}
\le
\epsilon_{\mathrm{store}}.
\label{eq:storage-budget-condition}
\end{equation}

Solving this condition yields the minimum admissible distance
\begin{equation}
d_{\min}
=
\left\lceil
2
\frac{
\log \!\left(
A N_{\mathrm{mem}} D_{\mathrm{mem}}
/
\epsilon_{\mathrm{store}}
\right)
}{
\log \!\left(
p_{\mathrm{th}}^{\mathrm{sc}}/p
\right)
}
-1
\right\rceil .
\label{eq:min-distance-allocator}
\end{equation}

This is only one possible sizing rule, but it illustrates how reliability
requirements determine code parameters.

The operation budget is evaluated separately. For a candidate logical program,
the Allocator estimates
\begin{equation}
\epsilon_{\mathrm{op}}^{\mathrm{est}}
=
\sum_{g\in G_{\mathrm{FT}}}
N_g\,p_g(d)
+
\epsilon_{\mathrm{fac}}
+
\epsilon_{\mathrm{inj}}
+
\epsilon_{\mathrm{approx}},
\label{eq:operation-error}
\end{equation}
where $N_g$ is the number of logical operations of type $g$,
$p_g(d)$ is the corresponding logical failure probability,
$\epsilon_{\mathrm{fac}}$ is the resource-state production contribution,
$\epsilon_{\mathrm{inj}}$ is the injection contribution, and
$\epsilon_{\mathrm{approx}}$ is the synthesis approximation error. The
candidate is operation-feasible only if
\begin{equation}
\epsilon_{\mathrm{op}}^{\mathrm{est}}
\le
\epsilon_{\mathrm{op}}.
\label{eq:operation-budget-condition}
\end{equation}

The same candidate also induces a non-Clifford resource demand. Let $M_T$
denote the number of $T$-state consumptions implied by the current candidate
$\mathrm{ISA}_L$, and let $T_{\mathrm{exec}}$ be the target execution duration measured
in fault-tolerant cycles. The required throughput is
\begin{equation}
\lambda_T
=
\frac{M_T}{T_{\mathrm{exec}}}.
\label{eq:t-throughput}
\end{equation}

If a factory type $f$ has footprint $n_f(d)$, output rate $r_f(d)$, and output
error $p_f(d)$, and if $k_f$ copies of that factory are instantiated, then the
throughput condition is
\begin{equation}
\sum_f
k_f r_f(d)
\ge
\lambda_T,
\label{eq:factory-throughput}
\end{equation}
with corresponding factory footprint
\begin{equation}
N_{\mathrm{fac}}(\mu,d)
=
\sum_f
k_f n_f(d).
\label{eq:factory-footprint}
\end{equation}

The same logic extends naturally to direct-$R_z$ resource factories and other
specialized non-Clifford resource-generation mechanisms.

The Allocator must also account for the physical-qubit footprint induced by the
layout. For a rotated surface-code patch with dedicated syndrome ancillas, a
distance-$d$ logical patch occupies
\begin{equation}
n_{\mathrm{patch}}(d)
=
d^2+(d^2-1)
=
2d^2-1.
\label{eq:patch-qubits}
\end{equation}

If the execution uses $N_{\mathrm{patch}}$ logical data or ancilla patches,
incurs routing overhead $N_{\mathrm{route}}(\ell,d)$, and reserves factory
footprint $N_{\mathrm{fac}}(\mu,d)$, then the total physical-qubit footprint is
\begin{equation}
N_{\mathrm{phys}}(\mu,d,\ell)
=
N_{\mathrm{patch}}
n_{\mathrm{patch}}(d)
+
N_{\mathrm{route}}(\ell,d)
+
N_{\mathrm{fac}}(\mu,d).
\label{eq:physical-footprint}
\end{equation}

The execution duration contains both useful computation and waiting overhead,
\begin{equation}
D_{\mathrm{exec}}(\mu,d,\ell)
=
D_{\mathrm{gate}}
+
D_{\mathrm{QEC}}
+
D_{\mathrm{wait}},
\label{eq:execution-depth}
\end{equation}
where $D_{\mathrm{gate}}$ counts logical-operation windows,
$D_{\mathrm{QEC}}$ counts syndrome-extraction cycles, and
$D_{\mathrm{wait}}$ accounts for stalls caused by limited factory throughput or
buffer availability. Consequently, factory allocation affects both footprint
and execution duration: adding factories increases qubit usage, whereas
insufficient throughput increases waiting time.

A candidate realization is feasible only if both storage and operation budgets
are satisfied:
\begin{equation}
\Omega
=
\left\{
(\mu,c,\ell)
\;\middle|\;
\epsilon_{\mathrm{store}}^{\mathrm{est}}
\le
\epsilon_{\mathrm{store}},
\;
\epsilon_{\mathrm{op}}^{\mathrm{est}}
\le
\epsilon_{\mathrm{op}}
\right\}.
\label{eq:alloc-feasible-set}
\end{equation}

Among feasible candidates, the Allocator selects the realization with minimum
spacetime volume,
\begin{equation}
(\mu^\ast,c^\ast,\ell^\ast)
=
\arg\min_{(\mu,c,\ell)\in\Omega}
V_{\mathrm{ST}}(\mathrm{ISA}_L;\mu,c,\ell),
\label{eq:alloc-opt-objective}
\end{equation}
where
\begin{equation}
V_{\mathrm{ST}}
=
N_{\mathrm{phys}}(\mu,d,\ell)
\,
D_{\mathrm{exec}}(\mu,d,\ell).
\label{eq:spacetime-cost}
\end{equation}

This objective makes the role of the previous quantities explicit: the error
budget determines the required code parameters and allowable factory
configurations, the layout determines the physical-qubit footprint, and the
factory throughput determines whether non-Clifford demand appears as additional
hardware resources, additional waiting time, or both.

Once the Allocator confirms that a candidate satisfies the error budget, it packages the candidate into a \textit{ResourcePlan}
\begin{equation}
\mathcal{R}
=
\mathcal{R}
\!\left(
\mathrm{ISA}_L,V_{ST};
\mu^\ast,
c^\ast,
d^\ast,
\ell^\ast
;
\kappa
\right),
\label{eq:resource-plan}
\end{equation}
where $\mu^\ast$ denotes the selected factory configuration,
$c^\ast$ the selected code realization,
$d^\ast$ the corresponding code distance (or analogous protection parameter),
,$\ell^\ast$ the selected layout organization and $\kappa$ the associated resource overhead. The \textit{ResourcePlan} specifies how the logical instruction stream
$\mathrm{ISA}_L$ should be realized and how much resource it will consume. It returned to the Synthesizer for optimality checking. If the Synthesizer determines that the resource consumption of this candidate reaches a minimum, it sets the $\texttt{Min\_Res\_Tag}$ flag to $\texttt{TRUE}$ and sends the finalized plan back to the Allocator, which then forwards it to the Transpiler.

If no feasible candidate exists in the current search space, the Allocator does
not force an invalid realization downstream. Instead, it returns infeasibility
or tightened budget information to the Synthesizer, thereby triggering another
round of threshold adjustment and logical re-synthesis. This design keeps
logical synthesis and resource allocation separate while allowing them to
cooperate through a well-defined budget-feedback interface.

In addition to this compile-time role, the Resource Allocator may receive
\textit{RuntimeStatus} updates from the Controller at predefined
reconfiguration boundaries. These updates do not interrupt the currently
executing \textit{ResourcePlan}. Instead, they are used to revise the plan for
subsequent execution phases, for example by changing active resource pools,
altering region assignments, or rebalancing factory usage under evolving
hardware conditions. The Allocator is therefore responsible both for
compile-time resource planning and for slower-timescale resource
reconfiguration during execution. Finally, execution outcomes and runtime summaries contained in \textit{FinalResults} are aggregated by the Resource Allocator and reported to the Application Layer through the \textsc{AppReport} interface.
\subsection{Transpiler}
\label{subsec:transpiler}

\begin{figure}[t]
    \centering
    \includegraphics[width=\linewidth]{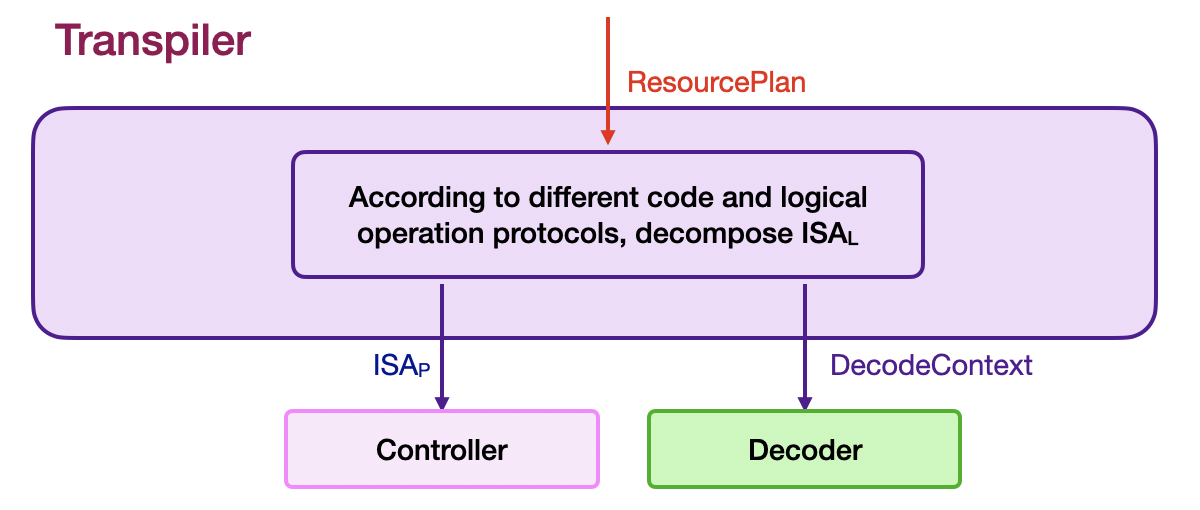}
    \caption{
    Internal logic of the Transpiler. The module receives a \textit{ResourcePlan} from the Resource Allocator. This plan determines the active code realization, fault-tolerant resources organization, and scheduled logical operations to be executed. The Transpiler then lowers the corresponding logical instruction stream according to the selected code family and logical-operation protocol, producing two outputs: a backend-executable physical instruction stream $\mathrm{ISA}_P$ for the Controller and a \textit{DecodeContext} object for the Decoder.
    }
    \label{fig:transpiler}
\end{figure}

The Transpiler is the main lowering boundary between logical execution
and backend-specific realization. It receives \textit{ResourcePlan}
$\mathcal{R}$ from the Resource Allocator and produces two outputs: a
backend-executable physical instruction stream $\mathrm{ISA}_P$ sent to the Controller,
and a \textit{DecodeContext} object sent to the Decoder.

From an architectural perspective, the Transpiler is not solely responsible for translating logical instructions into lower-level commands. What matters fundamentally is to enforce this translation within \emph{the specific code family and logical-operation protocol designated by the resource layer}.

Specifically, the lowering is determined jointly by the active code family and the corresponding logical-operation protocol through which that code is implemented. For one realization, a logical instruction may
be lowered into lattice-surgery operations; for another, into transversal
operations, gauge-fixing steps, teleportation-style primitives, or other
code-dependent execution subroutines. In this sense, the Transpiler is the
principal code-specific and backend-specific component of the Fault-Tolerance
Layer. Its architectural importance lies precisely in absorbing these details so
that the modules above it can continue to reason in terms of code-agnostic
fault-tolerant execution plans.

Beyond native gate operations, the physical instruction stream $\mathrm{ISA}_P$ produced by the Transpiler may incorporate timing constraints, measurement ordering, synchronization boundaries, reset operations, syndrome-extraction windows, transport primitives, as well as additional backend-specific execution directives required for faithful physical execution.

At the same time, the Transpiler exports a \textit{DecodeContext} object to the
Decoder. This object provides the execution-side context required to interpret
incoming syndrome and measurement data correctly under the current fault-tolerance
realization. Depending on the code and protocol, it may include stabilizer or
check layout, measurement ordering, region assignment, active logical-operation
context, or other code-specific decoding-side information.

As a result, the lowering of a given fault-tolerant computation may vary significantly across code families, layout settings, magic factories and hardware platforms, while the remaining components of the Fault-Tolerance Layer uniformly interact through a common abstraction boundary.
This design principle reflects the core architectural mandate of the Transpiler: to confine code- and backend-dependent execution logic within a well-defined lowering boundary, while maintaining stable and consistent higher-level interfaces for the layers above.

\subsection{Decoder}
\label{subsec:decoder}

\begin{figure}[t]
    \centering
    \includegraphics[width=\linewidth]{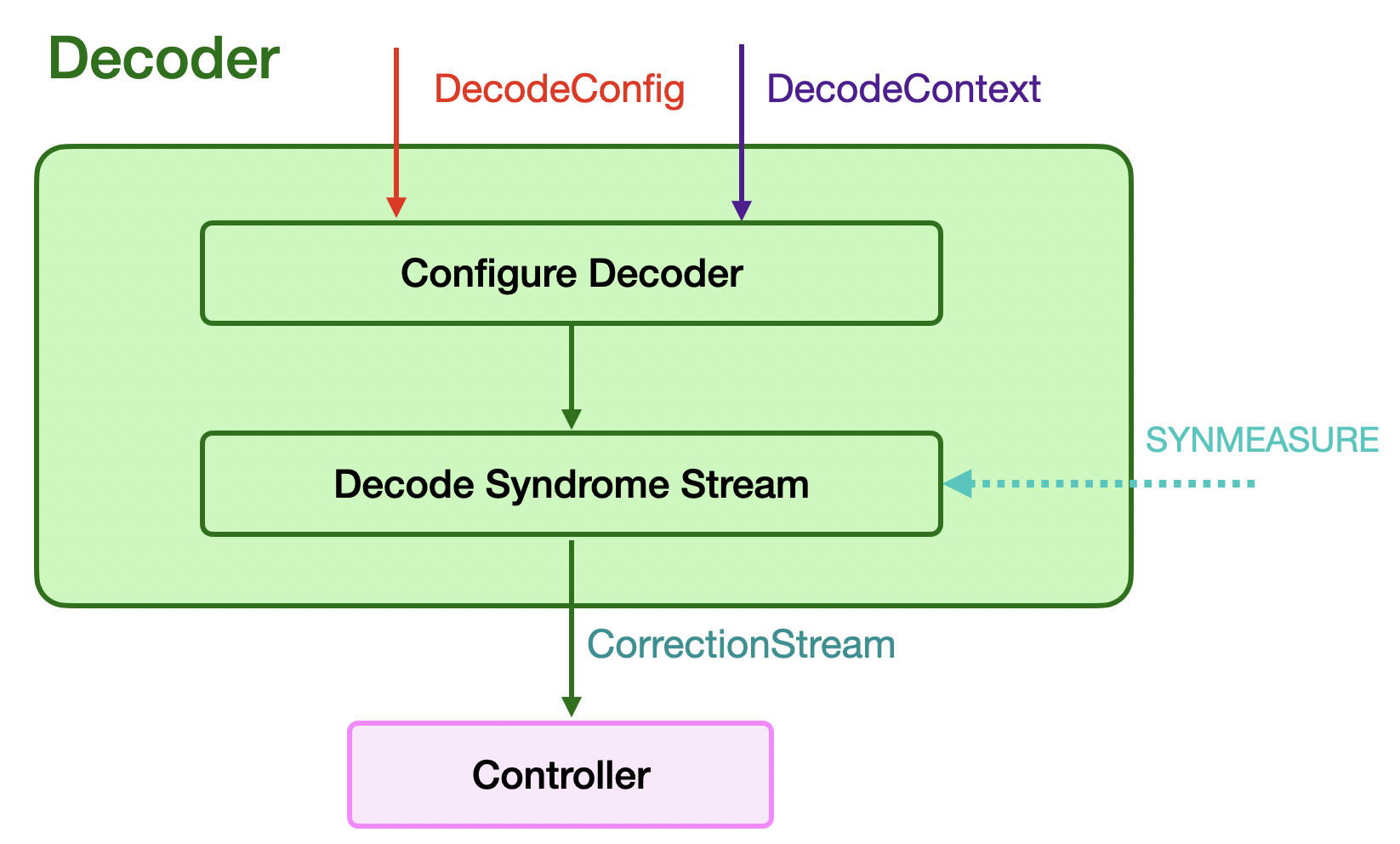}
    \caption{
    Internal logic of the Decoder. The Decoder first receives a \textit{DecodeConfig} object from the Resource Allocator and a \textit{DecodeContext} object from the Transpiler, which together determine the active decoding mode, code-specific interpretation rules, and execution context. It then consumes \textsc{SynMeasure} from the Hardware Layer and runs the configured decoding procedure on the incoming syndrome and measurement stream. The resulting \textit{CorrectionStream} is sent to the Controller for runtime execution.
    }
    \label{fig:decoder}
\end{figure}

The Decoder is intentionally narrow in scope. Its role is to translate hardware-facing
syndrome and measurement information into correction information that can be consumed
by the runtime control path. In the present architecture, the Decoder receives three
inputs: \textit{DecodeConfig} from the Resource Allocator, \textit{DecodeContext}
from the Transpiler, and \textsc{SynMeasure} from the Hardware Layer. Its primary output is a fast-path \textit{CorrectionStream} sent to the Controller.

Architecturally, the Decoder is best understood as a configurable decoding service.
The \textit{DecodeConfig} object specifies the decoder-side configuration implied by
the currently selected fault-tolerance realization, such as the active code family, decoder mode, or any code-dependent parameter settings. The \textit{DecodeContext} object provides the execution-side context needed to interpret the syndrome stream correctly, such as stabilizer layout, measurement ordering, region assignment, or other ayout-dependent decoding information. These two inputs serve to establish the conditions governing the interpretation of the incoming \textsc{SynMeasure} stream.

Once configured, the Decoder processes the syndrome and measurement stream emitted by
the Hardware Layer and applies the active decoding procedure. Depending on the code
family and deployment setting, this procedure may operate in block, sliding-window,
or streaming mode. However, the architectural contract remains unchanged: the Decoder
consumes measurement information and emits correction information  \textit{CorrectionStream} which may represent a physical correction, a Pauli-frame update, or an equivalent runtime correction object to controller. 

What matters architecturally is not the particular code family, resource allocation, decode algorithm or their resulting various syndrome formats. Instead, the design focus of this module is to establish a clear interface that specifies precisely what information is required for decoding and exactly where the Decoder sits within the overall fault‑tolerance stack. This separation allows the Decoder backend to be tailored to the active code, latency target, throughput requirement, and available classical resources, while preserving its fixed architectural role.
\subsection{Controller}
\label{subsec:controller}

\begin{figure}[t]
    \centering
    \includegraphics[width=\linewidth]{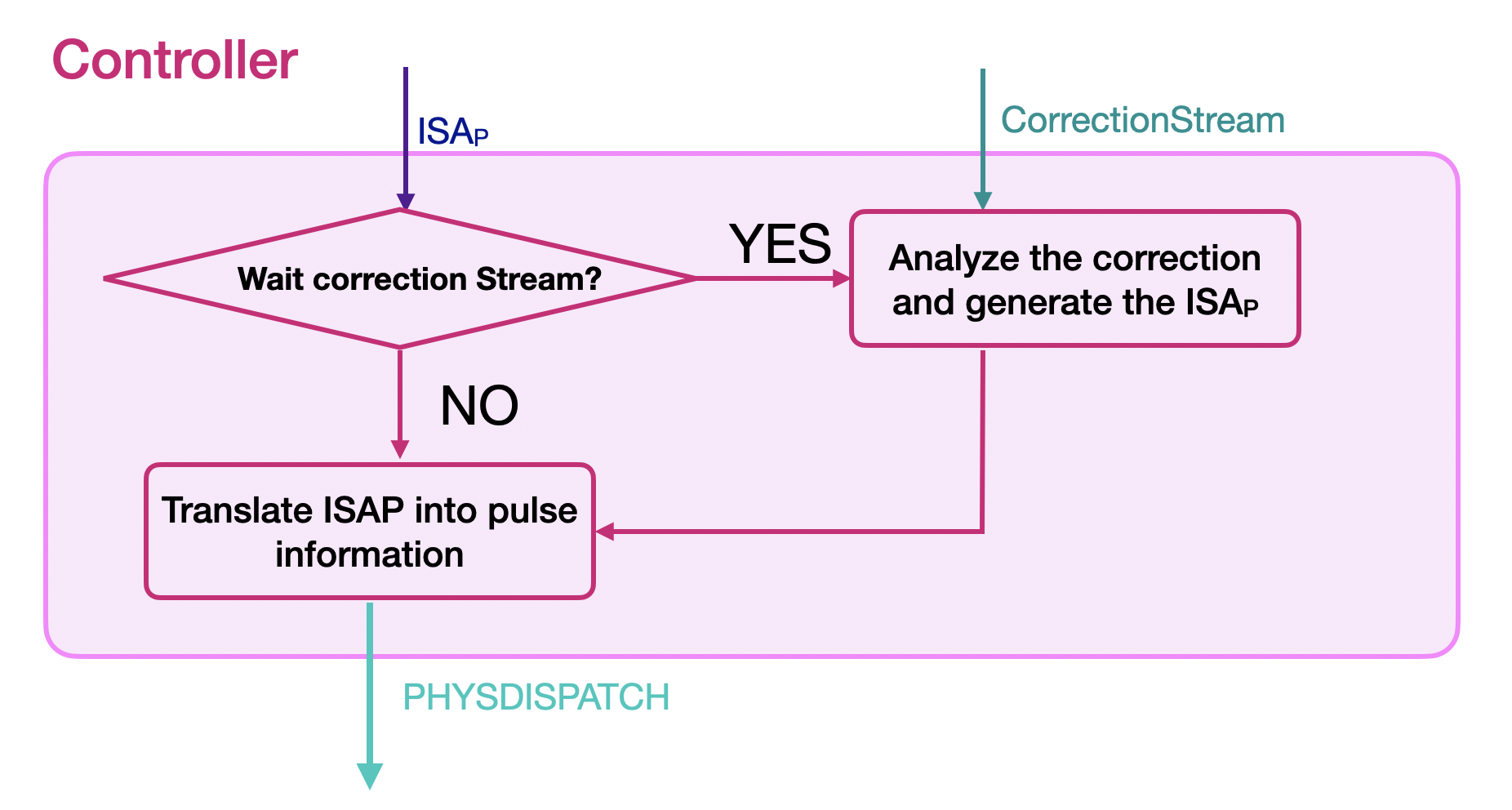}
    \caption{
    Internal logic of the Controller. The Controller receives a backend-executable
    physical instruction stream $\mathrm{ISA}_P$ from the Transpiler and a
    \textit{CorrectionStream} from the Decoder. At runtime, it decides whether the
    current execution point must wait for correction information before proceeding.
    If no correction-dependent synchronization is required, it directly translates
    the current $\mathrm{ISA}_P$ batch into device-level control information and
    dispatches it through \textsc{PhysDispatch}. If correction information must be
    incorporated, the Controller first analyzes the incoming correction stream and
    updates the effective execution state before generating the next dispatchable
    physical command batch.
    }
    \label{fig:controller}
\end{figure}

The Controller is the runtime coordination and adaptation authority of the
Fault-Tolerance Layer. It receives the backend-executable physical instruction
stream $\mathrm{ISA}_P$ from the Transpiler and the fast-path \textit{CorrectionStream}
from the Decoder, and it is responsible for turning these inputs into concrete
device-level execution commands delivered to the Hardware Layer through
\textsc{PhysDispatch}. In addition, it summarizes execution and device conditions
into \textit{RuntimeStatus} updates for the Resource Allocator and, when needed,
may propagate execution status or escalation signals upward through
\textsc{APPREPORT}. In addition, the Controller collects execution outcomes and runtime information from the hardware backend and forwards them to the Resource Allocator, which aggregates them into \textit{FinalResults} before reporting them to the Application Layer.

Architecturally, the Controller sits at the boundary between fault-tolerant execution
planning and live hardware control. Its key role is to combine the precomputed
physical instruction stream with runtime correction information before dispatch.
Note that the Controller need to maintain the ongoing physical instruction stream, enabling incremental updates in response to incoming syndrome information, thereby avoiding a full re‑transpilation. The detailed execution is illustrated in Fig.~\ref{fig:controller}. When the current execution point does not require correction-dependent synchronization, it directly translates the current $\mathrm{ISA}_P$ batch into lower-level
pulse or control information and sends the resulting commands to the hardware.
When the execution reaches a synchronization boundary at which correction
information matters, the Controller first waits for the relevant
\textit{CorrectionStream}, then analyzes that correction information, updates the
effective execution state, and only afterwards generates the next dispatchable
physical command batch. In this way, the Controller need not wait after every
physical gate; it waits only when the execution semantics of the current step
require correction to be incorporated.

A central mechanism inside the Controller is Pauli-frame integration. Rather than
physically applying every correction produced by the Decoder, the Controller may
maintain a classical record of pending Pauli updates and propagate them through
the active execution stream until a measurement, non-Clifford dependency, or other
synchronization point requires explicit resolution~\cite{chamberlandFaultTolerantQuantum2018,onMultilayeredPauli2023,riesebosPauliFrames2017}. This keeps the fast correction
path close to execution while avoiding unnecessary physical correction operations.

On longer timescales, the Controller also serves as the source of runtime status
for the rest of the fault-tolerant stack. By monitoring hardware status,
correction pressure, and execution progress, it can emit \textit{RuntimeStatus}
summaries back to the Resource Allocator, thereby enabling resource-level
reconfiguration for subsequent execution phases. Representative adaptation actions
include changing active resource pools, revising scheduling priorities, or
triggering fault-tolerant region rebinding when supported by the backend. In this
sense, the Controller bridges the immediate execution path and the slower
adaptation path: it is both the dispatch engine for active execution and the
runtime control point through which feedback becomes feasible.

\subsection{Execution and Feedback Paths inside the Fault-Tolerance Layer}
\label{subsec:ft-layer-dataflow}


\begin{figure}[htbp]
\centering

\includegraphics[width=.8\columnwidth]{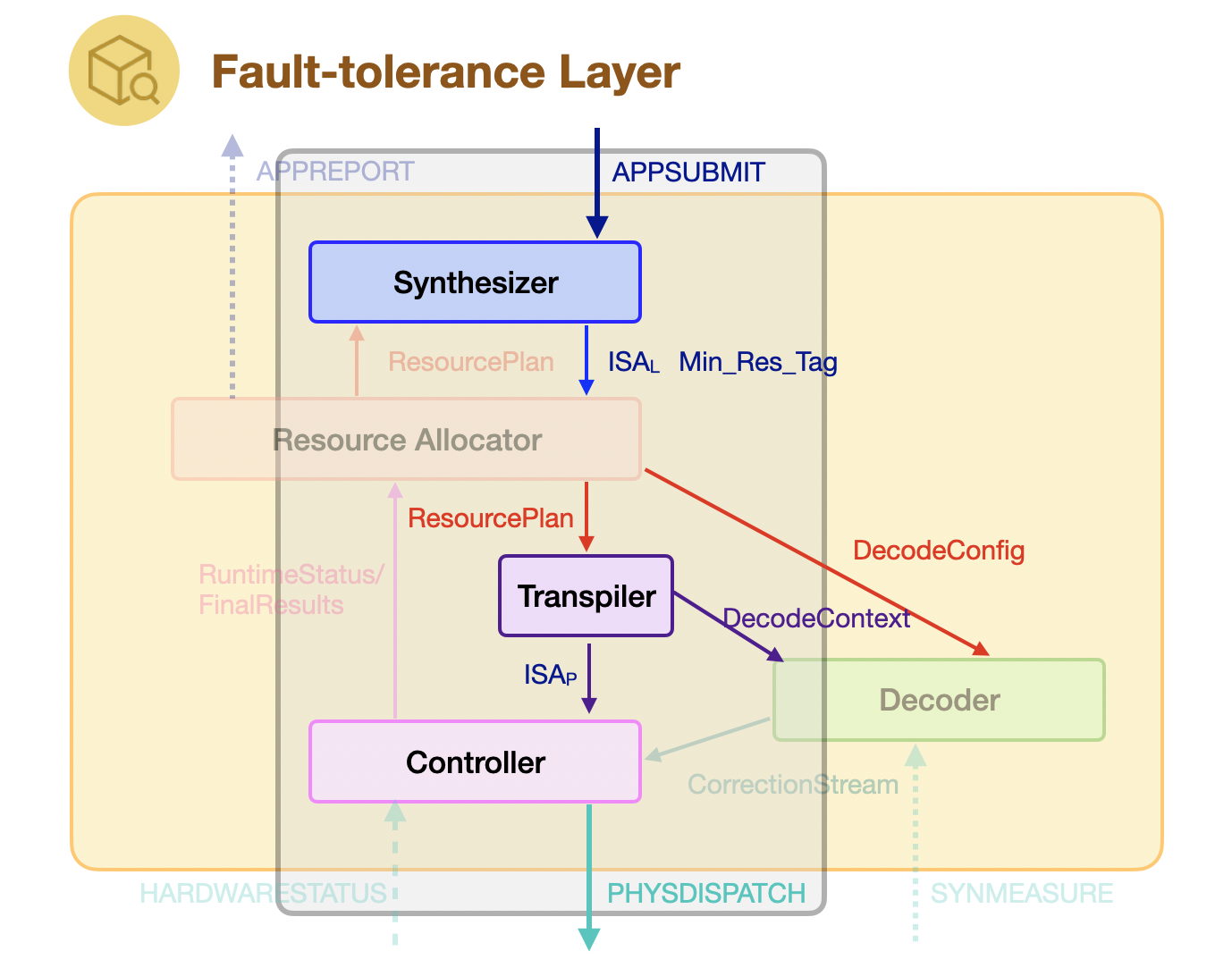}

\vspace{0.5em}

\includegraphics[width=.8\columnwidth]{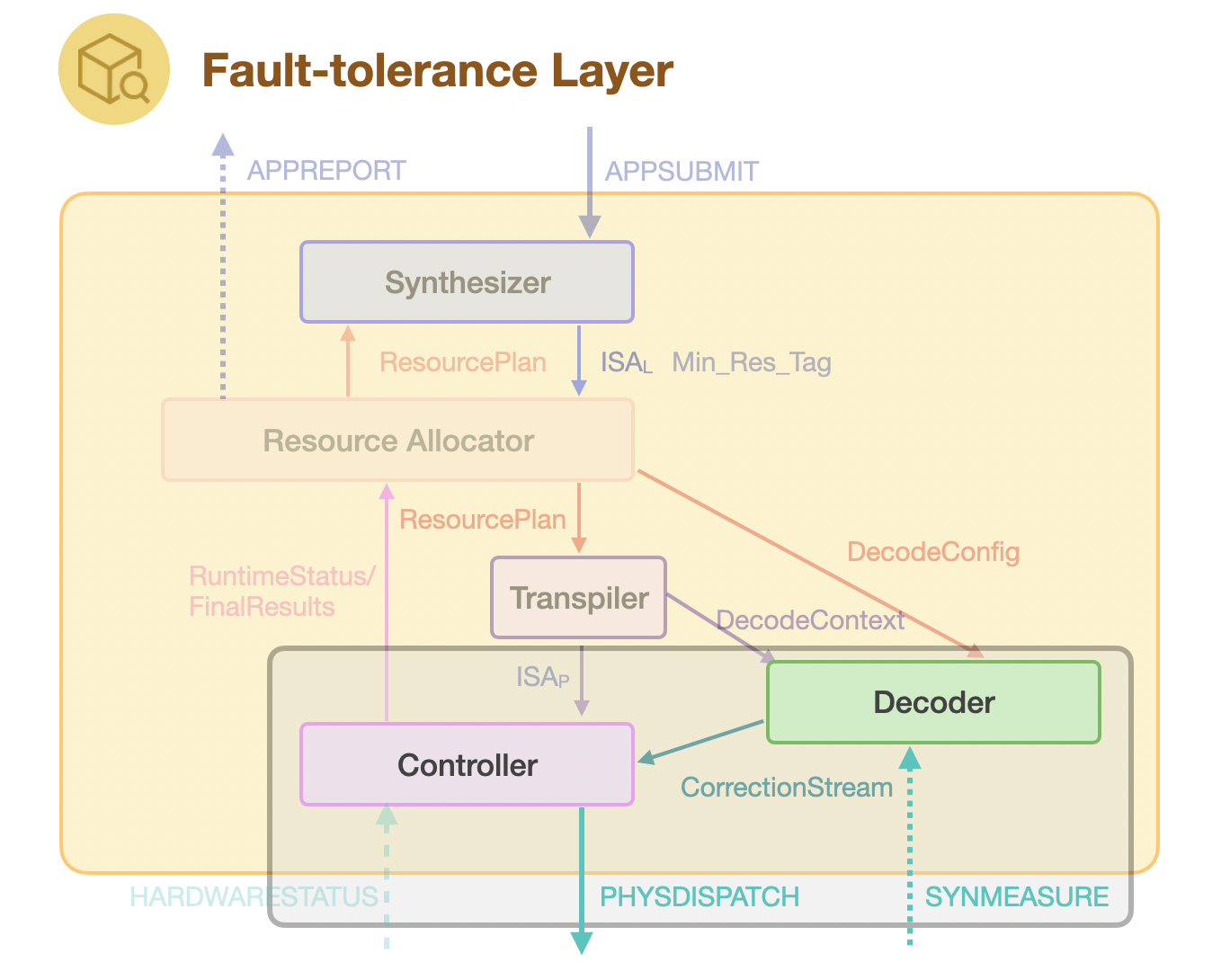}

\vspace{0.5em}

\includegraphics[width=.8\columnwidth]{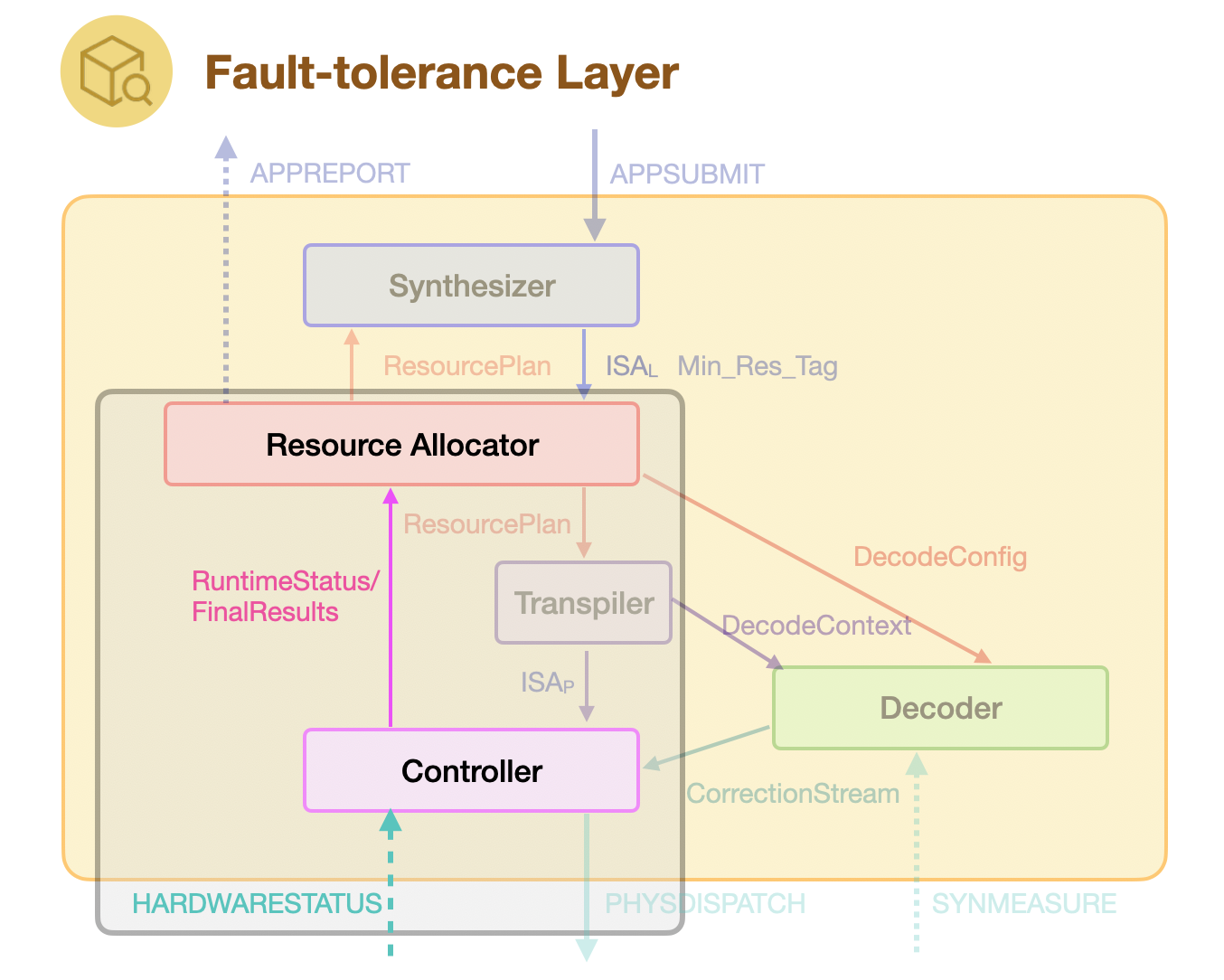}

\caption{
Three complementary execution paths inside the Fault-Tolerance Layer.\\
\textbf{Top:} forward refinement path, in which the logical program is progressively transformed into fault-tolerant physical execution through the sequence
$\mathcal{C}_{\mathrm{log}}
\rightarrow
\mathrm{ISA}_L
\rightarrow
\mathcal{R}
\rightarrow
\mathrm{ISA}_P
\rightarrow
\textsc{PhysDispatch}$.\\
\textbf{Middle:} fast correction path, where syndrome and measurement information returned through \textsc{SynMeasure} is processed by the Decoder and converted into a \textit{CorrectionStream} consumed by the Controller for low-latency correction and Pauli-frame management.\\
\textbf{Bottom:} slow adaptation path, where device-level status information returned through \textsc{HardwareStatus} is aggregated by the Controller and propagated as \textit{RuntimeStatus} to the Resource Allocator, enabling resource-level reconfiguration and runtime adaptation.\\
Together, the three paths illustrate how the architecture combines planning, execution, correction, and adaptation within a unified fault-tolerance layer.
}
\label{fig:threeflow-ftlayer}
\end{figure}

The module interaction graph in Fig.~\ref{fig:ftlayer-internal} can also be understood in terms of three higher-level execution paths. These paths do not introduce new modules or interfaces. Rather, they expose three distinct execution semantics that coexist within the same modular architecture: forward refinement, fast correction, and slow adaptation.

The first is the \emph{forward refinement path}, which progressively converts a hardware-agnostic logical program into backend-executable fault-tolerant execution. In this path, the application-level circuit $\mathcal{C}_{\mathrm{log}}$ is synthesized into the logical instruction stream $\mathrm{ISA}_L$, organized into a fault-tolerant \textit{ResourcePlan} $\mathcal{R}$, lowered into backend-level instructions $\mathrm{ISA}_P$, and ultimately dispatched to the Hardware Layer through \textsc{PhysDispatch}. This path captures the top-down transformation from logical intent to physical execution.

The second is the \emph{fast correction path}, which supports low-latency reaction during active execution. Here, syndrome and measurement data returned through \textsc{SynMeasure} are consumed by the Decoder, which emits a \textit{CorrectionStream} to the Controller. The Controller then incorporates this correction information at synchronization boundaries during ongoing execution. This path is intentionally kept close to the hardware-facing execution loop, since its purpose is immediate correction and execution continuity rather than global replanning.

The third is the \emph{slow adaptation path}, which supports runtime recalibration over longer timescales. In this path, device-level status information returned through \textsc{HardwareStatus} is consumed by the Controller, which summarizes the relevant execution and hardware conditions into \textit{RuntimeStatus} updates for the Resource Allocator. The Allocator can then revise fault-tolerant resources sizing, scheduling, or related execution parameters for subsequent phases of execution. This path therefore captures adaptive reconfiguration at the resource-management level rather than immediate error correction.

These three paths should be understood as complementary views of the same architecture. Figure~\ref{fig:ftlayer-internal} emphasizes the module interaction structure, whereas Fig.~\ref{fig:threeflow-ftlayer} emphasizes the execution semantics that emerge from those interactions.

\section{End-to-End Hamiltonian-Simulation Example}
\label{sec:hamiltonian_example}

To illustrate the proposed architecture in operation, we consider a small digital
Hamiltonian-simulation workload and trace its execution through the entire stack.
Hamiltonian-simulation workloads typically estimate observables from many
repeated executions of the same logical circuit. The example below therefore
describes one representative fault-tolerant run of the circuit, while the same
compiled workload may be executed repeatedly to accumulate measurement
statistics.

This distinction is important for interpreting the feedback paths. Within a
single run, the forward execution path and the fast correction path are active:
the circuit is synthesized, allocated, lowered, executed, and corrected using
syndrome information. Across repeated runs, the slow adaptation path can also
act: runtime information collected from earlier executions can be summarized
through \textit{RuntimeStatus} and used by the Resource Allocator when
constructing subsequent \textit{ResourcePlan}s.

The first part of the example is platform independent: the application circuit,
fault-tolerant logical normalization, non-Clifford synthesis choices, resource
allocation, and logical scheduling are shared by the two backend models. The
Transpiler then lowers the selected \textit{ResourcePlan} into a backend-specific
$\mathrm{ISA}_P$ stream. Here $\mathrm{ISA}_P$ denotes a physical-instruction
interface inside the Fault-Tolerance Layer, not the final pulse-level hardware
control language. The platform-specific layer below $\mathrm{ISA}_P$ realizes
these primitives using superconducting lattice-surgery commands or neutral-atom
transport, Rydberg interaction, and readout controls.

\subsection{Common Application Workload}
\label{sec:example_common_workload}

The target workload is a $4$-qubit simulation of the one-dimensional transverse-field Ising model (TFIM) with two first-order Trotter steps:
\begin{equation}
H=J \sum_{i=0}^{2} Z_i Z_{i+1} + h \sum_{i=0}^{3} X_i .
\end{equation}
We use coupling parameters $J=0.85$ and $h=0.55$, total evolution time $t=0.4$, and $r=2$ Trotter steps. With $\delta=t/r=0.2$, the Trotterized unitary is
\begin{equation}
U(t)\approx
\left[
e^{-i \delta J Z_0 Z_1}
e^{-i \delta J Z_1 Z_2}
e^{-i \delta J Z_2 Z_3}
\prod_{k=0}^{3} e^{-i \delta h X_k}
\right]^2 .
\end{equation}
The two rotation angles are therefore
\begin{equation}
\theta_{ZZ}=2J\delta=0.34,\qquad
\theta_X=2h\delta=0.22 .
\end{equation}
Both angles are non-Clifford at the fault tolerant logical level.

At the application layer, we first canonicalize the circuit into a Clifford+$R_z$ form. The $ZZ$ terms are represented by the standard Clifford compute--$R_z$--uncompute pattern, and each transverse-field rotation is expanded as $R_x(\theta)=H R_z(\theta)H$. Thus the only parameterized non-Clifford gates that enter the fault-tolerance layer are logical $R_z$ rotations.

\begin{lstlisting}[language=QASM3, caption={Application-layer canonical Clifford+$R_z$ input for the 4-qubit TFIM simulation}]
OPENQASM 3.0;
include "stdgates.inc";

qubit[4] q;
bit[4] c;

// Prepare |++++>
h q[0]; h q[1]; h q[2]; h q[3];

// Trotter step 1
cx q[0], q[1]; rz(0.34) q[1]; cx q[0], q[1];
cx q[1], q[2]; rz(0.34) q[2]; cx q[1], q[2];
cx q[2], q[3]; rz(0.34) q[3]; cx q[2], q[3];
h q[0]; rz(0.22) q[0]; h q[0];
h q[1]; rz(0.22) q[1]; h q[1];
h q[2]; rz(0.22) q[2]; h q[2];
h q[3]; rz(0.22) q[3]; h q[3];

// Trotter step 2
cx q[0], q[1]; rz(0.34) q[1]; cx q[0], q[1];
cx q[1], q[2]; rz(0.34) q[2]; cx q[1], q[2];
cx q[2], q[3]; rz(0.34) q[3]; cx q[2], q[3];
h q[0]; rz(0.22) q[0]; h q[0];
h q[1]; rz(0.22) q[1]; h q[1];
h q[2]; rz(0.22) q[2]; h q[2];
h q[3]; rz(0.22) q[3]; h q[3];

measure q[0] -> c[0];
measure q[1] -> c[1];
measure q[2] -> c[2];
measure q[3] -> c[3];
\end{lstlisting}

The canonical QASM circuit has a simple count. The four initial $H$ gates prepare the state $|++++\rangle$ and are Clifford operations. Each Trotter step contains three nearest-neighbor $ZZ$ interactions, and each interaction is expressed as
\begin{equation}
\mathrm{CNOT}\;-\;R_z(0.34)\;-\;\mathrm{CNOT}.
\end{equation}
Thus two Trotter steps contain $3\times 2\times 2=12$ application-level CNOT gates and $6$ $R_z$ rotations with angle $0.34$. The transverse-field part contributes four $H$--$R_z(0.22)$--$H$ blocks per Trotter step, giving $8$ additional $R_z$ rotations and $16$ basis-change $H$ gates across the full circuit. The total parameterized non-Clifford content is therefore $6+8=14$ logical $R_z$ sites, but these sites use only two distinct angles, $0.34$ and $0.22$. The four final measurements read out the four simulated qubits. The Application Layer passes this canonical Clifford+$R_z$ circuit to the Fault-Tolerance Layer through the \texttt{LogicalCircuit} API, together with the target tolerance and the repeated-angle metadata $\{0.34,0.22\}$. The synthesis problem below is therefore deliberately narrow: decide how each logical $R_z$ should be implemented.

\subsection{Common Fault-Tolerant Synthesis}
\label{sec:example_common_synthesis}

The Fault-Tolerance Layer receives the canonical Clifford+$R_z$ circuit from the Application Layer. At this point, the application-level preprocessing has already expressed the workload as a fixed Clifford skeleton together with a set of logical rotation sites. The Fault-Tolerance Layer therefore does not need to decompose $R_x$ gates, nor does it need to treat the $ZZ$ interactions as independent physical primitives. Instead, the Clifford gates surrounding each rotation are kept as part of the fixed logical circuit skeleton, and the Synthesizer focuses on the $14$ logical $R_z$ sites exposed by the circuit.

In the general architecture, each logical $R_z(\theta)$ site is processed by the threshold-based synthesis rule introduced in Sec.~\ref{subsec:circuit-synthesizer}. The threshold $\theta_{\mathrm{th}}$ is optimized through the synthesis-allocation loop rather than fixed in advance. Intermediate values of $\theta_{\mathrm{th}}$ generate hybrid logical programs: some rotation sites are compiled into Clifford+$T$ sequences, while the remaining sites are preserved as direct angle-resource injections. In this example, however, we do not run the full threshold-optimization loop. We instead expose two endpoint synthesis candidates to the Resource Allocator.

The first endpoint corresponds to $\theta_{\mathrm{th}}=0$, under which every logical $R_z$ rotation is compiled into a Clifford+$T$ sequence. This endpoint can be instantiated using Ross-Selinger-type single-qubit synthesis~\cite{rossOptimalAncillafree2016}, with the required $T$ states supplied by magic-state distillation factories~\cite{litinskiMagicState2019,gidneyEfficientMagic2019} or by magic-state cultivation~\cite{gidneyMagicState2024a}. The second endpoint corresponds to $\theta_{\mathrm{th}}\rightarrow\infty$, under which every logical $R_z$ rotation is preserved as a direct angle-resource injection, as in error-structure-tailored direct-rotation schemes~\cite{zengErrorstructuretailoredEarly2025}. Hybrid strategies between these endpoints, including star-mutation-style methods~\cite{toshioSTARMagicMutation2026} and related peephole rewrites, are also compatible with the same interface, but are not instantiated in this small demonstrator.

The Synthesizer therefore outputs an fault-tolerant-ready logical program consisting of a fixed Clifford skeleton together with fourteen logical rotation sites.
For the present workload, the rotation angles belong to two families,
$0.34$ and $0.22$,
corresponding respectively to the $ZZ$ interaction blocks and transverse-field blocks. The output passed to the Resource Allocator is an  fault-tolerant-ready logical program with $14$ logical $R_z$ sites, two repeated angle families, two endpoint non-Clifford synthesis candidates, and $4$ terminal logical measurements. The Synthesizer does not decide which candidate is finally used; that choice is made by the Resource Allocator under the error-budget and resource constraints described next.

\subsection{Common Resource Allocation and Logical Scheduling}
\label{sec:example_common_allocation}

Resource allocation is performed after the Synthesizer has exposed the two endpoint candidates and before the Transpiler lowers the selected plan to $\mathrm{ISA}_P$. The Allocator combines the candidate logical programs with the available factory models, the target run-level failure probability, and the protected-layout constraints. The relevant failure-budget condition is
\begin{equation}
\epsilon_{\mathrm{mem}}
+
\epsilon_{\mathrm{Cliff}}
+
\epsilon_{\mathrm{NC}}
\leq
\epsilon_{\mathrm{tot}} .
\end{equation}
Here $\epsilon_{\mathrm{mem}}$ is the accumulated memory contribution, $\epsilon_{\mathrm{Cliff}}$ is the contribution from fault-tolerant Clifford operations, and $\epsilon_{\mathrm{NC}}$ includes non-Clifford approximation error, resource-state infidelity, and injection failure.

The code distance is chosen by relating the target failure probability to the logical error rate. For a rotated surface-code demonstrator, we use the eq.~\eqref{eq:surface-logical-error} to obtain
\begin{equation}
p_L(d)
\approx
C
\left(
\frac{p}{p_{\mathrm{th}}}
\right)^{\lfloor(d+1)/2\rfloor},
\end{equation}
where $p$ is the physical error rate, $p_{\mathrm{th}}$ is the threshold, and $C$ is a code- and decoder-dependent prefactor. According to eq.~\eqref{eq:min-distance-allocator}, requiring $D p_L(d)\leq\epsilon_{\mathrm{tot}}$ over $D$ fault-tolerant operation windows gives
\begin{equation}
d
\geq
\left\lceil
\frac{
2\log(D C/\epsilon_{\mathrm{tot}})
}{
\log(p_{\mathrm{th}}/p)
}
-1
\right\rceil .
\label{eq:min-distance}
\end{equation}
Thus $d=3$ in this example is a small demonstrator choice for a short circuit and benign target parameters, not an architecture-wide constant. For the neutral-atom comparison below, we keep the same rotated-surface-code logical abstraction; the platform difference lies in protected-block organization and control, rather than in changing the logical code.

The Allocator then estimates the non-Clifford resource cost for each endpoint candidate. For the Clifford+$T$ endpoint, each logical $R_z(\theta)$ is approximated by a Clifford+$T$ sequence. For per-rotation approximation tolerance $\epsilon_{\mathrm{rot}}$, the typical Ross-Selinger scaling is
\begin{equation}
N_T(\theta,\epsilon_{\mathrm{rot}})
\approx
3\log_2(1/\epsilon_{\mathrm{rot}}).
\end{equation}
The required $T$ gates are supplied by distilled $|T\rangle$ states. As a concrete surface-code factory scale, the catalyzed $|CCZ\rangle\rightarrow 2|T\rangle$ construction of Ref.~\cite{gidneyEfficientMagic2019} uses a $12d\times6d$ footprint and outputs one $|CCZ\rangle$ every $5.5d$ code cycles. This gives the approximate per-$T$ production cost
\begin{equation}
C_T^{\mathrm{fac}}(d)
\approx
\frac{12d\cdot 6d\cdot 5.5d}{2}
=
198d^3
\end{equation}
physical qubit-cycles per distilled $|T\rangle$ state, before routing and buffering overhead. The corresponding non-Clifford production cost is estimated as
\begin{equation}
C_T^{\mathrm{tot}}
\sim
N_{\mathrm{rot}},
N_T(\epsilon_{\mathrm{rot}}),
C_T^{\mathrm{fac}}(d).
\end{equation}
For $\epsilon_{\mathrm{rot}}\sim10^{-4}$, this gives roughly $40$ to $45$ $T$ gates per rotation, or on the order of $6\times10^2$ distilled $|T\rangle$ states for the $14$ rotations.

For the direct-angle endpoint, each logical $R_z(\theta)$ site requests an angle resource state
\begin{equation}
|A_\theta\rangle_L
=
\frac{|0\rangle_L+e^{i\theta}|1\rangle_L}{\sqrt{2}},
\end{equation}
which is consumed through an angle-injection primitive. Since the workload only uses the repeated angles $0.34$ and $0.22$, the resource layer can specialize to two angle-state families. For a demonstrator-level estimate, we model the preparation cost as
\begin{equation}
C_{A_\theta}^{\mathrm{prep}}(d)
\approx
\kappa_\theta d^3,
\qquad
\kappa_\theta=O(1).
\end{equation}
This should be interpreted as a resource-state factory cost, not as a free logical $R_z$. The error-structure-tailored direct-rotation/projection scheme of Ref.~\cite{zengErrorstructuretailoredEarly2025} reports large physical qubit-cycle reductions relative to 15-to-1 magic-state distillation and cultivation in a Hamiltonian-simulation benchmark; because that result is implementation-dependent, we keep the coefficient $\kappa_\theta$ explicit. 

The total direct-angle resource cost is estimated as
\begin{equation}
C_A^{\mathrm{tot}}
\sim
N_{0.34}
C_{A_{0.34}}^{\mathrm{prep}}(d)
+
N_{0.22}
C_{A_{0.22}}^{\mathrm{prep}}(d),
\end{equation}
where $N_{0.34}$ and $N_{0.22}$ denote the numbers of angle-state injections associated with the two angle families appearing in the workload.

\begin{table}[t]
\caption{Endpoint non-Clifford resource models evaluated by the Allocator.}
\label{tab:tgate_choices_example}
\centering
\scriptsize
\begin{ruledtabular}
\begin{tabular}{lcc}
Candidate & Resource factory & Dominant cost model \\
\hline
Clifford+$T$ & distilled $|T\rangle$
& $N_{\rm rot}N_TC_T^{\rm fac}(d)$ \\
Angle injection & $|A_{0.34}\rangle, |A_{0.22}\rangle$
& $N_{\rm inj}\bar{C}{A\theta}(d)$ 
\end{tabular}
\end{ruledtabular}
\end{table}

Under the demonstrator parameters used here, both endpoint candidates can be made to satisfy the failure budget, but the direct angle-state candidate has lower online injection depth and lower estimated resource-state production volume. The Allocator therefore selects direct angle-state injection for the fault-tolerant execution and records this decision in the \textit{ResourcePlan} and \textit{DecodeConfig}.

The logical dependency graph is platform independent. For scheduling, we keep the Hamiltonian labels $ZZ_e$ and $ZZ_c$ as shorthand for Clifford+$R_z$ subcircuits. The two edge blocks on $(Q_0,Q_1)$ and $(Q_2,Q_3)$ are disjoint, so their internal $R_z(0.34)$ sites can be executed in the same $ZZ_e$ window. The central block on $(Q_1,Q_2)$ shares data with both edge blocks and therefore forms a separate $ZZ_c$ window. The four transverse-field $H$-$R_z(0.22)$-$H$ blocks commute and are logically independent; their online depth depends on the number of available rotation buffers.

With two online rotation buffers, each Trotter step requires
\begin{equation}
D_{\mathrm{inj}}^{(1)}
=
\left\lceil\frac{2}{2}\right\rceil
+
\left\lceil\frac{1}{2}\right\rceil
+
\left\lceil\frac{4}{2}\right\rceil
=
4
\end{equation}
injection windows. The three numerators count, respectively, two edge-disjoint $ZZ$-context rotations, one central $ZZ$-context rotation, and four transverse-field rotations. The full two-step workload therefore requires $D_{\mathrm{inj}}=8$ online injection windows in the selected two-buffer baseline. With four online rotation buffers, the per-step injection depth becomes
\begin{equation}
D_{\mathrm{inj}}^{(1)} = 
\left\lceil\frac{2}{4}\right\rceil
+
\left\lceil\frac{1}{4}\right\rceil
+
\left\lceil\frac{4}{4}\right\rceil =
3,
\end{equation}
reducing the two-step injection depth from $8$ to $6$ windows at the cost of additional buffers, routing, and factory-output bandwidth.

The output of this stage is a backend-neutral fault-tolerant \textit{ResourcePlan},
rather than a platform-specific physical program. The architectural flow is
\begin{equation}
\mathrm{ISA}_L
\longrightarrow
\mathcal{R}
\longrightarrow
\mathrm{ISA}_P
\longrightarrow
\mathrm{native\ controls}.
\end{equation}
Here $\mathrm{ISA}_L$ names logical operations such as \texttt{CNOT},
\texttt{Rz}$(\theta)$, and \texttt{Meas}; $\mathcal{R}$ records the selected
non-Clifford implementation, code distance, fault-tolerant resources organization,
injection-window schedule, and decoder configuration; and $\mathrm{ISA}_P$ is produced
later by the Transpiler as a backend-executable physical instruction
stream. The native-control layer below $\mathrm{ISA}_P$ contains platform-specific
objects such as calibrated pulses, transport waveforms, measurement timing, and
feedback-control details. The remainder of the example shows how the same
logical workload and selected synthesis strategy are realized through
platform-specific fault-tolerant resources organization and physical execution.
\subsection{Common Transpilation to Protected Physical Instructions}
\label{sec:example_common_transpilation}

After resource allocation, the selected output is still a logical plan rather than a hardware-control program. The role of the Transpiler is to lower this plan into a backend-executable
physical instruction stream $\mathrm{ISA}_P$ and to emit the corresponding
\textit{DecodeContext} for the Decoder. In the present example, the selected
plan fixes the rotated-surface-code abstraction, code distance, direct
angle-state injection path, two online rotation buffers, and the logical
injection-window schedule. The Transpiler turns these choices into
physical primitives such as
\texttt{PrepareBlock}, \texttt{Stabilize}, \texttt{RouteForInteraction},
\texttt{ApplyEntangler}, \texttt{InjectRzResource}, \texttt{UpdateFrame},
\texttt{MeasureBlock}, and \texttt{ReleaseOrRecycle}.

This step is still above native hardware control. The $\mathrm{ISA}_P$ stream specifies physical operations, timing constraints, measurement ordering,
synchronization boundaries, and decoding context, but it does not yet prescribe
microwave pulses, tweezer waveforms, transport trajectories, or detector-control
electronics. Those platform-specific realizations are supplied by the Hardware
Layer. Thus the same selected logical workload and fault-tolerant resources plan can
lead to different backend realizations:
\begin{equation}
\mathrm{ISA}_L
\longrightarrow
\mathcal{R}
\longrightarrow
\mathrm{ISA}_P^{(\mathrm{backend})}
\longrightarrow
\mathrm{native\ controls}.
\end{equation}
The two subsections below illustrate this final lowering for superconducting
circuits and neutral-atom arrays.

\subsection{\texorpdfstring{Superconducting Backend: $\mathrm{ISA}_P$ Realization and Hardware Controls}{Superconducting Backend: ISA-P Realization and Hardware Controls}}
\label{sec:superconducting_example}

After the Transpiler has lowered the selected \textit{ResourcePlan} into a
superconducting-specific $\mathrm{ISA}_P$ stream, the superconducting backend realizes the instructions as fixed-grid surface-code operations. In this demonstrator,
we model a distance-$3$ rotated surface-code patch as $9$ data qubits and $8$
syndrome ancillas, giving $17$ physical qubits per logical patch. For the selected
direct angle-injection path, the fault-tolerant layout contains four data patches
$Q_0,\ldots,Q_3$, a bridge/parity-measurement patch $Q_a$, two online rotation
buffers $Q_{r,1},Q_{r,2}$, spare routing or relocation capacity $B_1,B_2$, and a
shared repeated-angle factory region $F_\theta$. This should be read as an
illustrative few-hundred-qubit fault-tolerant footprint rather than a hardware-optimized
layout.

\begin{figure}
    \centering
\includegraphics[width=0.9\linewidth]{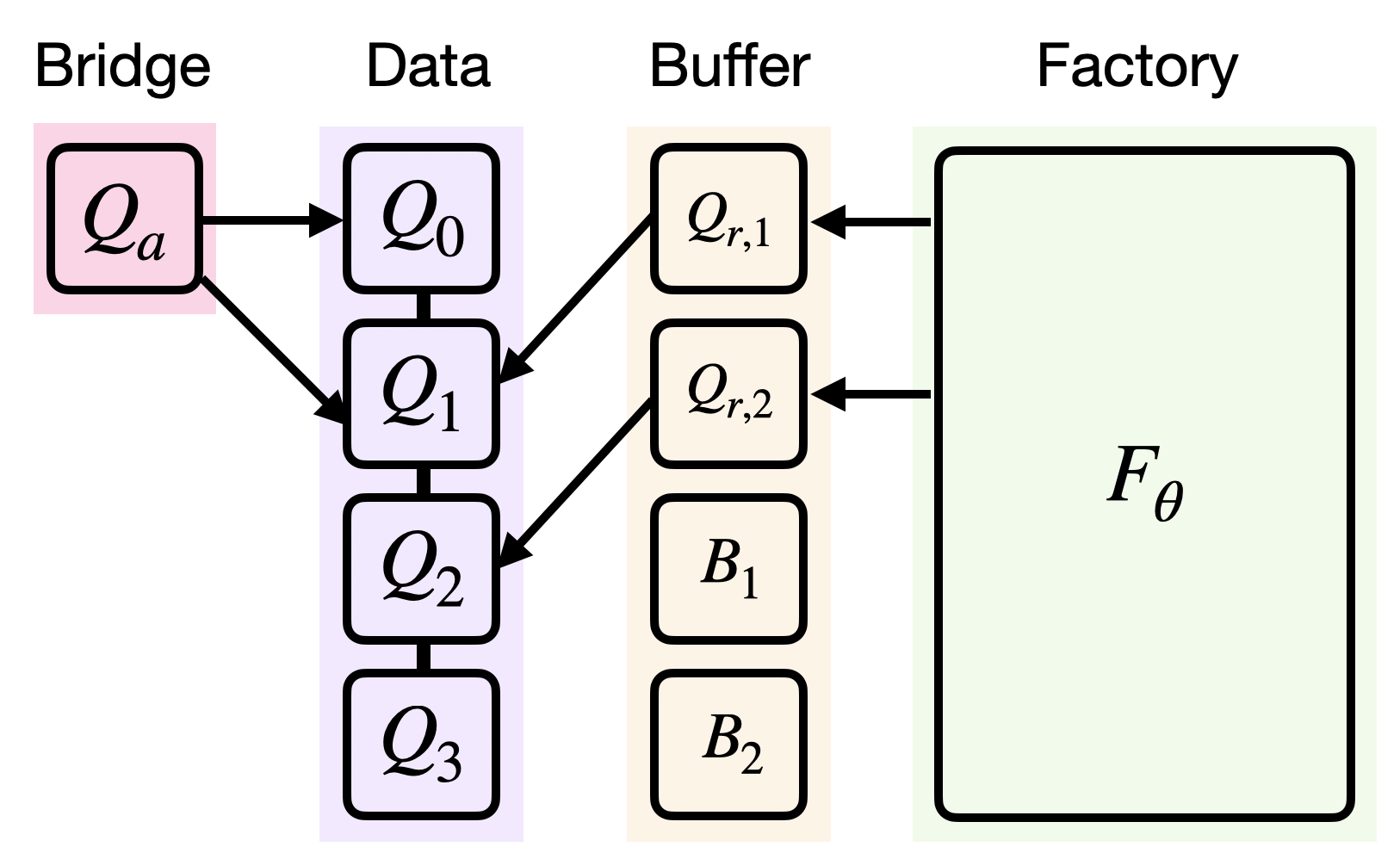}
\caption{Schematic superconducting fault-tolerant layout. Blue $Q_0,\ldots,Q_3$ are
fault-tolerant data patches encoding the four simulated TFIM qubits. The green
$F_\theta$ block denotes the repeated-angle resource factory. Orange
$Q_{r,1}$ and $Q_{r,2}$ are online rotation buffers, gray $B_1$ and $B_2$
represent spare routing or relocation capacity, and purple $Q_a$ is a
bridge/parity-measurement ancilla. Solid links denote dominant resource-delivery
or lattice-surgery paths, while the horizontal data chain follows the
nearest-neighbor TFIM interaction graph.}
\label{fig:sc_layout}
\end{figure}

Figure~\ref{fig:sc_layout} shows the corresponding superconducting fault-tolerant
layout. The backend is modeled as static surface-code patches on a fixed
nearest-neighbor grid. The data patches form a horizontal chain matching the TFIM
interaction graph, while the factory, buffers, and bridge patch are placed nearby
to reduce lattice-surgery routing overhead.

For the first $ZZ_e$ window, the fault-tolerant operation stream contains two disjoint
Clifford entangling contexts and two logical $R_z(0.34)$ injections. In the
superconducting realization, \texttt{RouteForInteraction} and
\texttt{ApplyEntangler} are implemented by bridge-patch allocation and
lattice-surgery templates, \texttt{InjectRzResource} is implemented by
angle-state surgery on the target logical patch, and \texttt{UpdateFrame} is
implemented by real-time Pauli-frame tracking. The following backend-control
sketch is schematic: it illustrates the ordering of fault-tolerant operations and
syndrome windows, not a literal pulse-level program.

\begin{lstlisting}[caption={Schematic superconducting backend-control sketch for the first parallel $ZZ$-rotation window}]
WAIT_UNTIL_READY(Qr1, angle=0.34)
WAIT_UNTIL_READY(Qr2, angle=0.34)

BEGIN_PARALLEL
  LOGICAL_CNOT_SURGERY(control=Q0, target=Q1)
  LOGICAL_CNOT_SURGERY(control=Q2, target=Q3)
END_PARALLEL

REPEAT d TIMES:
  SURFACE_CODE_CYCLE(active_patches=[Q0,Q1,Q2,Q3,Qr1,Qr2,Qa,B1,B2])
END

BEGIN_PARALLEL
  INJECT_LOGICAL_RZ(target=Q1, resource=Qr1, angle=0.34)
  INJECT_LOGICAL_RZ(target=Q3, resource=Qr2, angle=0.34)
END_PARALLEL

MEASURE(Qr1, key="inj_01")
MEASURE(Qr2, key="inj_23")
FRAME_UPDATE(target=Q1, branch=inj_01, angle=0.34)
FRAME_UPDATE(target=Q3, branch=inj_23, angle=0.34)

BEGIN_PARALLEL
  LOGICAL_CNOT_SURGERY(control=Q0, target=Q1)
  LOGICAL_CNOT_SURGERY(control=Q2, target=Q3)
END_PARALLEL

RELEASE(Qr1)
RELEASE(Qr2)
\end{lstlisting}
The syndrome-extraction rounds appearing in the execution sketch generate
\textsc{SynMeasure} streams that are consumed by the Decoder. Using the
\textit{DecodeConfig} and \textit{DecodeContext} generated earlier by the
Resource Allocator and Transpiler, the Decoder converts the syndrome history
into a \textit{CorrectionStream}. The resulting correction information is
consumed by the Controller, which updates the Pauli frame through
\texttt{FRAME\_UPDATE} operations rather than inserting explicit physical
correction gates. In this example, the decoder therefore remains off the main
resource-planning path but participates directly in runtime execution through
syndrome interpretation and frame tracking.

Hardware-native $R_z$ phases that appear inside calibrated superconducting
single-qubit controls are implemented as virtual frame updates, for example
$\texttt{SHIFT\_PHASE}(q,\varphi)$. This pulse-level operation sits below
$\mathrm{ISA}_P$ and should not be confused with a fault-tolerant arbitrary-angle rotation on
an encoded qubit. In the fault-tolerant execution considered here, the logical
arbitrary-angle rotation is realized through angle-state injection, syndrome
extraction, measurement, and Pauli-frame tracking.

\subsection{\texorpdfstring{Neutral-Atom Backend: $\mathrm{ISA}_P$ Realization and Hardware Controls}{Neutral-Atom Backend: ISA-P Realization and Hardware Controls}}
\label{sec:atom_example}

The neutral-atom backend receives a neutral-atom-specific $\mathrm{ISA}_P$ stream obtained
by lowering the same selected fault-tolerant plan through the Transpiler. The logical
workload, selected direct angle-state injection path, and code-distance rule in
Eq.~\eqref{eq:min-distance} are the same as in the superconducting example, but
the physical realization is different. In this demonstrator, we model the logical
objects as rotated-surface-code protected blocks for comparison, while the
backend organization is zoned and reconfigurable rather than fixed-grid. The
allocator therefore reserves storage capacity, entangling-zone slots, readout
bandwidth, transport bandwidth, and spare-block capacity.

\begin{figure}
    \centering
    \includegraphics[width=.8\linewidth]{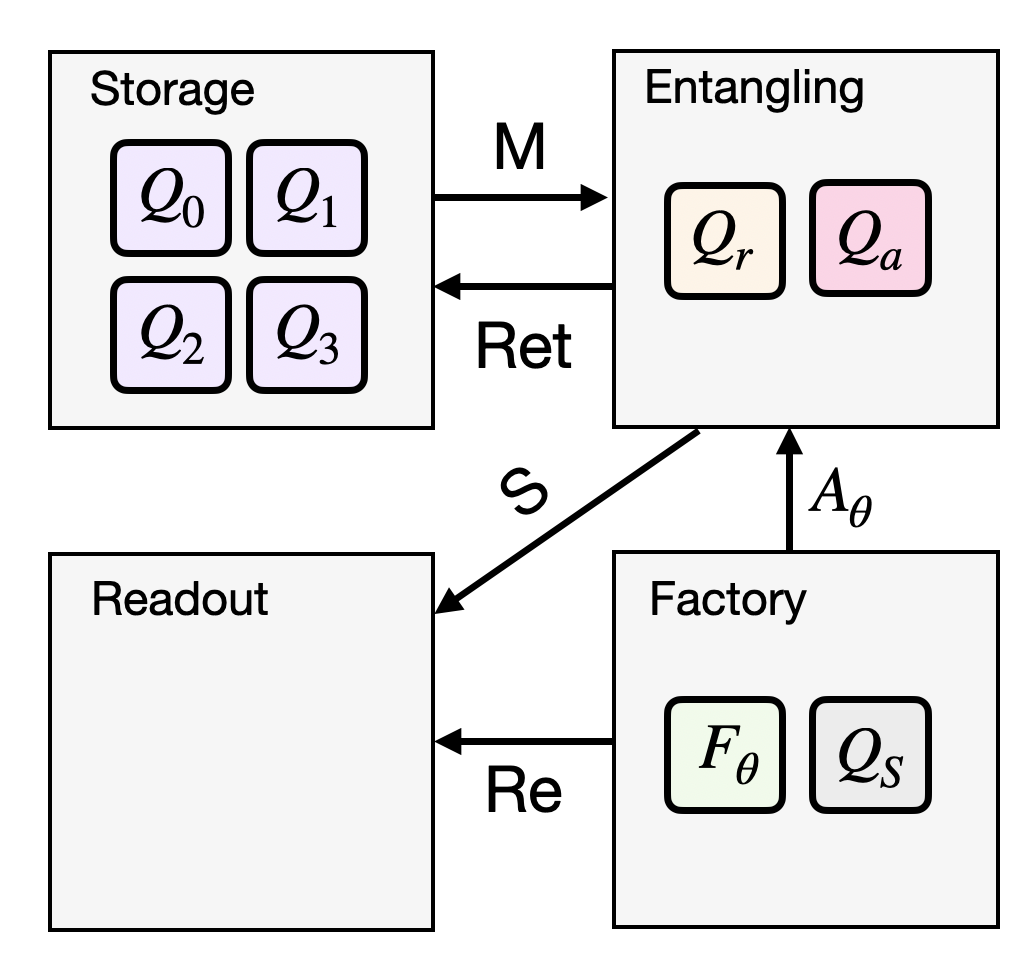}
\caption{Schematic neutral-atom zoned execution model. Blue $Q_i$ are fault-tolerant
data, orange $Q_r$ is a rotation buffer, purple $Q_a$ is an interaction
ancilla, green $F_\theta$ is the repeated-angle factory, and gray $Q_s$ is spare
capacity for reload or relocation. Arrow labels denote movement from storage to
the entangling zone (M), return transport (Ret), syndrome/readout flow (S),
angle-resource delivery ($A_\theta$), and reload/recycle flow (Re). Compared
with Fig.~\ref{fig:sc_layout}, the logical workload is unchanged, but allocation
is dominated by movement, zone occupancy, readout capacity, and spare-block
relocation.}
\label{fig:atom_zones}
\end{figure}

Figure~\ref{fig:atom_zones} summarizes this zoned organization. It is drawn as a
zone diagram rather than a patch layout because the relevant backend resource is
not a static adjacency graph, but the ability to move protected blocks between
storage, interaction, factory, and readout zones while respecting transport,
readout, and interaction-slot constraints.

The neutral-atom realization of $\mathrm{ISA}_P$ should be read as a representative
fault-tolerant-control template rather than as a claim that all devices expose the
same low-level opcodes. Reconfigurable optical-tweezer arrays support atom
rearrangement and coherent transport of entangled atoms~\cite{bluvsteinQuantumProcessor2022a};
Rydberg interactions provide programmable entangling operations, including
parallel high-fidelity two-qubit gates~\cite{everedHighfidelityParallel2023b}; and
recent logical-processor demonstrations use zoned operation, logical-level
control, mid-circuit readout, and reconfigurable connectivity~\cite{bluvsteinLogicalQuantum2024a}.
Accordingly, \texttt{RouteForInteraction} is realized by transport templates,
\texttt{ApplyEntangler} by fault-tolerant Rydberg-mediated entangling routines,
\texttt{InjectRzResource} by angle-state delivery and injection,
\texttt{MeasureBlock} by fluorescence-based readout, and
\texttt{ReleaseOrRecycle} by reservoir or reload operations.

A $ZZ$-context $R_z(0.34)$ site is schematically realized by moving the
participating protected blocks and an interaction ancilla into the entangling
zone, applying a fault-tolerant entangling routine, injecting the angle resource on
the target block, applying the second fault-tolerant entangling routine to uncompute
the Clifford context, and returning the participating blocks to storage or
readout. The transverse-field $R_z(0.22)$ sites use the same angle-resource path
but only require one data block and local basis changes.

At the backend-control level below $\mathrm{ISA}_P$, the first parallel Ising window may
be represented schematically as
\begin{lstlisting}[caption={Schematic neutral-atom backend-control sketch for the first parallel $ZZ$-rotation window}]
WAIT_UNTIL_READY(Qr1, angle=0.34)
WAIT_UNTIL_READY(Qr2, angle=0.34)

MOVE_BLOCK(Q0, storage -> entangling_slot_0)
MOVE_BLOCK(Q1, storage -> entangling_slot_0)
MOVE_BLOCK(Q2, storage -> entangling_slot_1)
MOVE_BLOCK(Q3, storage -> entangling_slot_1)
MOVE_BLOCK(Qr1, factory_output -> entangling_slot_0)
MOVE_BLOCK(Qr2, factory_output -> entangling_slot_1)

BEGIN_PARALLEL
  APPLY_PROTECTED_ENTANGLER(control=Q0, target=Q1)
  APPLY_PROTECTED_ENTANGLER(control=Q2, target=Q3)
END_PARALLEL

BEGIN_PARALLEL
  INJECT_LOGICAL_RZ(target=Q1, resource=Qr1, angle=0.34)
  INJECT_LOGICAL_RZ(target=Q3, resource=Qr2, angle=0.34)
END_PARALLEL

READ_SYNDROME(blocks=[Q0,Q1,Q2,Q3,Qr1,Qr2])
MEASURE_LOGICAL(Qr1, key="inj_01")
MEASURE_LOGICAL(Qr2, key="inj_23")
FRAME_UPDATE(target=Q1, branch=inj_01, angle=0.34)
FRAME_UPDATE(target=Q3, branch=inj_23, angle=0.34)

BEGIN_PARALLEL
  APPLY_PROTECTED_ENTANGLER(control=Q0, target=Q1)
  APPLY_PROTECTED_ENTANGLER(control=Q2, target=Q3)
END_PARALLEL

MOVE_BLOCK(Q0, entangling_slot_0 -> storage)
MOVE_BLOCK(Q1, entangling_slot_0 -> storage)
MOVE_BLOCK(Q2, entangling_slot_1 -> storage)
MOVE_BLOCK(Q3, entangling_slot_1 -> storage)
RESET_OR_RECYCLE(Qr1)
RESET_OR_RECYCLE(Qr2)
\end{lstlisting}

The resulting neutral-atom execution is dominated by zone allocation, coherent
transport, interaction-slot contention, readout throughput, erasure handling, and
spare-block relocation. A representative runtime feedback event is atom loss on the block hosting
$Q_2$. The resulting detector outcomes and loss flags are delivered through
\textsc{SynMeasure} and interpreted by the Decoder as erasure information.
The Controller can then report the event through \textit{RuntimeStatus},
allowing the Resource Allocator to rebind the logical block to spare capacity
in a subsequent execution segment.

\subsection{Cross-Platform fault-tolerant Operation Comparison}
\label{sec:example_comparison}

Although the fault-tolerant layouts differ substantially, both realizations use the
same Decoder–Controller feedback structure: syndrome information is converted
into correction updates through the Decoder, while runtime events are surfaced
through RuntimeStatus for resource-level adaptation.

Table~\ref{tab:isa-lowering-platforms} summarizes how the same logical operations
and fault-tolerant primitive classes are realized differently on the two backends. The important ordering is that resource allocation first fixes the code distance,
non-Clifford resource path, buffers, and fault-tolerant scheduling constraints. The
Transpiler then emits backend-specific $\mathrm{ISA}_P$ streams, and the Hardware Layer expands those fault-tolerant primitives into superconducting or neutral-atom control
templates.

\begin{table*}[t]
\centering
\caption{Fault-tolerant primitive classes and example backend realizations. SC denotes superconducting circuits, and NA denotes neutral-atom arrays. $\mathrm{ISA}_P$ remains above native hardware controls; calibrated pulses, transport waveforms, and device-control timing are one layer lower.}
\label{tab:isa-lowering-platforms}
\small
\begin{ruledtabular}
\begin{tabular}{lll}
Logical operation & Fault-tolerant primitive & Example backend realization \\
\hline
\texttt{Init}
&
\texttt{PrepareBlock}, \texttt{Stabilize}
&
SC: patch reset and stabilizer rounds \\
&&NA: block loading and syndrome refresh\\

\texttt{H}, \texttt{S}
&
\texttt{ApplyClifford} or \texttt{UpdateFrame}
&
SC: patch deformation or frame update \\
&& NA: basis change or frame update\\

\texttt{CNOT}
&
\texttt{RouteForInteraction}, \texttt{ApplyEntangler}
&
SC: lattice-surgery template \\
&&NA: transport plus Rydberg-mediated entangler\\
\texttt{T}
&
\texttt{InjectResource}$(|T\rangle)$
&
SC: $|T\rangle$ surgery injection \\
&&NA: transport resource block and inject\\

\texttt{Rz}$(\theta)$
&
\texttt{InjectRzResource}
&
SC: angle-state surgery\\
&&NA: angle-state delivery and injection \\

\texttt{Meas}
&
\texttt{MeasureBlock}, \texttt{FinalizeSyndrome}
&
SC: patch readout and syndrome finalization\\
&&NA: transport to readout zone and fluorescence readout \\
\end{tabular}
\end{ruledtabular}
\end{table*}

\begin{figure*}
    \centering
\includegraphics[width=.9\linewidth]{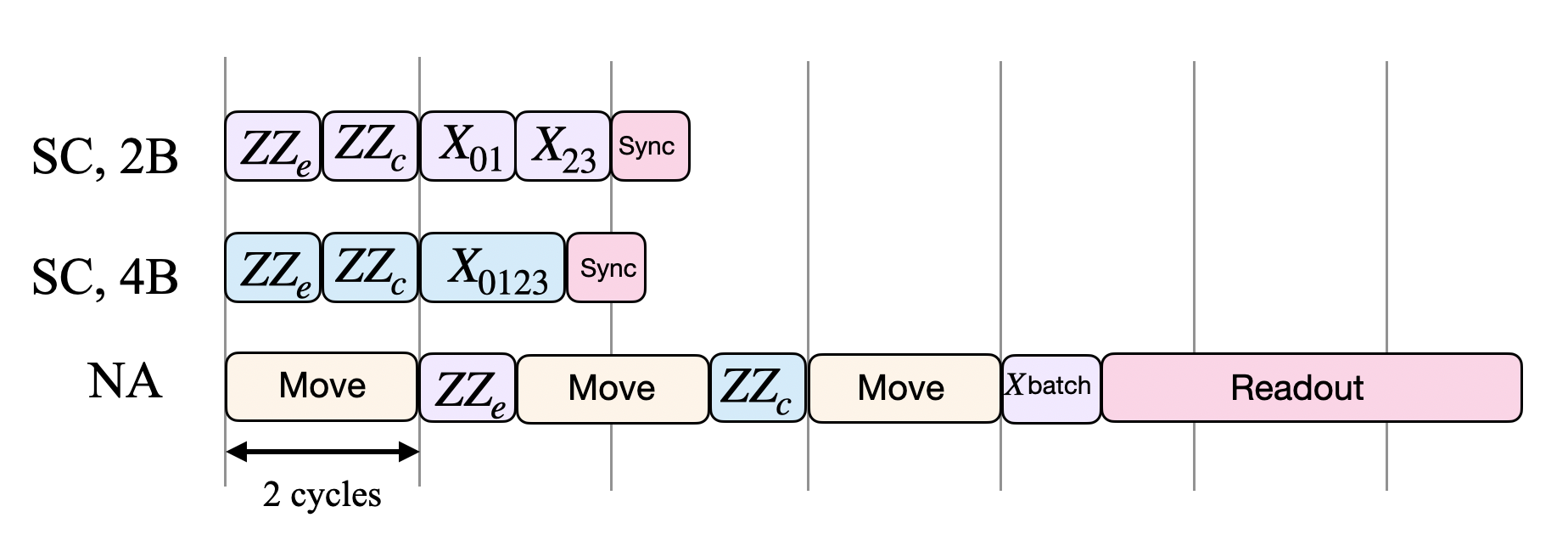}
\caption{Representative timing-aware schedule for one Trotter step on the two backend models. The superconducting schedules are shown for two online rotation buffers (SC, 2B) and four online rotation buffers (SC, 4B). The neutral-atom schedule includes explicit movement and readout stages, which are drawn longer to indicate their larger representative timing cost. The horizontal scale is schematic rather than calibrated. The marked two-cycle interval serves only as a scale reference, and its physical duration should be interpreted as platform-dependent.}
\label{fig:timeline_compare}
\end{figure*}

Figure~\ref{fig:timeline_compare} compares one Trotter step under three scheduling
views: the selected two-buffer superconducting schedule, the latency-optimized
four-buffer superconducting schedule, and the neutral-atom zoned schedule. The
full workload repeats this pattern for the second Trotter step.

The two examples share the same application-layer workload, Clifford+$R_z$
canonicalization, non-Clifford candidate comparison, resource-allocation logic,
and run-level code-distance rule. What changes after resource allocation is the
backend-specific realization: superconducting hardware emphasizes static patch
placement, lattice-surgery routing, bridge allocation, and Pauli-frame tracking
on a fixed grid, whereas neutral-atom hardware emphasizes movable fault-tolerant
blocks, zoned scheduling, readout capacity, and erasure-aware relocation.

\section{Discussion and Future Work}\label{sec:discussion}

The value of the proposed fault-tolerance layer is not merely that it inserts another box into the stack, but that it identifies a missing architectural boundary. Much of today's fault-tolerant quantum computing effort is already implicitly organized around this boundary: compilation must anticipate code-specific logical operations, resource allocation must reason about non-Clifford resources and ancilla supply, execution must incorporate decoding and measurement feedback, and hardware behavior must influence runtime decisions. Yet these responsibilities are often distributed across tools, codes, and platform-specific workflows without a clear systems abstraction. The present work argues that making this boundary explicit is itself an architectural advance.

Viewed in this way, the fault-tolerance layer should not be understood as a replacement for code design, decoding, or hardware control. Rather, it is the layer that determines how these components are composed into a coherent execution path. Its architectural role is therefore similar to that of an intermediate systems layer: it does not eliminate lower-level specialization, but instead provides a structured point at which specialization can be coordinated. This distinction is important. The main question is not whether fault tolerance exists in a system, but whether it exists as an explicit and programmable part of the architecture.

This viewpoint also clarifies an important design tension. On the one hand, future large-scale systems will need stronger modularity, cleaner interfaces, and some degree of portability across toolchains and hardware backends. On the other hand, fault-tolerant execution remains deeply shaped by code-specific operations, measurement structure, hardware connectivity, and classical reaction latency. The purpose of the fault-tolerance layer is therefore not to erase hardware dependence, but to organize it. In other words, the layer should expose the constraints that matter architecturally, while hiding those that do not need to propagate upward. This is a subtler goal than classical abstraction, but it is precisely what makes the layer meaningful.

A second implication is that the proposed architecture is naturally compatible with a future in which quantum computing becomes more modular, heterogeneous, and distributed. Large-scale systems are unlikely to remain single-region machines with a fixed and uniform execution model. Instead, they may involve multiple regions serving different purposes, such as encoded memory, computation, communication, and non-Clifford resource generation. Once that happens, the central problem is no longer only how to protect a logical qubit, but how to orchestrate interactions among multiple fault-tolerant subsystems. In such a setting, the fault-tolerance layer becomes not just a software--hardware mediator, but also a coordinator among fault-tolerant modules.

At the same time, this work remains intentionally architectural. It identifies and structures the layer, but does not yet claim an optimal decomposition, or a quantitatively validated implementation. Several questions therefore remain open. What should count as the right logical instruction abstraction? Which responsibilities belong inside the fault-tolerance layer, and which should remain in the compiler, decoder, or hardware runtime? How much of the layer can be made platform-independent, and how much should remain code- or hardware-specific? These are not peripheral engineering details; they are part of what it means to mature the architecture.

An especially important direction for future work is to study the role of the fault-tolerance layer in modular and distributed systems. If future quantum computers are assembled from multiple protected modules, then the layer will need to coordinate not only local execution within each module, but also inter-module communication, synchronization, code conversion, and shared resource management. This would extend its role from an intermediate abstraction within a single stack to a broader orchestration layer for large-scale fault-tolerant quantum computing.

Overall, the present work should be read as a proposal to make an emerging systems layer explicit. Its contribution is to identify this layer, characterize its role, and argue that it deserves to be treated as a first-class architectural object. The next stage is to make that object precise enough to implement, evaluate, and eventually standardize.

\section{Conclusion}\label{sec:conclusion}


This work argues that practical fault-tolerant quantum computing requires more than advances in algorithms, error-correcting codes, and hardware platforms considered in isolation. Its central systems challenge is architectural: fault-tolerant execution must be coordinated across synthesis, logical resource management, logical-to-physical translation, decoding, and runtime adaptation. To make this coordination explicit, we introduced a layered architecture centered around a dedicated fault-tolerance layer between the application stack and the underlying quantum hardware.

The central contribution of this paper is therefore not a new code or a standalone decoding method, but the identification and architectural characterization of fault tolerance as a first-class systems layer in the quantum computing stack. We defined the functional role of this layer, organized it into explicit internal modules, established formally defined cross-layer interfaces to both the application layer and the hardware layer, and gave a concrete end-to-end example showing how universal quantum computation can be carried through a top-down fault-tolerant execution flow. In this sense, the proposed architecture turns fault tolerance from an implicit collection of backend mechanisms into an explicit architectural object.

More broadly, the proposed architecture is code-agnostic and platform-extensible, and provides a unified framework for reasoning about cross-layer trade-offs that are often left implicit in code-centric, compiler-centric, or hardware-centric approaches. By separating concerns while preserving runtime feedback and resource coordination, it offers a structured way to connect high-level programs, fault-tolerant resources, and hardware execution within a single systems model.

We hope this work helps frame fault-tolerant quantum computing as an architectural problem in its own right. As the field moves toward practical large-scale systems, the design of interfaces, abstractions, runtimes, and coordination layers for fault-tolerant execution should become a central concern not only for the quantum

\begin{acknowledgments}
We thank Qiming Ding, Ming Gong, Pai Peng for fruitful discussion. This work is supported by Beijing Natural Science Foundation Z250004, 
the National Natural Science Foundation of China Grant (No.~12361161602), 
NSAF (Grant No.~U2330201), 
Beijing Science and Technology Planning Project (Grant No.~Z25110100810000), National Natural Science Foundation of China Grant (No.~12225507), 
and the High-performance Computing Platform of Peking University.

\end{acknowledgments}

\bibliography{refs}

@article{shor1995scheme,
  title={Scheme for reducing decoherence in quantum computer memory},
  author={Shor, Peter W},
  journal={Physical review A},
  volume={52},
  number={4},
  pages={R2493},
  year={1995},
  publisher={APS}
}

@article{bravyi2012magic,
  title={Magic-state distillation with low overhead},
  author={Bravyi, Sergey and Haah, Jeongwan},
  journal={Physical Review A—Atomic, Molecular, and Optical Physics},
  volume={86},
  number={5},
  pages={052329},
  year={2012},
  publisher={APS}
}

@article{terhal2015quantum,
  title={Quantum error correction for quantum memories},
  author={Terhal, Barbara M},
  journal={Reviews of Modern Physics},
  volume={87},
  number={2},
  pages={307--346},
  year={2015},
  publisher={APS}
}

@article{preskill1998reliable,
  title={Reliable quantum computers},
  author={Preskill, John},
  journal={Proceedings of the Royal Society of London. Series A: Mathematical, Physical and Engineering Sciences},
  volume={454},
  number={1969},
  pages={385--410},
  year={1998},
  publisher={The Royal Society}
}

@article{mohseni2024build,
  title={How to build a quantum supercomputer: Scaling from hundreds to millions of qubits},
  author={Mohseni, Masoud and Scherer, Artur and Johnson, K Grace and Wertheim, Oded and Otten, Matthew and Anand, Namit and Aadit, Navid Anjum and Alexeev, Yuri and Ben-Shach, Gilad and Bresniker, Kirk M and others},
  journal={arXiv preprint arXiv:2411.10406},
  year={2024}
}

@book{david2017computer,
  title={Computer organization and design: the hardware/software interface},
  author={David A, David A Patterson and John, L and others},
  year={2017},
  publisher={ELSEVEIR}
}

@article{fowler2012surface,
  title={Surface codes: Towards practical large-scale quantum computation},
  author={Fowler, Austin G and Mariantoni, Matteo and Martinis, John M and Cleland, Andrew N},
  journal={Physical Review A—Atomic, Molecular, and Optical Physics},
  volume={86},
  number={3},
  pages={032324},
  year={2012},
  publisher={APS}
}

@article{gidney2021factor,
  title={How to factor 2048 bit RSA integers in 8 hours using 20 million noisy qubits},
  author={Gidney, Craig and Eker{\aa}, Martin},
  journal={Quantum},
  volume={5},
  pages={433},
  year={2021},
  publisher={Verein zur F{\"o}rderung des Open Access Publizierens in den Quantenwissenschaften}
}

@article{litinski2019game,
  title={A game of surface codes: Large-scale quantum computing with lattice surgery},
  author={Litinski, Daniel},
  journal={Quantum},
  volume={3},
  pages={128},
  year={2019},
  publisher={Verein zur F{\"o}rderung des Open Access Publizierens in den Quantenwissenschaften}
}

@article{ryan2021realization,
  title={Realization of real-time fault-tolerant quantum error correction},
  author={Ryan-Anderson, Ciaran and Bohnet, Justin G and Lee, Kenny and Gresh, Daniel and Hankin, Aaron and Gaebler, John P and Francois, David and Chernoguzov, Alexander and Lucchetti, Dominic and Brown, Natalie C and others},
  journal={Physical Review X},
  volume={11},
  number={4},
  pages={041058},
  year={2021},
  publisher={APS}
}

@article{jones2012layered,
  title={Layered architecture for quantum computing},
  author={Jones, N Cody and Van Meter, Rodney and Fowler, Austin G and McMahon, Peter L and Kim, Jungsang and Ladd, Thaddeus D and Yamamoto, Yoshihisa},
  journal={Physical Review X},
  volume={2},
  number={3},
  pages={031007},
  year={2012},
  publisher={APS}
}

@article{koch2007charge,
  title={Charge-insensitive qubit design derived from the Cooper pair box},
  author={Koch, Jens and Yu, Terri M and Gambetta, Jay and Houck, Andrew A and Schuster, David I and Majer, Johannes and Blais, Alexandre and Devoret, Michel H and Girvin, Steven M and Schoelkopf, Robert J},
  journal={Physical Review A—Atomic, Molecular, and Optical Physics},
  volume={76},
  number={4},
  pages={042319},
  year={2007},
  publisher={APS}
}

@article{acharyaQuantumError2025,
  title = {Quantum Error Correction below the Surface Code Threshold},
  author = {Acharya, Rajeev and Abanin, Dmitry A. and {Aghababaie-Beni}, Laleh and Aleiner, Igor and Andersen, Trond I. and Ansmann, Markus and Arute, Frank and Arya, Kunal and Asfaw, Abraham and Astrakhantsev, Nikita and Atalaya, Juan and Babbush, Ryan and Bacon, Dave and Ballard, Brian and Bardin, Joseph C. and Bausch, Johannes and Bengtsson, Andreas and Bilmes, Alexander and Blackwell, Sam and Boixo, Sergio and Bortoli, Gina and Bourassa, Alexandre and Bovaird, Jenna and Brill, Leon and Broughton, Michael and Browne, David A. and Buchea, Brett and Buckley, Bob B. and Buell, David A. and Burger, Tim and Burkett, Brian and Bushnell, Nicholas and Cabrera, Anthony and Campero, Juan and Chang, Hung-Shen and Chen, Yu and Chen, Zijun and Chiaro, Ben and Chik, Desmond and Chou, Charina and Claes, Jahan and Cleland, Agnetta Y. and Cogan, Josh and Collins, Roberto and Conner, Paul and Courtney, William and Crook, Alexander L. and Curtin, Ben and Das, Sayan and Davies, Alex and De Lorenzo, Laura and Debroy, Dripto M. and Demura, Sean and Devoret, Michel and Di Paolo, Agustin and Donohoe, Paul and Drozdov, Ilya and Dunsworth, Andrew and Earle, Clint and Edlich, Thomas and Eickbusch, Alec and Elbag, Aviv Moshe and Elzouka, Mahmoud and Erickson, Catherine and Faoro, Lara and Farhi, Edward and Ferreira, Vinicius S. and Burgos, Leslie Flores and Forati, Ebrahim and Fowler, Austin G. and Foxen, Brooks and Ganjam, Suhas and Garcia, Gonzalo and Gasca, Robert and Genois, {\'E}lie and Giang, William and Gidney, Craig and Gilboa, Dar and Gosula, Raja and Dau, Alejandro Grajales and Graumann, Dietrich and Greene, Alex and Gross, Jonathan A. and Habegger, Steve and Hall, John and Hamilton, Michael C. and Hansen, Monica and Harrigan, Matthew P. and Harrington, Sean D. and Heras, Francisco J. H. and Heslin, Stephen and Heu, Paula and Higgott, Oscar and Hill, Gordon and Hilton, Jeremy and Holland, George and Hong, Sabrina and Huang, Hsin-Yuan and Huff, Ashley and Huggins, William J. and Ioffe, Lev B. and Isakov, Sergei V. and Iveland, Justin and Jeffrey, Evan and Jiang, Zhang and Jones, Cody and Jordan, Stephen and Joshi, Chaitali and Juhas, Pavol and Kafri, Dvir and Kang, Hui and Karamlou, Amir H. and Kechedzhi, Kostyantyn and Kelly, Julian and Khaire, Trupti and Khattar, Tanuj and Khezri, Mostafa and Kim, Seon and Klimov, Paul V. and Klots, Andrey R. and Kobrin, Bryce and Kohli, Pushmeet and Korotkov, Alexander N. and Kostritsa, Fedor and Kothari, Robin and Kozlovskii, Borislav and Kreikebaum, John Mark and Kurilovich, Vladislav D. and Lacroix, Nathan and Landhuis, David and {Lange-Dei}, Tiano and Langley, Brandon W. and Laptev, Pavel and Lau, Kim-Ming and Le Guevel, Lo{\"i}ck and Ledford, Justin and Lee, Joonho and Lee, Kenny and Lensky, Yuri D. and Leon, Shannon and Lester, Brian J. and Li, Wing Yan and Li, Yin and Lill, Alexander T. and Liu, Wayne and Livingston, William P. and Locharla, Aditya and Lucero, Erik and Lundahl, Daniel and Lunt, Aaron and Madhuk, Sid and Malone, Fionn D. and Maloney, Ashley and Mandr{\`a}, Salvatore and Manyika, James and Martin, Leigh S. and Martin, Orion and Martin, Steven and Maxfield, Cameron and McClean, Jarrod R. and McEwen, Matt and Meeks, Seneca and Megrant, Anthony and Mi, Xiao and Miao, Kevin C. and Mieszala, Amanda and Molavi, Reza and Molina, Sebastian and Montazeri, Shirin and Morvan, Alexis and Movassagh, Ramis and Mruczkiewicz, Wojciech and Naaman, Ofer and Neeley, Matthew and Neill, Charles and Nersisyan, Ani and Neven, Hartmut and Newman, Michael and Ng, Jiun How and Nguyen, Anthony and Nguyen, Murray and Ni, Chia-Hung and Niu, Murphy Yuezhen and O'Brien, Thomas E. and Oliver, William D. and Opremcak, Alex and Ottosson, Kristoffer and Petukhov, Andre and Pizzuto, Alex and Platt, John and Potter, Rebecca and Pritchard, Orion and Pryadko, Leonid P. and Quintana, Chris and Ramachandran, Ganesh and Reagor, Matthew J. and Redding, John and Rhodes, David M. and Roberts, Gabrielle and Rosenberg, Eliott and Rosenfeld, Emma and Roushan, Pedram and Rubin, Nicholas C. and Saei, Negar and Sank, Daniel and Sankaragomathi, Kannan and Satzinger, Kevin J. and Schurkus, Henry F. and Schuster, Christopher and Senior, Andrew W. and Shearn, Michael J. and Shorter, Aaron and Shutty, Noah and Shvarts, Vladimir and Singh, Shraddha and Sivak, Volodymyr and Skruzny, Jindra and Small, Spencer and Smelyanskiy, Vadim and Smith, W. Clarke and Somma, Rolando D. and Springer, Sofia and Sterling, George and Strain, Doug and Suchard, Jordan and Szasz, Aaron and Sztein, Alex and Thor, Douglas and Torres, Alfredo and Torunbalci, M. Mert and Vaishnav, Abeer and Vargas, Justin and Vdovichev, Sergey and Vidal, Guifre and Villalonga, Benjamin and Heidweiller, Catherine Vollgraff and Waltman, Steven and Wang, Shannon X. and Ware, Brayden and Weber, Kate and Weidel, Travis and White, Theodore and Wong, Kristi and Woo, Bryan W. K. and Xing, Cheng and Yao, Z. Jamie and Yeh, Ping and Ying, Bicheng and Yoo, Juhwan and Yosri, Noureldin and Young, Grayson and Zalcman, Adam and Zhang, Yaxing and Zhu, Ningfeng and Zobrist, Nicholas and {Google Quantum AI and Collaborators}},
  year = 2025,
  journal = {Nature},
  volume = {638},
  number = {8052},
  pages = {920--926},
  publisher = {Nature Publishing Group},
  doi = {10.1038/s41586-024-08449-y}
}

@article{anHighFidelity2022,
  title = {High {{Fidelity State Preparation}} and {{Measurement}} of {{Ion Hyperfine Qubits}} with \${{I}}{$>\backslash$}frac\textbraceleft 1\textbraceright\textbraceleft 2\textbraceright\$},
  author = {An, Fangzhao Alex and Ransford, Anthony and Schaffer, Andrew and Sletten, Lucas R. and Gaebler, John and Hostetter, James and Vittorini, Grahame},
  year = 2022,
  journal = {Physical Review Letters},
  volume = {129},
  number = {13},
  pages = {130501},
  publisher = {American Physical Society},
  doi = {10.1103/PhysRevLett.129.130501}
}

@misc{bergholmPennyLaneAutomatic2022,
  title = {{{PennyLane}}: {{Automatic}} Differentiation of Hybrid Quantum-Classical Computations},
  author = {Bergholm, Ville and Izaac, Josh and Schuld, Maria and Gogolin, Christian and Ahmed, Shahnawaz and Ajith, Vishnu and Alam, M. Sohaib and {Alonso-Linaje}, Guillermo and AkashNarayanan, B. and Asadi, Ali and Arrazola, Juan Miguel and Azad, Utkarsh and Banning, Sam and Blank, Carsten and Bromley, Thomas R. and Cordier, Benjamin A. and Ceroni, Jack and Delgado, Alain and Matteo, Olivia Di and Dusko, Amintor and Garg, Tanya and Guala, Diego and Hayes, Anthony and Hill, Ryan and Ijaz, Aroosa and Isacsson, Theodor and Ittah, David and Jahangiri, Soran and Jain, Prateek and Jiang, Edward and Khandelwal, Ankit and Kottmann, Korbinian and Lang, Robert A. and Lee, Christina and Loke, Thomas and Lowe, Angus and McKiernan, Keri and Meyer, Johannes Jakob and {Monta{\~n}ez-Barrera}, J. A. and Moyard, Romain and Niu, Zeyue and O'Riordan, Lee James and Oud, Steven and Panigrahi, Ashish and Park, Chae-Yeun and Polatajko, Daniel and Quesada, Nicol{\'a}s and Roberts, Chase and S{\'a}, Nahum and Schoch, Isidor and Shi, Borun and Shu, Shuli and Sim, Sukin and Singh, Arshpreet and Strandberg, Ingrid and Soni, Jay and Sz{\'a}va, Antal and Thabet, Slimane and {Vargas-Hern{\'a}ndez}, Rodrigo A. and Vincent, Trevor and Vitucci, Nicola and Weber, Maurice and Wierichs, David and Wiersema, Roeland and Willmann, Moritz and Wong, Vincent and Zhang, Shaoming and Killoran, Nathan},
  year = 2022,
  eprint = {1811.04968},
  archivePrefix = {arXiv}
}

@article{bluvsteinLogicalQuantum2024a,
  title = {Logical Quantum Processor Based on Reconfigurable Atom Arrays},
  author = {Bluvstein, Dolev and Evered, Simon J. and Geim, Alexandra A. and Li, Sophie H. and Zhou, Hengyun and Manovitz, Tom and Ebadi, Sepehr and Cain, Madelyn and Kalinowski, Marcin and Hangleiter, Dominik and Bonilla Ataides, J. Pablo and Maskara, Nishad and Cong, Iris and Gao, Xun and Sales Rodriguez, Pedro and Karolyshyn, Thomas and Semeghini, Giulia and Gullans, Michael J. and Greiner, Markus and Vuleti{\'c}, Vladan and Lukin, Mikhail D.},
  year = 2024,
  journal = {Nature},
  volume = {626},
  number = {7997},
  pages = {58--65},
  publisher = {Nature Publishing Group},
  doi = {10.1038/s41586-023-06927-3}
}

@article{bluvsteinQuantumProcessor2022a,
  title = {A Quantum Processor Based on Coherent Transport of Entangled Atom Arrays},
  author = {Bluvstein, Dolev and Levine, Harry and Semeghini, Giulia and Wang, Tout T. and Ebadi, Sepehr and Kalinowski, Marcin and Keesling, Alexander and Maskara, Nishad and Pichler, Hannes and Greiner, Markus and Vuleti{\'c}, Vladan and Lukin, Mikhail D.},
  year = 2022,
  journal = {Nature},
  volume = {604},
  number = {7906},
  pages = {451--456},
  publisher = {Nature Publishing Group},
  doi = {10.1038/s41586-022-04592-6}
}

@article{bravyiHighthresholdLowoverhead2024a,
  title = {High-Threshold and Low-Overhead Fault-Tolerant Quantum Memory},
  author = {Bravyi, Sergey and Cross, Andrew W. and Gambetta, Jay M. and Maslov, Dmitri and Rall, Patrick and Yoder, Theodore J.},
  year = 2024,
  journal = {Nature},
  volume = {627},
  number = {8005},
  pages = {778--782},
  publisher = {Nature Publishing Group},
  doi = {10.1038/s41586-024-07107-7}
}

@article{bravyiUniversalQuantum2005,
  title = {Universal Quantum Computation with Ideal {{Clifford}} Gates and Noisy Ancillas},
  author = {Bravyi, Sergey and Kitaev, Alexei},
  year = 2005,
  journal = {Physical Review A},
  volume = {71},
  number = {2},
  pages = {022316},
  publisher = {American Physical Society},
  doi = {10.1103/PhysRevA.71.022316}
}

@article{calderbankGoodQuantum1996,
  title = {Good Quantum Error-Correcting Codes Exist},
  author = {Calderbank, A. R. and Shor, Peter W.},
  year = 1996,
  journal = {Physical Review A},
  volume = {54},
  number = {2},
  pages = {1098--1105},
  publisher = {American Physical Society},
  doi = {10.1103/PhysRevA.54.1098}
}

@article{campbellRoadsFaulttolerant2017,
  title = {Roads towards Fault-Tolerant Universal Quantum Computation},
  author = {Campbell, Earl T. and Terhal, Barbara M. and Vuillot, Christophe},
  year = 2017,
  journal = {Nature},
  volume = {549},
  number = {7671},
  pages = {172--179},
  doi = {10.1038/nature23460}
}

@article{chamberlandTopologicalSubsystem2020,
  title = {Topological and {{Subsystem Codes}} on {{Low-Degree Graphs}} with {{Flag Qubits}}},
  author = {Chamberland, Christopher and Zhu, Guanyu and Yoder, Theodore J. and Hertzberg, Jared B. and Cross, Andrew W.},
  year = 2020,
  journal = {Physical Review X},
  volume = {10},
  number = {1},
  pages = {011022},
  publisher = {American Physical Society},
  doi = {10.1103/PhysRevX.10.011022}
}

@article{chowCircuitBasedLeakagetoErasure2024,
  title = {Circuit-{{Based Leakage-to-Erasure Conversion}} in a {{Neutral-Atom Quantum Processor}}},
  author = {Chow, Matthew N. H. and Buchemmavari, Vikas and Omanakuttan, Sivaprasad and Little, Bethany J. and Pandey, Saurabh and Deutsch, Ivan H. and Jau, Yuan-Yu},
  year = 2024,
  journal = {PRX Quantum},
  volume = {5},
  number = {4},
  pages = {040343},
  publisher = {American Physical Society},
  doi = {10.1103/PRXQuantum.5.040343}
}

@article{ciracQuantumComputations1995,
  title = {Quantum {{Computations}} with {{Cold Trapped Ions}}},
  author = {Cirac, J. I. and Zoller, P.},
  year = 1995,
  journal = {Physical Review Letters},
  volume = {74},
  number = {20},
  pages = {4091--4094},
  publisher = {American Physical Society},
  doi = {10.1103/PhysRevLett.74.4091}
}

@article{daguerreExperimentalDemonstration2025,
  title = {Experimental {{Demonstration}} of {{High-Fidelity Logical Magic States}} from {{Code Switching}}},
  author = {Daguerre, Lucas and {Blume-Kohout}, Robin and Brown, Natalie C. and Hayes, David and Kim, Isaac H.},
  year = 2025,
  journal = {Physical Review X},
  volume = {15},
  number = {4},
  pages = {041008},
  publisher = {American Physical Society},
  doi = {10.1103/dck4-x9c2}
}

@book{dingQuantumComputer2022,
  title = {Quantum {{Computer Systems}}: {{Research}} for {{Noisy Intermediate-Scale Quantum Computers}}},
  author = {Ding, Yongshan and Chong, Frederic T.},
  year = 2022,
  publisher = {Springer Nature},
  googlebooks = {k4ZyEAAAQBAJ}
}

@article{eastinRestrictionsTransversal2009,
  title = {Restrictions on {{Transversal Encoded Quantum Gate Sets}}},
  author = {Eastin, Bryan and Knill, Emanuel},
  year = 2009,
  journal = {Physical Review Letters},
  volume = {102},
  number = {11},
  pages = {110502},
  publisher = {American Physical Society},
  doi = {10.1103/PhysRevLett.102.110502}
}

@article{eganFaulttolerantControl2021,
  title = {Fault-Tolerant Control of an Error-Corrected Qubit},
  author = {Egan, Laird and Debroy, Dripto M. and Noel, Crystal and Risinger, Andrew and Zhu, Daiwei and Biswas, Debopriyo and Newman, Michael and Li, Muyuan and Brown, Kenneth R. and Cetina, Marko and Monroe, Christopher},
  year = 2021,
  journal = {Nature},
  volume = {598},
  number = {7880},
  pages = {281--286},
  publisher = {Nature Publishing Group},
  doi = {10.1038/s41586-021-03928-y}
}

@article{everedHighfidelityParallel2023b,
  title = {High-Fidelity Parallel Entangling Gates on a Neutral-Atom Quantum Computer},
  author = {Evered, Simon J. and Bluvstein, Dolev and Kalinowski, Marcin and Ebadi, Sepehr and Manovitz, Tom and Zhou, Hengyun and Li, Sophie H. and Geim, Alexandra A. and Wang, Tout T. and Maskara, Nishad and Levine, Harry and Semeghini, Giulia and Greiner, Markus and Vuleti{\'c}, Vladan and Lukin, Mikhail D.},
  year = 2023,
  journal = {Nature},
  volume = {622},
  number = {7982},
  pages = {268--272},
  publisher = {Nature Publishing Group},
  doi = {10.1038/s41586-023-06481-y}
}

@article{fowlerSurfaceCodes2012,
  title = {Surface Codes: {{Towards}} Practical Large-Scale Quantum Computation},
  author = {Fowler, Austin G. and Mariantoni, Matteo and Martinis, John M. and Cleland, Andrew N.},
  year = 2012,
  journal = {Physical Review A},
  volume = {86},
  number = {3},
  pages = {032324},
  publisher = {American Physical Society},
  doi = {10.1103/PhysRevA.86.032324}
}

@article{gaoEstablishingNew2025,
  title = {Establishing a {{New Benchmark}} in {{Quantum Computational Advantage}} with 105-Qubit {{Zuchongzhi}} 3.0 {{Processor}}},
  author = {Gao, Dongxin and Fan, Daojin and Zha, Chen and Bei, Jiahao and Cai, Guoqing and Cai, Jianbin and Cao, Sirui and Chen, Fusheng and Chen, Jiang and Chen, Kefu and Chen, Xiawei and Chen, Xiqing and Chen, Zhe and Chen, Zhiyuan and Chen, Zihua and Chu, Wenhao and Deng, Hui and Deng, Zhibin and Ding, Pei and Ding, Xun and Ding, Zhuzhengqi and Dong, Shuai and Dong, Yupeng and Fan, Bo and Fu, Yuanhao and Gao, Song and Ge, Lei and Gong, Ming and Gui, Jiacheng and Guo, Cheng and Guo, Shaojun and Guo, Xiaoyang and Han, Lianchen and He, Tan and Hong, Linyin and Hu, Yisen and Huang, He-Liang and Huo, Yong-Heng and Jiang, Tao and Jiang, Zuokai and Jin, Honghong and Leng, Yunxiang and Li, Dayu and Li, Dongdong and Li, Fangyu and Li, Jiaqi and Li, Jinjin and Li, Junyan and Li, Junyun and Li, Na and Li, Shaowei and Li, Wei and Li, Yuhuai and Li, Yuan and Liang, Futian and Liang, Xuelian and Liao, Nanxing and Lin, Jin and Lin, Weiping and Liu, Dailin and Liu, Hongxiu and Liu, Maliang and Liu, Xinyu and Liu, Xuemeng and Liu, Yancheng and Lou, Haoxin and Ma, Yuwei and Meng, Lingxin and Mou, Hao and Nan, Kailiang and Nie, Binghan and Nie, Meijuan and Ning, Jie and Niu, Le and Peng, Wenyi and Qian, Haoran and Rong, Hao and Rong, Tao and Shen, Huiyan and Shen, Qiong and Su, Hong and Su, Feifan and Sun, Chenyin and Sun, Liangchao and Sun, Tianzuo and Sun, Yingxiu and Tan, Yimeng and Tan, Jun and Tang, Longyue and Tu, Wenbing and Wan, Cai and Wang, Jiafei and Wang, Biao and Wang, Chang and Wang, Chen and Wang, Chu and Wang, Jian and Wang, Liangyuan and Wang, Rui and Wang, Shengtao and Wang, Xiaomin and Wang, Xinzhe and Wang, Xunxun and Wang, Yeru and Wei, Zuolin and Wei, Jiazhou and Wu, Dachao and Wu, Gang and Wu, Jin and Wu, Shengjie and Wu, Yulin and Xie, Shiyong and Xin, Lianjie and Xu, Yu and Xue, Chun and Yan, Kai and Yang, Weifeng and Yang, Xinpeng and Yang, Yang and Ye, Yangsen and Ye, Zhenping and Ying, Chong and Yu, Jiale and Yu, Qinjing and Yu, Wenhu and Zeng, Xiangdong and Zhan, Shaoyu and Zhang, Feifei and Zhang, Haibin and Zhang, Kaili and Zhang, Pan and Zhang, Wen and Zhang, Yiming and Zhang, Yongzhuo and Zhang, Lixiang and Zhao, Guming and Zhao, Peng and Zhao, Xianhe and Zhao, Xintao and Zhao, Youwei and Zhao, Zhong and Zheng, Luyuan and Zhou, Fei and Zhou, Liang and Zhou, Na and Zhou, Naibin and Zhou, Shifeng and Zhou, Shuang and Zhou, Zhengxiao and Zhu, Chengjun and Zhu, Qingling and Zou, Guihong and Zou, Haonan and Zhang, Qiang and Lu, Chao-Yang and Peng, Cheng-Zhi and Zhu, Xiaobo and Pan, Jian-Wei},
  year = 2025,
  journal = {Physical Review Letters},
  volume = {134},
  number = {9},
  pages = {090601},
  publisher = {American Physical Society},
  doi = {10.1103/PhysRevLett.134.090601}
}

@article{gidneyEfficientMagic2019,
  title = {Efficient Magic State Factories with a Catalyzed $\vert\mathrm{CCZ}\rangle$ to $2\vert \mathrm{T}\rangle$ Transformation},
  author = {Gidney, Craig and Fowler, Austin G.},
  year = 2019,
  journal = {Quantum},
  volume = {3},
  pages = {135},
  publisher = {Verein zur F\"orderung des Open Access Publizierens in den Quantenwissenschaften},
  doi = {10.22331/q-2019-04-30-135}
}

@misc{gidneyMagicState2024a,
  title = {Magic State Cultivation: Growing {{T}} States as Cheap as {{CNOT}} Gates},
  author = {Gidney, Craig and Shutty, Noah and Jones, Cody},
  year = 2024,
  eprint = {2409.17595},
  archivePrefix = {arXiv}
}

@article{gottesmanClassQuantum1996,
  title = {Class of Quantum Error-Correcting Codes Saturating the Quantum {{Hamming}} Bound},
  author = {Gottesman, Daniel},
  year = 1996,
  journal = {Physical Review A},
  volume = {54},
  number = {3},
  pages = {1862--1868},
  publisher = {American Physical Society},
  doi = {10.1103/PhysRevA.54.1862}
}

@article{heExperimentalQuantum2025,
  title = {Experimental {{Quantum Error Correction}} below the {{Surface Code Threshold}} via {{All-Microwave Leakage Suppression}}},
  author = {He, Tan and Lin, Weiping and Wang, Rui and Li, Yuan and Bei, Jiahao and Cai, Jianbin and Cao, Sirui and Chen, Danning and Chen, Kefu and Chen, Xiawei and Chen, Zhe and Chen, Zhiyuan and Chen, Zihua and Chu, Wenhao and Deng, Hui and Ding, Xun and Ding, Zhuzhengqi and Fan, Bo and Fan, Daojin and Fu, Yuanhao and Gao, Dongxin and Gong, Ming and Gui, Jiacheng and Guo, Cheng and Guo, Shaojun and Han, Lianchen and Hong, Linyin and Hu, Yisen and Huang, He-Liang and Huo, Yong-Heng and Jiang, Chenyan and Jiang, Lei and Jiang, Tao and Jiang, Zuokai and Jin, Honghong and Li, Dayu and Li, Dongdong and Li, Jiaqi and Li, Jinjin and Li, Junyan and Li, Junyun and Li, Na and Li, Shaowei and Li, Yuhuai and Liang, Futian and Liao, Nanxing and Lin, Jin and Liu, Ke and Liu, Maliang and Liu, Yancheng and Lou, Haoxin and Ma, Yuwei and Nan, Kailiang and Nie, Meijuan and Niu, Le and Peng, Wenyi and Qian, Haoran and Rong, Hao and Rong, Tao and Shen, Huiyan and Shen, Qiong and Su, Hong and Su, Feifan and Sun, Chenyin and Sun, Liangchao and Sun, Tianzuo and Sun, Yingxiu and Tan, Yimeng and Tan, Jun and Tu, Wenbing and Wang, Jiafei and Wang, Biao and Wang, Chang and Wang, Chen and Wang, Chu and Wang, Jian and Wang, Shengtao and Wang, Xinzhe and Wei, Zuolin and Wu, Dachao and Wu, Gang and Wu, Yulin and Xu, Yu and Xue, Chun and Yan, Kai and Yan, Xin and Yang, Weifeng and Yang, Xinpeng and Yang, Yang and Ye, Yangsen and Ye, Zhenping and Yi, Zhengzhong and Ying, Chong and Yu, Jiale and Yu, Qinjing and Zeng, Xiangdong and Zha, Chen and Zhan, Shaoyu and Zhang, Haibin and Zhang, He and Zhang, Kaili and Zhang, Wen and Zhang, Yiming and Zhang, Yongzhuo and Zhang, Ziying and Zhao, Guming and Zhao, Xintao and Zhao, Youwei and Zhao, Zhong and Zheng, Luyuan and Zhou, Fei and Zhou, Liang and Zhou, Na and Zhou, Naibin and Zhu, Chengjun and Zhu, Qingling and Zou, Guihong and Zou, Haonan and Zhang, Qiang and Lu, Chao-Yang and Peng, Cheng-Zhi and Chen, Fusheng and Zhu, XiaoBo and Pan, Jian-Wei},
  year = 2025,
  journal = {Physical Review Letters},
  volume = {135},
  number = {26},
  pages = {260601},
  publisher = {American Physical Society},
  doi = {10.1103/rqkg-dw31}
}

@misc{ivezicIBMUnveils2023,
  title = {{{IBM Unveils Condor}}: 1,121-{{Qubit Quantum Processor}}},
  author = {Ivezic, Marin},
  year = 2023,
  url = {https://postquantum.com/industry-news/ibm-condor/},
}

@misc{ivezicIBMUnveils2024,
  title = {{{IBM Unveils}} 156-{{Qubit}} `{{Heron R2}}' {{Quantum Processor}}},
  author = {Ivezic, Marin},
  year = 2024,
  url = {https://postquantum.com/industry-news/ibm-heron-r2-quantum/},
}

@article{javadi2024quantum,
  title={Quantum computing with Qiskit},
  author={Javadi-Abhari, Ali and Treinish, Matthew and Krsulich, Kevin and Wood, Christopher J and Lishman, Jake and Gacon, Julien and Martiel, Simon and Nation, Paul D and Bishop, Lev S and Cross, Andrew W and others},
  journal={arXiv preprint arXiv:2405.08810},
  year={2024}
}

@article{kitaevFaulttolerantQuantum2003,
  title = {Fault-Tolerant Quantum Computation by Anyons},
  author = {Kitaev, A. {\relax Yu}.},
  year = 2003,
  journal = {Annals of Physics},
  volume = {303},
  number = {1},
  pages = {2--30},
  doi = {10.1016/S0003-4916(02)00018-0}
}

@inproceedings{koboriLSQCAResourceEfficient2025,
  title = {{{LSQCA}}: {{Resource-Efficient Load}}/{{Store Architecture}} for {{Limited-Scale Fault-Tolerant Quantum Computing}}},
  booktitle = {2025 {{IEEE International Symposium}} on {{High Performance Computer Architecture}} ({{HPCA}})},
  author = {Kobori, Takumi and Suzuki, Yasunari and Ueno, Yosuke and Tanimoto, Teruo and Todo, Synge and Tokunaga, Yuuki},
  year = 2025,
  pages = {304--320},
  doi = {10.1109/HPCA61900.2025.00033}
}

@article{krinnerRealizingRepeated2022,
  title = {Realizing Repeated Quantum Error Correction in a Distance-Three Surface Code},
  author = {Krinner, Sebastian and Lacroix, Nathan and Remm, Ants and Di Paolo, Agustin and Genois, Elie and Leroux, Catherine and Hellings, Christoph and Lazar, Stefania and Swiadek, Francois and Herrmann, Johannes and Norris, Graham J. and Andersen, Christian Kraglund and M{\"u}ller, Markus and Blais, Alexandre and Eichler, Christopher and Wallraff, Andreas},
  year = 2022,
  journal = {Nature},
  volume = {605},
  number = {7911},
  pages = {669--674},
  publisher = {Nature Publishing Group},
  doi = {10.1038/s41586-022-04566-8}
}

@article{litinskiMagicState2019,
  title = {Magic {{State Distillation}}: {{Not}} as {{Costly}} as {{You Think}}},
  author = {Litinski, Daniel},
  year = 2019,
  journal = {Quantum},
  volume = {3},
  pages = {205},
  publisher = {Verein zur F\"orderung des Open Access Publizierens in den Quantenwissenschaften},
  doi = {10.22331/q-2019-12-02-205}
}

@article{löschnauerScalableHighFidelity2025,
  title = {Scalable, {{High-Fidelity All-Electronic Control}} of {{Trapped-Ion Qubits}}},
  author = {L{\"o}schnauer, C.M. and Mosca Toba, J. and Hughes, A.C. and King, S.A. and Weber, M.A. and Srinivas, R. and Matt, R. and Nourshargh, R. and Allcock, D.T.C. and Ballance, C.J. and Matthiesen, C. and Malinowski, M. and Harty, T.P.},
  year = 2025,
  journal = {PRX Quantum},
  volume = {6},
  number = {4},
  pages = {040313},
  publisher = {American Physical Society},
  doi = {10.1103/h4wk-v31j}
}

@article{marxer999Fidelity2026,
  title = {Above 99.9\% {{Fidelity Single-Qubit Gates}}, {{Two-Qubit Gates}}, and {{Readout}} in a {{Single Superconducting Quantum Device}}},
  author = {Marxer, Fabian and Mro{\.z}ek, Jakub and Andersson, Joona and Abdurakhimov, Leonid and Adam, Janos and Bergholm, Ville and Beriwal, Rohit and Chan, Chun Fai and Dahl, Saga and Das, Soumya Ranjan and Deppe, Frank and Fedorets, Olexiy and Gao, Zheming and Gomez Frieiro, Alejandro and Gusenkova, Daria and Guthrie, Andrew and Hiltunen, Tuukka and Hsu, Hao and Hyypp{\"a}, Eric and Ikonen, Joni and Inel, Sinan and Jolin, Shan W. and Karis, Azad and Kim, Seung-Goo and Kindel, William and Komlev, Anton and Koistinen, Miikka and Kokkoniemi, Roope and Kumar, Snigdha and Ku, Hsiang-Sheng and Lamprich, Julia and Laine, Sami and Landra, Alessandro and Lee, Lan-Hsuan and Lethif, Nizar and Liebermann, Per and Liu, Wei and Mitra, Kunal and Myll{\"a}ri, Tuomas and {Ockeloen-Korppi}, Caspar and Orell, Tuure and Plyshch, Alexander and R{\"a}bin{\"a}, Jukka and Rebello, Arthur and Renger, Michael and Reentil{\"a}, Outi and Ritvas, Jussi and Saarinen, Sampo and Salmenkivi, Otto and Sarsby, Matthew and Savytskyi, Mykhailo and Selinmaa, Ville and Steggles, Matthew and Takala, Eelis and Takmakov, Ivan and Tarasinski, Brian and Tuorila, Jani and V{\"a}limaa, Alpo and Verjauw, Jeroen and Wesdorp, Jaap and Wurz, Nicola and Qiu, Wei and Zhu, Lihuang and Hassel, Juha and Heinsoo, Johannes and Geresdi, Attila and Veps{\"a}l{\"a}inen, Antti},
  year = 2026,
  journal = {PRX Quantum},
  volume = {7},
  number = {2},
  pages = {020333},
  publisher = {American Physical Society},
  doi = {10.1103/n86s-2b88}
}

@article{monroeDemonstrationFundamental1995,
  title = {Demonstration of a {{Fundamental Quantum Logic Gate}}},
  author = {Monroe, C. and Meekhof, D. M. and King, B. E. and Itano, W. M. and Wineland, D. J.},
  year = 1995,
  journal = {Physical Review Letters},
  volume = {75},
  number = {25},
  pages = {4714--4717},
  publisher = {American Physical Society},
  doi = {10.1103/PhysRevLett.75.4714}
}

@article{mosesRaceTrackTrappedIon2023,
  title = {A {{Race-Track Trapped-Ion Quantum Processor}}},
  author = {Moses, S. A. and Baldwin, C. H. and Allman, M. S. and Ancona, R. and Ascarrunz, L. and Barnes, C. and Bartolotta, J. and Bjork, B. and Blanchard, P. and Bohn, M. and Bohnet, J. G. and Brown, N. C. and Burdick, N. Q. and Burton, W. C. and Campbell, S. L. and Campora, J. P. and Carron, C. and Chambers, J. and Chan, J. W. and Chen, Y. H. and Chernoguzov, A. and Chertkov, E. and Colina, J. and Curtis, J. P. and Daniel, R. and DeCross, M. and Deen, D. and Delaney, C. and Dreiling, J. M. and Ertsgaard, C. T. and Esposito, J. and Estey, B. and Fabrikant, M. and Figgatt, C. and Foltz, C. and {Foss-Feig}, M. and Francois, D. and Gaebler, J. P. and Gatterman, T. M. and Gilbreth, C. N. and Giles, J. and Glynn, E. and Hall, A. and Hankin, A. M. and Hansen, A. and Hayes, D. and Higashi, B. and Hoffman, I. M. and Horning, B. and Hout, J. J. and Jacobs, R. and Johansen, J. and Jones, L. and Karcz, J. and Klein, T. and Lauria, P. and Lee, P. and Liefer, D. and Lu, S. T. and Lucchetti, D. and Lytle, C. and Malm, A. and Matheny, M. and Mathewson, B. and Mayer, K. and Miller, D. B. and Mills, M. and Neyenhuis, B. and Nugent, L. and Olson, S. and Parks, J. and Price, G. N. and Price, Z. and Pugh, M. and Ransford, A. and Reed, A. P. and Roman, C. and Rowe, M. and {Ryan-Anderson}, C. and Sanders, S. and Sedlacek, J. and Shevchuk, P. and Siegfried, P. and Skripka, T. and Spaun, B. and Sprenkle, R. T. and Stutz, R. P. and Swallows, M. and Tobey, R. I. and Tran, A. and Tran, T. and Vogt, E. and Volin, C. and Walker, J. and Zolot, A. M. and Pino, J. M.},
  year = 2023,
  journal = {Physical Review X},
  volume = {13},
  number = {4},
  pages = {041052},
  publisher = {American Physical Society},
  doi = {10.1103/PhysRevX.13.041052}
}

@article{nesterovCnotGates2022,
  title = {Cnot {{Gates}} for {{Fluxonium Qubits}} via {{Selective Darkening}} of {{Transitions}}},
  author = {Nesterov, Konstantin N. and Wang, Chen and Manucharyan, Vladimir E. and Vavilov, Maxim G.},
  year = 2022,
  journal = {Physical Review Applied},
  volume = {18},
  number = {3},
  pages = {034063},
  publisher = {American Physical Society},
  doi = {10.1103/PhysRevApplied.18.034063}
}

@article{pecorariHighrateQuantum2025,
  title = {High-Rate Quantum {{LDPC}} Codes for Long-Range-Connected Neutral Atom Registers},
  author = {Pecorari, Laura and Jandura, Sven and Brennen, Gavin K. and Pupillo, Guido},
  year = 2025,
  journal = {Nature Communications},
  volume = {16},
  number = {1},
  pages = {1111},
  publisher = {Nature Publishing Group},
  doi = {10.1038/s41467-025-56255-5}
}

@article{pinoDemonstrationTrappedion2021,
  title = {Demonstration of the Trapped-Ion Quantum {{CCD}} Computer Architecture},
  author = {Pino, J. M. and Dreiling, J. M. and Figgatt, C. and Gaebler, J. P. and Moses, S. A. and Allman, M. S. and Baldwin, C. H. and {Foss-Feig}, M. and Hayes, D. and Mayer, K. and {Ryan-Anderson}, C. and Neyenhuis, B.},
  year = 2021,
  journal = {Nature},
  volume = {592},
  number = {7853},
  pages = {209--213},
  doi = {10.1038/s41586-021-03318-4}
}

@article{postlerDemonstrationFaulttolerant2022,
  title = {Demonstration of Fault-Tolerant Universal Quantum Gate Operations},
  author = {Postler, Lukas and Heu{$\beta$}en, Sascha and Pogorelov, Ivan and Rispler, Manuel and Feldker, Thomas and Meth, Michael and Marciniak, Christian D. and Stricker, Roman and Ringbauer, Martin and Blatt, Rainer and Schindler, Philipp and M{\"u}ller, Markus and Monz, Thomas},
  year = 2022,
  journal = {Nature},
  volume = {605},
  number = {7911},
  pages = {675--680},
  publisher = {Nature Publishing Group},
  doi = {10.1038/s41586-022-04721-1}
}

@article{raussendorfFaultTolerantQuantum2007,
  title = {Fault-{{Tolerant Quantum Computation}} with {{High Threshold}} in {{Two Dimensions}}},
  author = {Raussendorf, Robert and Harrington, Jim},
  year = 2007,
  journal = {Physical Review Letters},
  volume = {98},
  number = {19},
  pages = {190504},
  publisher = {American Physical Society},
  doi = {10.1103/PhysRevLett.98.190504}
}

@misc{reichardtLogicalComputation2024,
  title = {Logical Computation Demonstrated with a Neutral Atom Quantum Processor},
  author = {Reichardt, Ben W. and Paetznick, Adam and Aasen, David and Basov, Ivan and {Bello-Rivas}, Juan M. and Bonderson, Parsa and Chao, Rui and van Dam, Wim and Hastings, Matthew B. and Paz, Andres and da Silva, Marcus P. and Sundaram, Aarthi and Svore, Krysta M. and Vaschillo, Alexander and Wang, Zhenghan and Zanner, Matt and Cairncross, William B. and Chen, Cheng-An and Crow, Daniel and Kim, Hyosub and Kindem, Jonathan M. and King, Jonathan and McDonald, Michael and Norcia, Matthew A. and Ryou, Albert and Stone, Mark and Wadleigh, Laura and Barnes, Katrina and Battaglino, Peter and Bohdanowicz, Thomas C. and Booth, Graham and Brown, Andrew and Brown, Mark O. and Cassella, Kayleigh and Coxe, Robin and Epstein, Jeffrey M. and Feldkamp, Max and Griger, Christopher and Halperin, Eli and Heinz, Andre and Hummel, Frederic and Jaffe, Matthew and Jones, Antonia M. W. and Kapit, Eliot and Kotru, Krish and Lauigan, Joseph and Li, Ming and Marjanovic, Jan and Megidish, Eli and Meredith, Matthew and Morshead, Ryan and Muniz, Juan A. and Narayanaswami, Sandeep and Nishiguchi, Ciro and Paule, Timothy and Pawlak, Kelly A. and Pudenz, Kristen L. and P{\'e}rez, David Rodr{\'i}guez and Simon, Jon and Smull, Aaron and Stack, Daniel and Urbanek, Miroslav and van de Veerdonk, Ren{\'e} J. M. and Vendeiro, Zachary and Weverka, Robert T. and Wilkason, Thomas and Wu, Tsung-Yao and Xie, Xin and {Zalys-Geller}, Evan and Zhang, Xiaogang and Bloom, Benjamin J.},
  year = 2024,
  eprint = {2411.11822},
  archivePrefix = {arXiv}
}

@article{rossOptimalAncillafree2016,
  title = {Optimal Ancilla-Free {{Clifford}}+{{T}} Approximation of z-Rotations},
  author = {Ross, Neil J. and Selinger, Peter},
  year = 2016,
  journal = {Quantum Information and Computation},
  volume = {16},
  number = {11\&12},
  pages = {901--953},
  doi = {10.26421/QIC16.11-12-1}
}

@misc{ryan-andersonHighfidelityFaulttolerant2024,
  title = {High-Fidelity and {{Fault-tolerant Teleportation}} of a {{Logical Qubit}} Using {{Transversal Gates}} and {{Lattice Surgery}} on a {{Trapped-ion Quantum Computer}}},
  author = {{Ryan-Anderson}, C. and Brown, N. C. and Baldwin, C. H. and Dreiling, J. M. and Foltz, C. and Gaebler, J. P. and Gatterman, T. M. and Hewitt, N. and Holliman, C. and Horst, C. V. and Johansen, J. and Lucchetti, D. and Mengle, T. and Matheny, M. and Matsuoka, Y. and Mayer, K. and Mills, M. and Moses, S. A. and Neyenhuis, B. and Pino, J. and Siegfried, P. and Stutz, R. P. and Walker, J. and Hayes, D.},
  year = 2024,
  eprint = {2404.16728},
  archivePrefix = {arXiv}
}

@article{ryan-andersonRealizationRealTime2021,
  title = {Realization of {{Real-Time Fault-Tolerant Quantum Error Correction}}},
  author = {{Ryan-Anderson}, C. and Bohnet, J. G. and Lee, K. and Gresh, D. and Hankin, A. and Gaebler, J. P. and Francois, D. and Chernoguzov, A. and Lucchetti, D. and Brown, N. C. and Gatterman, T. M. and Halit, S. K. and Gilmore, K. and Gerber, J. A. and Neyenhuis, B. and Hayes, D. and Stutz, R. P.},
  year = 2021,
  journal = {Physical Review X},
  volume = {11},
  number = {4},
  pages = {041058},
  publisher = {American Physical Society},
  doi = {10.1103/PhysRevX.11.041058}
}

@article{salesrodriguezExperimentalDemonstration2025,
  title = {Experimental Demonstration of Logical Magic State Distillation},
  author = {Sales Rodriguez, Pedro and Robinson, John M. and Jepsen, Paul Niklas and He, Zhiyang and Duckering, Casey and Zhao, Chen and Wu, Kai-Hsin and Campo, Joseph and Bagnall, Kevin and Kwon, Minho and Karolyshyn, Thomas and Weinberg, Phillip and Cain, Madelyn and Evered, Simon J. and Geim, Alexandra A. and Kalinowski, Marcin and Li, Sophie H. and Manovitz, Tom and {Amato-Grill}, Jesse and Basham, James I. and Bernstein, Liane and Braverman, Boris and Bylinskii, Alexei and Choukri, Adam and DeAngelo, Robert J. and Fang, Fang and Fieweger, Connor and Frederick, Paige and Haines, David and Hamdan, Majd and Hammett, Julian and Hsu, Ning and Hu, Ming-Guang and Huber, Florian and Jia, Ningyuan and Kedar, Dhruv and Kornja{\v c}a, Milan and Liu, Fangli and Long, John and Lopatin, Jonathan and Lopes, Pedro L. S. and Luo, Xiu-Zhe and Macr{\`i}, Tommaso and Markovi{\'c}, Ognjen and {Mart{\'i}nez-Mart{\'i}nez}, Luis A. and Meng, Xianmei and Ostermann, Stefan and Ostroumov, Evgeny and Paquette, David and Qiang, Zexuan and Shofman, Vadim and Singh, Anshuman and Singh, Manuj and Sinha, Nandan and Thoreen, Henry and Wan, Noel and Wang, Yiping and {Waxman-Lenz}, Daniel and Wong, Tak and Wurtz, Jonathan and Zhdanov, Andrii and Zheng, Laurent and Greiner, Markus and Keesling, Alexander and Gemelke, Nathan and Vuleti{\'c}, Vladan and Kitagawa, Takuya and Wang, Sheng-Tao and Bluvstein, Dolev and Lukin, Mikhail D. and Lukin, Alexander and Zhou, Hengyun and Cant{\'u}, Sergio H.},
  year = 2025,
  journal = {Nature},
  volume = {645},
  number = {8081},
  pages = {620--625},
  publisher = {Nature Publishing Group},
  doi = {10.1038/s41586-025-09367-3}
}

@inproceedings{schererAutomatedDesign2026,
  title = {Automated {{Design}}, {{Compilation}}, and {{Performance Benchmarking}} for {{Fault-Tolerant Quantum Computer Architectures Using TopQAD}}},
  booktitle = {{{APS Global Physics Summit}} 2026},
  author = {Scherer, Artur and Balaniuk, Severyn and Dagnew, Gebremedhin A. and Gabbassov, Einar and Gera, Saneer and Kavaki, Ali H. and Khalid, Abdullah and Kong, Xiangzhou and Kramer, Mia and Kulchytskyy, Bohdan and Lotfi, Pouria and Nguyen, Huy-Anh and Nguyen, Kevin and Olfert, Katiemarie and Silva, Allyson and Torosov, Boyan and Wang, Yumeng and Webb, Zak and Yang, Chan-Woo and Zhang, Xiangyi and Ronagh, Pooya},
  year = 2026,
  url = {https://meetings-archive.aps.org/smt/2026/mar-p11/11/}
}

@inproceedings{shorFaulttolerantQuantum1996,
  title = {Fault-Tolerant Quantum Computation},
  booktitle = {Proceedings of 37th {{Conference}} on {{Foundations}} of {{Computer Science}}},
  author = {Shor, P.W.},
  year = 1996,
  pages = {56--65},
  doi = {10.1109/SFCS.1996.548464}
}

@article{somoroffMillisecondCoherence2023,
  title = {Millisecond {{Coherence}} in a {{Superconducting Qubit}}},
  author = {Somoroff, Aaron and Ficheux, Quentin and Mencia, Raymond A. and Xiong, Haonan and Kuzmin, Roman and Manucharyan, Vladimir E.},
  year = 2023,
  journal = {Physical Review Letters},
  volume = {130},
  number = {26},
  pages = {267001},
  publisher = {American Physical Society},
  doi = {10.1103/PhysRevLett.130.267001}
}

@article{steaneSimpleQuantum1996,
  title = {Simple Quantum Error-Correcting Codes},
  author = {Steane, A. M.},
  year = 1996,
  journal = {Physical Review A},
  volume = {54},
  number = {6},
  pages = {4741--4751},
  publisher = {American Physical Society},
  doi = {10.1103/PhysRevA.54.4741}
}

@article{sundaresanDemonstratingMultiround2023,
  title = {Demonstrating Multi-Round Subsystem Quantum Error Correction Using Matching and Maximum Likelihood Decoders},
  author = {Sundaresan, Neereja and Yoder, Theodore J. and Kim, Youngseok and Li, Muyuan and Chen, Edward H. and Harper, Grace and Thorbeck, Ted and Cross, Andrew W. and C{\'o}rcoles, Antonio D. and Takita, Maika},
  year = 2023,
  journal = {Nature Communications},
  volume = {14},
  number = {1},
  pages = {2852},
  publisher = {Nature Publishing Group},
  doi = {10.1038/s41467-023-38247-5}
}

@inproceedings{svoreEnablingScalable2018,
  title = {Q\#: {{Enabling Scalable Quantum Computing}} and {{Development}} with a {{High-level DSL}}},
  booktitle = {Proceedings of the {{Real World Domain Specific Languages Workshop}} 2018},
  author = {Svore, Krysta and Geller, Alan and Troyer, Matthias and Azariah, John and Granade, Christopher and Heim, Bettina and Kliuchnikov, Vadym and Mykhailova, Mariia and Paz, Andres and Roetteler, Martin},
  year = 2018,
  series = {{{RWDSL2018}}},
  pages = {1--10},
  publisher = {Association for Computing Machinery},
  address = {New York, NY, USA},
  doi = {10.1145/3183895.3183901}
}

@misc{toshioSTARMagicMutation2026,
  title = {{{STAR-Magic Mutation}}: {{Even More Efficient Analog Rotation Gates}} for {{Early Fault-Tolerant Quantum Computer}}},
  author = {Toshio, Riki and Kanasugi, Shota and Fujisaki, Jun and Oshima, Hirotaka and Sato, Shintaro and Fujii, Keisuke},
  year = 2026,
  eprint = {2603.22891},
  archivePrefix = {arXiv}
}

@inproceedings{uenoQECOOLOnLine2021,
  title = {{{QECOOL}}: {{On-Line Quantum Error Correction}} with a {{Superconducting Decoder}} for {{Surface Code}}},
  booktitle = {2021 58th {{ACM}}/{{IEEE Design Automation Conference}} ({{DAC}})},
  author = {Ueno, Yosuke and Kondo, Masaaki and Tanaka, Masamitsu and Suzuki, Yasunari and Tabuchi, Yutaka},
  year = 2021,
  pages = {451--456},
  doi = {10.1109/DAC18074.2021.9586326}
}

@inproceedings{uenoQULATISQuantum2022a,
  title = {{{QULATIS}}: {{A Quantum Error Correction Methodology}} toward {{Lattice Surgery}}},
  booktitle = {2022 {{IEEE International Symposium}} on {{High-Performance Computer Architecture}} ({{HPCA}})},
  author = {Ueno, Yosuke and Kondo, Masaaki and Tanaka, Masamitsu and Suzuki, Yasunari and Tabuchi, Yutaka},
  year = 2022,
  pages = {274--287},
  doi = {10.1109/HPCA53966.2022.00028}
}

@article{wangSingleIon2021,
  title = {Single Ion Qubit with Estimated Coherence Time Exceeding One Hour},
  author = {Wang, Pengfei and Luan, Chun-Yang and Qiao, Mu and Um, Mark and Zhang, Junhua and Wang, Ye and Yuan, Xiao and Gu, Mile and Zhang, Jingning and Kim, Kihwan},
  year = 2021,
  journal = {Nature Communications},
  volume = {12},
  number = {1},
  pages = {233},
  publisher = {Nature Publishing Group},
  doi = {10.1038/s41467-020-20330-w}
}

@misc{wegmannChipmunqFaultTolerant2026,
  title = {Chipmunq: {{A Fault-Tolerant Compiler}} for {{Chiplet Quantum Architectures}}},
  author = {Wegmann, Peter and {\'S}wierkowska, Aleksandra and Giortamis, Emmanouil and Bhatotia, Pramod},
  year = 2026,
  eprint = {2603.16389},
  archivePrefix = {arXiv}
}

@article{winelandQuantumInformation2003,
  title = {Quantum Information Processing with Trapped Ions},
  author = {Wineland, D. J. and Barrett, M. and Britton, J. and Chiaverini, J. and DeMarco, B. and Itano, W. M. and Jelenkovi{\'c}, B. and Langer, C. and Leibfried, D. and Meyer, V. and Rosenband, T. and Sch{\"a}tz, T.},
  editor = {Knight, P. L. and Hinds, E. A. and Plenio, M. B.},
  year = 2003,
  journal = {Philosophical Transactions of the Royal Society A: Mathematical, Physical and Engineering Sciences},
  volume = {361},
  number = {1808},
  pages = {1349--1361},
  doi = {10.1098/rsta.2003.1205}
}

@article{wuErasureConversion2022,
  title = {Erasure Conversion for Fault-Tolerant Quantum Computing in Alkaline Earth {{Rydberg}} Atom Arrays},
  author = {Wu, Yue and Kolkowitz, Shimon and Puri, Shruti and Thompson, Jeff D.},
  year = 2022,
  journal = {Nature Communications},
  volume = {13},
  number = {1},
  pages = {4657},
  publisher = {Nature Publishing Group},
  doi = {10.1038/s41467-022-32094-6}
}

@article{xuConstantoverheadFaulttolerant2024,
  title = {Constant-Overhead Fault-Tolerant Quantum Computation with Reconfigurable Atom Arrays},
  author = {Xu, Qian and Bonilla Ataides, J. Pablo and Pattison, Christopher A. and Raveendran, Nithin and Bluvstein, Dolev and Wurtz, Jonathan and Vasi{\'c}, Bane and Lukin, Mikhail D. and Jiang, Liang and Zhou, Hengyun},
  year = 2024,
  journal = {Nature Physics},
  volume = {20},
  number = {7},
  pages = {1084--1090},
  publisher = {Nature Publishing Group},
  doi = {10.1038/s41567-024-02479-z}
}

@misc{yinFlexionAdaptive2025,
  title = {Flexion: {{Adaptive In-Situ Encoding}} for {{On-Demand QEC}} in {{Ion Trap Systems}}},
  author = {Yin, Keyi and Fang, Xiang and Chen, Zhuo and Li, Ang and Hayes, David and Kaur, Eneet and Nejabati, Reza and Haeffner, Hartmut and Campbell, Wes and Hudson, Eric and Palsberg, Jens and Humble, Travis and Ding, Yufei},
  year = 2025,
  eprint = {2504.16303},
  archivePrefix = {arXiv}
}

@misc{yoderTourGross2025,
  title = {Tour de Gross: {{A}} Modular Quantum Computer Based on Bivariate Bicycle Codes},
  author = {Yoder, Theodore J. and Schoute, Eddie and Rall, Patrick and Pritchett, Emily and Gambetta, Jay M. and Cross, Andrew W. and Carroll, Malcolm and Beverland, Michael E.},
  year = 2025,
  eprint = {2506.03094},
  archivePrefix = {arXiv}
}

@misc{zengErrorstructuretailoredEarly2025,
  title = {Error-Structure-Tailored Early Fault-Tolerant Quantum Computing},
  author = {Zeng, Pei and Zheng, Guo and Xu, Qian and Jiang, Liang},
  year = 2025,
  eprint = {2511.19983},
  archivePrefix = {arXiv}
}

@article{zhangClassicalArchitecture2023,
  title = {A {{Classical Architecture}} for {{Digital Quantum Computers}}},
  author = {Zhang, Fang and Zhu, Xing and Chao, Rui and Huang, Cupjin and Kong, Linghang and Chen, Guoyang and Ding, Dawei and Feng, Haishan and Gao, Yihuai and Ni, Xiaotong and Qiu, Liwei and Wei, Zhe and Yang, Yueming and Zhao, Yang and Shi, Yaoyun and Zhang, Weifeng and Zhou, Peng and Chen, Jianxin},
  year = 2023,
  journal = {ACM Transactions on Quantum Computing},
  volume = {5},
  number = {1},
  pages = {3:1--3:24},
  doi = {10.1145/3626199}
}

@article{chamberlandFaultTolerantQuantum2018,
  title = {Fault-{{Tolerant Quantum Computing}} in the {{Pauli}} or {{Clifford Frame}} with {{Slow Error Diagnostics}}},
  author = {Chamberland, Christopher and Iyer, Pavithran and Poulin, David},
  year = 2018,
  journal = {Quantum},
  volume = {2},
  pages = {43},
  doi = {10.22331/q-2018-01-04-43}
}

@article{onMultilayeredPauli2023,
  title = {A Multilayered {{Pauli}} Tracking Architecture for Lattice Surgery-based Logical Qubits},
  author = {On, Jin-Ho and Kim, Chei-Yol and Oh, Soo-Cheol and Lee, Sang-Min and Cha, Gyu-Il},
  year = 2023,
  journal = {ETRI Journal},
  volume = {45},
  number = {3},
  pages = {462--478},
  doi = {10.4218/etrij.2022-0037}
}

@inproceedings{riesebosPauliFrames2017,
  title = {Pauli {{Frames}} for {{Quantum Computer Architectures}}},
  booktitle = {Proceedings of the 54th {{Annual Design Automation Conference}} 2017},
  author = {Riesebos, L. and Fu, X. and Varsamopoulos, S. and Almudever, C. G. and Bertels, K.},
  year = 2017,
  pages = {1--6},
  publisher = {ACM},
  address = {Austin TX USA},
  doi = {10.1145/3061639.3062300}
}

@article{cross2022openqasm,
  title={OpenQASM 3: A broader and deeper quantum assembly language},
  author={Cross, Andrew and Javadi-Abhari, Ali and Alexander, Thomas and De Beaudrap, Niel and Bishop, Lev S and Heidel, Steven and Ryan, Colm A and Sivarajah, Prasahnt and Smolin, John and Gambetta, Jay M and others},
  journal={ACM Transactions on Quantum Computing},
  volume={3},
  number={3},
  pages={1--50},
  year={2022},
  publisher={ACM New York, NY}
}

@misc{qirbook-full,
  title={The QIR Book: A Comprehensive Guide to the Quantum Intermediate Representation},
  author={QIR Alliance Community Contributors},
  year={2023},
  month={4},
  howpublished={Open-access online technical handbook},
  url={https://www.qir-alliance.org/qir-book/},
  note={Last updated April 2023; accessed June 20, 2026}
}

@misc{caldwell2025platform,
      title={Platform Architecture for Tight Coupling of High-Performance Computing with Quantum Processors}, 
      author={Shane A. Caldwell and Moein Khazraee and Elena Agostini and Tom Lassiter and Corey Simpson and Omri Kahalon and Mrudula Kanuri and Jin-Sung Kim and Sam Stanwyck and Muyuan Li and Jan Olle and Christopher Chamberland and Ben Howe and Bruno Schmitt and Justin G. Lietz and Alex McCaskey and Jun Ye and Ang Li and Alicia B. Magann and Corey I. Ostrove and Kenneth Rudinger and Robin Blume-Kohout and Kevin Young and Nathan E. Miller and Yilun Xu and Gang Huang and Irfan Siddiqi and John Lange and Christopher Zimmer and Travis Humble},
      year={2025},
      eprint={2510.25213},
      archivePrefix={arXiv},
      primaryClass={quant-ph},
      url={https://arxiv.org/abs/2510.25213}, 
}

@article{ho2018announcing,
  title={Announcing Cirq: an open source framework for NISQ algorithms},
  author={Ho, Alan and Bacon, Dave},
  journal={Google AI Blog},
  volume={18},
  year={2018}
}

@inproceedings{green2013quipper,
  title={Quipper: a scalable quantum programming language},
  author={Green, Alexander S and Lumsdaine, Peter LeFanu and Ross, Neil J and Selinger, Peter and Valiron, Beno{\^\i}t},
  booktitle={Proceedings of the 34th ACM SIGPLAN conference on Programming language design and implementation},
  pages={333--342},
  year={2013}
}

@article{steiger2018projectq,
  title={ProjectQ: an open source software framework for quantum computing},
  author={Steiger, Damian S and H{\"a}ner, Thomas and Troyer, Matthias},
  journal={Quantum},
  volume={2},
  pages={49},
  year={2018},
  publisher={Verein zur F{\"o}rderung des Open Access Publizierens in den Quantenwissenschaften}
}

@inproceedings{burgholzer2026munich,
  title={The Munich Quantum Software Stack: Connecting End Users, Integrating Diverse Quantum Technologies, Accelerating HPC},
  author={Burgholzer, Lukas and Echavarria, Jorge and Hopf, Patrick and Stade, Yannick and Rovara, Damian and Schmid, Ludwig and Kaya, Erc{\"u}ment and Mete, Burak and Farooqi, Muhammad Nufail and Chung, Minh and others},
  booktitle={Proceedings of the Supercomputing Asia and International Conference on High Performance Computing in Asia Pacific Region},
  pages={55--67},
  year={2026}
}

@article{smith2020open,
  title={An open-source, industrial-strength optimizing compiler for quantum programs},
  author={Smith, Robert S and Peterson, Eric C and Skilbeck, Mark G and Davis, Erik J},
  journal={Quantum Science \& Technology},
  volume={5},
  number={4},
  pages={044001},
  year={2020},
  publisher={IOP Publishing}
}

@misc{zapatabenchqfull,
  title={BenchQ: Toolchain for Benchmarking Fault-Tolerant Quantum Computation Resources},
  author={Watkins, Daniel and Paler, Alexandru and Devitt, Simon J. and Zapata Computing},
  year={2023},
  month={9},
  howpublished={Open-source GitHub repository},
  url={https://github.com/zapatacomputing/benchq},
  note={Accessed June 20, 2026}
}

@article{watkins2024high,
  title={A high performance compiler for very large scale surface code computations},
  author={Watkins, George and Nguyen, Hoang Minh and Watkins, Keelan and Pearce, Steven and Lau, Hoi-Kwan and Paler, Alexandru},
  journal={Quantum},
  volume={8},
  pages={1354},
  year={2024},
  publisher={Verein zur F{\"o}rderung des Open Access Publizierens in den Quantenwissenschaften}
}

@article{wong2025cross,
  title={A cross-platform execution engine for the quantum intermediate representation: E. Wong et al.},
  author={Wong, Elaine and Leyton-Ortega, Vicente and Claudino, Daniel and Johnson, Seth R and Adams, Austin J and Afrose, Sharmin and Gowrishankar, Meenambika and Cabrera, Anthony and Humble, Travis S},
  journal={The Journal of Supercomputing},
  volume={81},
  number={16},
  pages={1521},
  year={2025},
  publisher={Springer}
}

@article{sivarajah2021tuket,
  title={t$|$ket$>$: A retargetable compiler for {NISQ} devices},
  author={Sivarajah, Seyon and Dilkes, Silas and Cowtan, Alexander and Simmons, Will and Edgington, Alec and Duncan, Ross},
  journal={Quantum Science and Technology},
  volume={6},
  pages={014003},
  year={2021},
  publisher={IOP Publishing},
  doi={10.1088/2058-9565/ab8e92}
}

@misc{liu2026scalableopensourceqecsubmicrosecond,
      title={A Scalable Open-Source QEC System with Sub-Microsecond Decoding-Feedback Latency}, 
      author={Junyi Liu and Yi Lee and Yilun Xu and Gang Huang and Xiaodi Wu},
      year={2026},
      eprint={2603.16203},
      archivePrefix={arXiv},
      primaryClass={quant-ph},
      url={https://arxiv.org/abs/2603.16203}, 
}

\clearpage
\newpage
\onecolumngrid
\appendix

\end{document}